\documentclass{emulateapj}

\usepackage{graphicx}
\usepackage{txfonts}
\usepackage{natbib}
\graphicspath{{./figs/}}
\usepackage{multirow}   %,amsmath}
\usepackage[colorlinks,linkcolor=blue,anchorcolor=blue,citecolor=blue]{hyperref}
\usepackage[hyperref]{backref}

\begin{document}


\title{JVLA 1.5GHz continuum observation of CLASH clusters I: \\radio properties of the BCGs}
\author{Heng Yu\altaffilmark{1}, Paolo Tozzi\altaffilmark{2,1}, Reinout van Weeren\altaffilmark{3},
Elisabetta Liuzzo\altaffilmark{4}, Gabriele Giovannini\altaffilmark{5}, Megan Donahue\altaffilmark{6}, 
Italo Balestra\altaffilmark{7}, Piero Rosati\altaffilmark{8}, Manuel Aravena\altaffilmark{9}.}

\altaffiltext{1}{Department of Astronomy, Beijing Normal University, 100875, Beijing, China}
\altaffiltext{2}{INAF - Osservatorio Astrofisico di Arcetri, Largo E. Fermi, I-50125 Firenze, Italy}
\altaffiltext{3}{Harvard-Smithsonian Center for Astrophysics, 60 Garden Streen, Cambridge, MA 02138, USA}
\altaffiltext{4}{INAF - Istituto di Radioastronomia, Via P. Gobetti 101, 40129 Bologna, Italy}
\altaffiltext{5}{Universit\`{a} di Bologna - Via Zamboni 33, 40126 Bologna, Italy}
\altaffiltext{6}{Michigan State University, 567 Wilson Rd. Rm 3272, East Lansing, MI 48824, USA}
\altaffiltext{7}{INAF - Osservatorio Astronomico di Trieste, via G. B. Tiepolo 11, I-34143, Trieste, Italy}
\altaffiltext{8}{Universit\`{a} di Ferrara, Via Saragat 1, 44122 Ferrara, Italy}
\altaffiltext{9}{N\'ucleo de Astronom\'{\i}a, Facultad de Ingenier\'{\i}a y Ciencias, Universidad Diego Portales, Av. Ej\'ercito 441, Santiago, Chile}

\begin{abstract}
We present high-resolution ($\sim 1"$), 1.5 GHz continuum observations of the brightest cluster galaxies 
(BCGs) of 13 CLASH (Cluster Lensing And Supernova survey with Hubble)
clusters at $0.18<z<0.69$ with the Karl G. Jansky Very Large Array (JVLA).  
Radio emission is clearly detected and characterized for 
11 BCGs, while for two of them we obtain only upper limits to their radio flux ($<0.1$ mJy at 5$\sigma$ confidence level).  
We also consider five additional clusters whose BCG is detected in FIRST or NVSS.
We find radio powers in the range from $2\times 10^{23}$ to $\sim 10^{26}$ $W~Hz^{-1}$ and radio spectral indices
$\alpha_{1.5}^{30}$ (defined as the slope between 1.5 and 30 GHz) distributed from $\sim -1$
to $-0.25$ around the central value $\langle \alpha \rangle= - 0.68$.  The radio emission from the BCGs is 
resolved in three cases (Abell 383, MACS J1931, and RX J2129), and unresolved or marginally resolved in the remaining eight cases
observed with JVLA. In all the cases the BCGs are consistent with being powered by active galactic nuclei (AGN).  The radio 
power shows a positive correlation with the BCG star formation rate, and a negative correlation with the central entropy of
the surrounding intracluster medium (ICM) except in two cases (MACS J1206 and CL J1226).  
Finally, over the restricted range in radio power sampled by the CLASH BCGs, 
we observe a significant scatter between the radio power and the average mechanical power stored in the ICM cavities. 
\end{abstract}


\section{Introduction}
\label{sec:intro}

Brightest cluster galaxies (BCGs) are among the most massive galaxies in the universe, and their 
formation and evolution are intimately linked to the evolution of the host cluster \citep[see][for a recent 
overview of properties of local BCGs]{2014Lauer,2015aHogan}.  They usually live in the most active 
central cluster regions, show a small peculiar velocity with respect to other cluster members, and are 
often surrounded by a cool core.  However, in a few cases, significant offset from the X-ray center 
and relatively large peculiar velocity may be observed \citep[see][]{2014Lauer}.  
Their star formation history and nuclear activity are reflected 
in the chemical and thermodynamic properties of the X-ray emitting intracluster medium (ICM).
In relaxed clusters, where the BCG is close to the X-ray center, the ICM is heavily affected by the 
feedback from the central active galactic nucleus (AGN), which prevents runaway cooling of the ICM and provides
a direct explanation for the cooling-flow problem \citep{1994Fabian,2012Fabian}.
The signature of such feedback can be investigated in the X-ray band in terms of gas entropy
structure, radio plasma-filled cavities in the ICM, and distribution of heavy elements in the ICM.
Despite the sense that physical mechanisms contributing
to the feedback are now well established, the detailed physics of the energy balance between 
the different baryonic components (stars, hot gas, and cold gas) and the 
regulation of nuclear activity and its duty cycle in the BCG are still under investigation.  

In fact, the largest contribution to the  feedback in terms of energy budget is  
associated with the "mechanical-mode" nuclear activity, which consists in the production of 
extremely energetic radio jets or AGN outflows and winds created during accretion onto the supermassive black 
hole hosted by the BCGs. The accretion mechanism and the AGN feeding in massive halos have been 
modeled recently by several 
studies \citep[see][]{2012Gaspari,2013Gaspari,2015bVoit}.  In addition, radiative cooling appears to be efficiently 
quenched by AGN activity in cool cores \citep[e.g.][]{2009Mittal}.  Mechanical-mode feedback from supermassive 
black holes is invoked to explain the quenching of the potential massive cooling flow and the 
non-detection of cold gas below $\sim 2$ keV in the cluster 
cores, despite the inferred cooling time being much shorter than the cluster lifetime in a subset of cluster cores 
\citep{2006Peterson}.  Star formation is also observed to be quenched or significantly
suppressed, although with a significant time delay \citep[e.g.][]{2016Molendi}.  
This picture is reinforced by the large fraction of radio-luminous galaxies  among BCGs, which has been well established for 
many years \citep{1990Burns}, and by the fact that virtually every strong cool core cluster hosts a radio-loud BCG 
\citep{2009Sun,2015aHogan}.  It is found that BCGs are 10 times more likely to host an AGN than any other cluster galaxy, and 
about 3 times more likely than other cluster galaxies with comparable K-band luminosity \citep[][]{2007Lin}. 

In more detail, the relativistic jets and/or outflows inject mechanical energy into the 
ICM, creating buoyantly rising bubbles or cavities filled by radio lobes 
\citep[e.g.][]{2000McNamara,2012Hlavacek-Larrondo}. 
A significant fraction of this mechanical energy is expected to be transformed into internal energy of the ICM in the form 
of shock heating, turbulent motions, dissipation of sound waves, and turbulent mixing \citep[e.g.][]{2017Lau}.
The total mechanical energy associated with the cavities can be roughly estimated as 
the enthalpy $4PV$ where $P$ and $V$ are the ICM pressure and the cavity volume, respectively, and it appears 
to be of the same order as that needed to stop the cooling 
\citep[see also][]{2001Blanton,2004Birzan,2006Dunn,2007Wise,2007Sanders,2009Sanders}.
These studies have been possible thanks to the unambiguous detection of cavities in the ICM observed as 
round-shaped depressions in the X-ray emission, spatially overlapping with AGN lobes.  
The energetics of the mechanical feedback have been  systematically investigated at low and medium
redshifts \citep{2007Jetha,2008Birzan,2008Dunn,2010Blanton,2010Dunn,2011Sullivan,
2012Hlavacek-Larrondo,2016Shin} and pushed to the limits of detectability of X-ray cavities up to 
$z\sim 1.2$ thanks to the {\sl Chandra} follow-up of a sample of SZ-selected clusters
\citep{2015Hlavacek-Larrondo}.  Cool cores are expected to be present  from an early epoch
\citep[see][]{2010Santos,2017McDonald} 
and a gentle feedback should be in place since then. However, while the average mechanical energy associated with
feedback is sufficient to offset cooling, the process is expected to be intermittent.  
For example, the multiphase condensation and rain toward the central AGN as envisaged in the
chaotic cold accretion scenario \citep[see][]{2017Gaspari} predicts a flicker noise variability with 
a logarithmic slope of the power spectrum of $-1$, characteristic of fractal and chaotic phenomena.
The mechanical mode of AGN feedback is expected to be tightly self-regulated in most - if not all - BCGs, 
with frequent but not destructive outbursts, which appear to have a duty cycle close to unity 
\citep{2009Mittal,2015aHogan,2017Lau}.  In this picture, feedback can probably always be  tracked by radio emission, 
but the detailed mechanism that is responsible for the transfer of 
the mechanical energy to the ICM is still not fully understood, and the evolution of the feedback 
with cosmic time is poorly constrained.  Both aspects are of paramount importance in the framework 
of galaxy formation and evolution of the large scale structures of the universe.

In this respect, in-depth studies of BCGs and their complex environment using vastly different 
wavelengths are crucial to reach a comprehensive picture of the feedback phenomena.
A unique opportunity for studying BCG properties and their evolution is provided by the Cluster 
Lensing And Supernovae survey with Hubble \citep[CLASH][]{2012Postman}.  CLASH is a Hubble 
Space Telescope (HST) 524-orbit Multi-Cycle Treasury program to use the gravitational lensing properties of 
25 galaxy clusters to accurately constrain the baryonic mass and dark matter distributions in the 
cluster core and in the outskirts, to exploit their lensing properties to find highly magnified 
high-z galaxies, and to search for Type Ia supernovae at $z > 1$ to improve constraints on the time
dependence of the dark energy equation of state and the evolution of supernovae.  A total of 16
broadband filters, spanning the near-UV to near-IR, are employed for a 20-orbit campaign on 
each cluster.  In addition, CLASH clusters are observed in the X-ray band with {\sl Chandra} and 
XMM-Newton.  In particular, 
all the CLASH clusters have {\sl Chandra} imaging with medium-deep exposures (from 20 to 130 ks, 
with an average of 60 ks).  We already know that  X--ray {\sl Chandra} data of CLASH clusters often 
show structures in the inner 30 kpc, which corresponds to 10 arcsec at $z \sim 0.2$ and to 5 arcsec at 
$z \sim 0.6$.  The detection of X-ray cavities has already been reported in the literature for some of 
them individually: RXJ1532 by \citet{2008Dunn}, MACSJ1931 by \citet{2011Ehlert}, and MACSJ1423 by
\citet{2008Birzan}.  A recent systematic investigation by \citet{2016Shin} reported cavity detection from beta-model 
subtracted images for seven CLASH clusters (MACSJ1720, Abell 383, MACSJ0329, MACSJ0744 
in addition to those already mentioned).  All the clusters are also observed in the mid-infrared (MIR) with Herschel, 
in the near-infrared (NIR) with Spitzer, and in the optical with Subaru/Superime-Cam, 
and are also intensively followed-up in the optical band to obtain detailed spectra and 
securely confirm member galaxies thanks to a VLT large program (PI P. Rosati) in addition to 
spectroscopy on 5 northern clusters with the Large Binocular Telescope.   CLASH is the 
first large and representative sample of X-ray-selected clusters consistently observed with 
HST in 16 optical and NIR bands, and therefore stands out as one of the most ambitious observational projects 
on galaxy clusters ever attempted, with a strong legacy value.   Similar efforts are currently underway with the 
HST follow-up of 41 massive clusters X-ray-selected from the RELICS survey (PI D. Coe) and of 
a similar number of X-ray selected clusters from the MACS survey \citep[the SNAPshot survey,][]{2017Repp}.

Given the unprecedented combination of space- and ground-based data of the CLASH project, radio observations are a key 
ingredient toward a comprehensive investigation of the feedback processes.  In this paper, we present the first part of an 
observational campaign in the 1-2 GHz radio continuum with the Karl G. Jansky Very Large Array (JVLA).  Our goal is to 
characterize the radio properties of member galaxies in CLASH clusters, with a strong emphasis on the radio properties of the 
BCG and the connection with the surrounding ICM, to pave the way for a detailed investigation of the feedback processes 
in massive clusters.  
% Eventually, the link between the radio nuclear activity of the CLASH BCG and the surrounding 
% medium will be investigated thanks to the synergy of radio and X-ray data, as well as the other multi-wavelength data.
The paper is organized as follows. In Section \ref{sec:sample} we describe the sample.  In Section \ref{sec:obs} we describe 
the observations and the data reduction.  We present our results in Section \ref{sec:results}, and our conclusions are 
summarized in Section \ref{sec:conclusions}.  Throughout this paper we adopt the 7 yr WMAP cosmology, with 
$\rm \Omega_{m}$= 0.272, $\rm \Omega_{\Lambda}$ = 0.728 and $H_{0}$ = 70.4 km $\rm s^{-1}$ $\rm Mpc^{-1}$
\citep{komatsu2011}.  Quoted error bars always correspond to the 1$\sigma$ confidence level.

\section{Sample selection} \label{sec:sample}

The sample of CLASH clusters, originally selected on the basis of their large mass and magnification power of gravitational 
lensing, populate the intermediate redshift range $0.18 < z < 0.9$, corresponding 
to a look-back time interval of 2.4-5.7 Gyr, a period that has been poorly investigated so far. This is also
the epoch when most of the effects of the feedback are visible in terms of evolution of the cluster X-ray 
luminosity-temperature relation of the cluster \citep{2007Branchesi}.

Among the 25 clusters of the CLASH sample, only 
20 clusters appear dynamically relaxed.  The other 5 CLASH clusters are, in fact, dynamically disturbed, 
and were selected because of their higher lensing magnification factor. Therefore, 
they do not show well defined cluster cores centered on a dominant BCG. A deep JVLA observation of the merging 
cluster MACSJ0717 is presented in \citet{2017vanWeeren}.  In this work we focus on the 20 relaxed CLASH clusters that 
have a well-defined dominant BCG coincident with or very close to the peak of the X-ray cluster emission. 
Since our primary science goal is to investigate the relation between the BCG and core properties in 
massive clusters, we postpone the observation of merging clusters.  All the 20 relaxed CLASH clusters are observable 
from the VLA except one (RXJ2248).  Also, the cluster CLJ1226, with the highest redshift $z=0.89$,
was not in our accepted VLA sample because of a conflict with another program.
Therefore, we proposed to observe 18 clusters in L band (20 cm) and A 
configuration (JVLA proposal VLA-14A-040, AT441, PI P. Tozzi) with the aim of reaching a noise level of $\sim 0.01-0.02$ 
mJy/beam. Therefore, assuming a nominal detection threshold corresponding to a $S/N =5$, we aim at 
fluxes fainter $\sim 50$ and $\sim 20$ times deeper than the NRAO VLA Sky Survey \citep[NVSS
\footnote{NVSS is complete above $\sim 2.5 \, $mJy at 1.5 GHz for $DEC >-40^{\circ}$ 
(see http://www.cv.nrao.edu/nvss/).},][]{1998Condon} and than the Faint Images of the Radio Sky at Twenty-cm 
\citep[FIRST\footnote{The FIRST catalog released in 2014 December covers about 10,575 square degrees of sky both in 
the northern and southern hemispheres, with a detection threshold  of $\sim  1 $ mJy at 1.5 GHz 
(see http://sundog.stsci.edu/).}][]{2015Helfand} for point-like sources, respectively. The requirement to achieve this 
sensitivity corresponds roughly to an observation time of about $80$ minutes per field with the JVLA, including
overheads.  We choose the A 
configuration to achieve the maximum angular resolution of $\sim 1.3$ arcsec in the L band.

\begin{table*}[tp]
\centering
\scriptsize
\caption{BCG Counterparts of the Relaxed CLASH Cluster Sample}
\begin{tabular}{|c|c|c|c|c|c|c|}
 \hline
Name & R.A. & Decl. & z & Optical & NVSS $20"$ match & FIRST $2"$ match \\%& PLANCK  & TGSS &  WENSS \\
\hline
Abell 383 &  02:48:03.36 &  -03:31:44.7 &   0.1887$^1$ & 6dF J0248034-033145 & J024803-033143 & J024803.3-033144 \\ %& PSZ1 G177.64-53.52 & J024803.4-033143 \\ 
Abell 209 &  01:31:52.57 &  -13:36:38.8 &  0.2098$^2$ & 2MASX J01315250-1336409 & J013152-133659 & no coverage \\%& PSZ1 G159.81-73.47 & J013152.8-133700\\ 
Abell 1423 &  11:57:17.35 &  +33:36:39.6 & 0.2140$^3$  & 2MASX J11571737+3336399 & J115716+333644 & J115716.8+333629 \\%& PSZ1 G180.56+76.66 & J115716.7+333648 & B1154.7+3353\\ 
RXJ2129 &  21:29:39.94 &  +00:05:18.8 &  0.2339$^3$ & WISE J212939.98+000521.9 & J212940+000522 & J212939.9+000521 \\%& PSZ1 G053.65-34.49 & J212939.8+000520 \\ 
Abell 611 &  08:00:56.83 &  +36:03:24.1 &   0.2873$^4$& 2MASX J08005684+3603234 & no detection & no detection \\%& PSZ1 G184.70+28.92 &-\\ 
MS2137 &  21:40:15.18 &  -23:39:40.7 &  0.3130$^5$ & 2MASX J21401517-2339398 & J214014-233939 & no coverage \\  % & & -
RXJ1532 &  15:32:53.78 &  +30:20:58.7 &  0.3620$^6$ & SDSS J153253.78+302059.3 & J153253+302059 & J153253.7+302059  \\  % & & J153253.8+302058 & B1530.8+3030B
MACSJ1931 &  19:31:49.66 &  -26:34:34.0 &   0.352$^{10}$ & WISE J193149.63-263433.0 & no detection & no coverage \\%& PSZ1 G012.58-20.07 & J193149.6-263432\\ 
MACSJ1720 &  17:20:16.95 &  +35:36:23.6 &   0.387$^7$ & WISE J172016.75+353626.1 & J172016+353628 & J172016.7+353625 \\%& PSZ1 G059.51+33.06 & J172016.5+353625 & B1718.4+3538\\ 
MACSJ0429 &  04:29:36.10 &  -02:53:08.0 &0.399$^{11}$ & 2MASX J04293604-0253073 & J042936-025306 & no coverage \\  % & & J042935.9-025306 
MACSJ0329 &  03:29:41.68 &  -02:11:47.7 &   0.450$^{11}$& WISE J032941.57-021146.6 & J032941-021152 & no coverage \\ 
MACSJ1423 &  14:23:47.76 &  +24:04:40.5 &   0.5457$^6$& SDSS J142347.87+240442.4 & J142347+240439 & J142347.9+240442 \\ % & & J142347.9+240442
MACSJ0744 &  07:44:52.80 &  +39:27:24.4 &  0.6986$^{6,7}$& SDSS J074452.81+392726.7 & no detection & no detection \\ 
\hline
Abell 2261 &  17:22:27.25 &  +32:07:58.6 &   0.2229$^3$ & SDSS J172227.18+320757.2 & J172227+320757 & J172227.0+320758 \\% & PSZ1 G055.58+31.87 & J172226.8+320757　&　B1720.5+3210\\ 
RXJ2248  & 22:48:44.29  & -44:31:48.4  &  0.3471$^{12}$& WISE J224844.05-443150.7 & no coverage & no coverage\\
MACSJ1115 &  11:15:52.05  & +01:29:56.6  & 0.3520$^6$& SDSS J111551.90+012955.0 & J111551+012955 & J111551.8+012955 \\ % & & J111551.9+012956
MACSJ1206 &  12:06:12.28  & -08:48:02.4  &  0.4398$^9$ & WISE J120612.16-084803.1 & J120612--084802 & no coverage \\ % & & J120612.2-084803
RXJ1347  & 13:47:30.59  & -11:45:10.1  &  0.4495$^{12}$& WISE J134730.61-114509.5 & J134730--114508 & no coverage \\ %& & J134730.6-114508
MACSJ1311 &  13:11:01.67  & -03:10:39.5  & 0.4917$^6$  & SDSS J131101.79-031039.7 & no detection & no detection \\
ClJ1226 & 12:26:58.37 & +33:32:47.4 & 0.8908$^8$ & SDSS J122658.24+333248.5 & J122658+333244  & J122658.1+333248 \\
\hline
\end{tabular}
\label{tab:sam}
\tablecomments{The first 13 clusters are observed in the program 
VLA-14A-040.  The other 7 relaxed clusters 
are also included for completeness. We list the position each cluster (second and third columns) 
from \citet{2012Postman}, the BCG redshift (fourth columns), the optical counterpart of BCG (fifth columns),
and the radio counterpart candidate in the NVSS and FIRST catalogs (sixth and seventh column).  The optical counterpart is 
unambiguously assigned thanks to a visual comparison with HST images, while the preliminary radio counterpart 
candidates are obtained with a simple distance criterion with a matching radius of 20 and 2 arcsec for NVSS and FIRST, 
respectively.  ``No detection" means the field is observed but no potential counterpart is found 
within the matching radius.  ``No coverage" means that the field is not observed.}
\tablerefs{[1]~\citet{2014Geller},~[2]~VLT-VIMOS,~[3]~\citet{2013Rines},~[4]~\citet{2013Lemze},
~[5]~\citet{2000Bauer} ~[6]~SDSS, DR12 ~\citet{2015Alam}, ~[7]~\citet{2010Stern}. ~[8]~\citet{2013Jorgensen},
~[9]~\citet{2015Girardi}, ~[10]~\citet{2004Allen}, ~[11]~\citet{2008Stott},~[12]~\citet{2009Guzzo}}
\end{table*}

In 2014 we obtained data for only 14 out of 18 clusters. One of these targets (Abell 2261) was seriously affected by
radio frequency interference (RFI). As a result, no useful image was obtained.  
Therefore, we will present new data for 13  targets only\footnote{MACSJ1720 is partially 
affected by the same type of interference; however, we were able to obtain useful data, despite this field shows the largest 
noise.}. The observed targets are listed in Table \ref{tab:sam}, together with the other CLASH clusters 
included in the relaxed sample.  We plan to complete the observation of the entire CLASH sample with a future proposal, 
including the 5 merging CLASH clusters observable from the JVLA site.  In Table \ref{tab:sam} we also identify the optical 
counterparts of each cluster BCG found in optical or IR surveys among 6dFGS \citep{2004Jones}, 2MASS 
\citep{2006Skrutskie}, SDSS \citep{2000York}, and WISE \citep{2010Wright}.  All our clusters have a well-defined BCG with 
no ambiguous cases (e.g., a cluster with two comparable galaxies). 
In the fourth column of Table \ref{tab:sam} we list the cluster redshift published in the literature. 
% Note that we will use CLASH spectroscopic data when comparing radio and optical properties of the BCG.
In the sixth and seventh columns of Table \ref{tab:sam} we list the radio counterpart
candidates from NVSS and FIRST, respectively, that would be associated with the BCG by assuming a 
simple matching criterion based on the optical and radio position. In detail, we select the NVSS and 
FIRST source closest to the position of the optical counterpart within a radius of 20 arcsec and 2 arcsec 
for NVSS and FIRST, respectively.  A large matching radius is suggested also for very bright sources in NVSS, where 
% 40\% of the FWHM beam size is 20 arcsec, and 
the FWHM is 45 arcsec\footnote{See discussion by R. L. White on the NRAO Science Forum https://science.nrao.edu/forums.}.  
Since FIRST resolution is 5 arcsec on average, a matching radius of 2 arcsec is 
chosen for consistency with the radius of 20 arcsec used for NVSS sources.
With this conservative choice, among the sources observed with the JVLA, 10 out of 13 BCGs in our sample 
have a radio counterpart either in the NVSS or FIRST survey or both, while five fields do not have FIRST coverage. 
Among the seven sources not observed in our program, five and three have radio counterparts in NVSS and FIRST, respectively.

\begin{table*}
\label{tab:general}
 \caption{Observation and Calibration Parameters of Program VLA-14A-040}
\begin{center}
\begin{tabular}{|c|c|}
 \hline
Central frequency & 1.5 GHz \\
Configuration & A \\
No. of antennas & 27 \\
No. of spectral windows & 16 \\
Total bandwidth (GHz) &1.0 \\
No. of channels/spw & 64 \\
Total no. of channels & 1024 \\
Spectral window bandwidth (MHz) & 64 \\
Channel bandwidth (MHz) & 1.0 \\
Channel separation (MHz) & 0.5 \\
\hline
\end{tabular}
\end{center}
\end{table*}

\section{Observations and Data Reduction}
\label{sec:obs}

We present here new data on 13 clusters observed with the JVLA in A configuration from February 24th to April 24th, 2014.  
The A configuration has a maximum baseline of 36.4 km.  We used a bandwidth of 1 GHz centered at 1.5 GHz (L band). 
The largest angular size of a radio source detectable at 1.5 GHz with the A configuration is about 36 arcsec.  
The full-width half-maximum of the primary beam is $\theta_{PB}= 30$  arcmin.  The observing setup is summarized in 
Table \ref{tab:general}. Total exposure time, useful spectral windows, phase and gain calibrators,
beam size and noise level for each target are listed in Table \ref{tab:obs}.  Each cluster in our sample was observed for 
about 1 hour or slightly more.  The typical angular resolution (synthesized beam size) is  $\sim 1.3$ arcsec. 
We note that the noise level reached in our images at the aimpoint is on average $0.022$ mJy, about twice as large as 
the value of $0.01$ mJy that was the goal of the proposal.  The main reason for this noise level is the 
geostationary satellite belt (which is around DEC= 0  $\pm 10$ deg) which introduces a significant amount of extra 
RFI for five of our targets not accounted for in the proposal.  Moreover, for the two fields with the 
highest noise, RXJ1532 and MACSJ1720, where the noise level at the aimpoint is of the order of $0.07$ and $0.05$ 
mJy, respectively, the flux calibrator used for the observations was not optimal, and this causes an uncertain bandpass 
calibration.  In addition, half of the observation of MACSJ1720 was carried out with 6 spectral windows (spws), and most 
of them had to be omitted from the analysis. Finally, RXJ2129 and again MACSJ1720, have very bright and complex 
off-axis sources, which are difficult to clean.  Overall, the average noise level achieved in the 13 fields is low enough to 
reach our science goals, despite being a factor of $\sim 2$ larger than expected, 
and two fields having exceptionally high noise (more than five times the goal {\sl rms}).

\begin{table*}
\label{tab:obs}
\begin{center}
\caption{Data quality of Program VLA-14A-040}
 \begin{tabular}{|c|c|c|c|c|c|c|r|}
 \hline
\multirow{2}{*}{Cluster} & $T_{obs}$  
& \multicolumn{2}{|c|}{Calibrator}  &  rms  & \multicolumn{2}{|c|}{Beam size}\\
\cline{3-4}
\cline{6-7}  & (min)  & flux & phase & ($\mu$Jy) & arcsec $\times$ arcsec & degree\\
\hline
Abell 383 & 62.4   & 3C48 &  J0241-0815 & 19 &1.11 $\times$  1.01 & -11  \\ 
Abell 209 &  62.4 & 3C48 & J0132-1654 & 22 & 1.64 $\times$ 1.13 & 27\\ 
Abell 1423 &  63.5 & 3C286 & J1215+3448 &  18  & 1.58 $\times$  1.07 & 39 \\ 
RXJ2129 &  61.2 & 3C48 & J2136+0041 & 42 & 1.12 $\times$ 1.07 & -11\\ 
Abell 611 &  63.0 & 3C147  & J0751+3313 & 14 & 1.03 $\times$  0.95 & -68\\ 
MS2137-2353 &  58.8 & 3C48 & J2138-2439 & 19 & 1.76 $\times$  0.86 & -3\\ 
RXJ1532 &  62.4 &  3C295$^*$ & J1602+3326 & 47 & 1.67 $\times$  1.11 & 52\\ 
MACSJ1931 &  61.2 & 3C48 & J1924-2914 & 29 & 2.05 $\times$  0.94 & -169\\ 
MACSJ1720 &  57.9 & 3C295$^*$ & J1721+3542 & 69 & 1.34 $\times$  1.09 & 28 \\ 
MACSJ0429 &  57.7 & 3C48 & J0423-0120& 29  & 1.06 $\times$  1.02 & -14\\ 
MACSJ0329 &  61.3 & 3C48 & J0339-0146& 17  & 1.10 $\times$  1.04 & -28\\ 
MACSJ1423 &  62.4 & 3C286  & J1436+2321& 29 & 1.17 $\times$  1.11 & 72\\ 
MACSJ0744 &  61.8 & 3C147  & J0753+4231& 15 & 1.03 $\times$  0.97 & 25\\ 
\hline
\end{tabular}
\end{center}
\tablecomments{Total exposure time, calibrators, {\sl rms} noise at the aimpoint, 
beam size and orientation for the radio data of all the clusters observed in the program VLT-14A-040.}
\vspace{1ex}
\raggedright {\footnotesize * Because 3C295 is not a suitable flux calibrator for VLA configuration A,
we adopt the phase calibrator for RXJ1532 and MACSJ1720.
The flux of the phase calibrator J1602 is set to 2.9 Jy with an index of 0.15,
while flux of J1721 is 0.3 Jy with an index of 0.  Both indexes are fitted with VLA measurements in bands less than 2GHz.}
\end{table*}

Data calibration is performed with the reduction package Common Astronomy Software Applications (CASA, version 4.7.0) 
following standard JVLA procedures for low frequency wide-band, wide-field imaging data.  After applying the standard antenna 
position correction and the gain curve and opacity correction, the original data are processed with the Hanning smoothing.  Then 
we apply the {\tt rflag} algorithm to remove strong RFI. The RFI at the spectral window 8 is mostly caused by satellite 
communication, and is always stronger than the signal from calibrators.  Therefore, we mask spectral window 8 in all our 
observations.  After the bandpass correction and the gain correction, the resulting images employ natural weighting of the 
visibility data.  We consider a square field of view (FOV) of 30 arcmin on a side. The size of each pixel is set to 0.3 arcsec.
With these choices, the FOV fully covers the X-ray emission in {\sl Chandra} ACIS-I and the resolution is comparable to that of 
{\sl Chandra} at the aimpoint.

\begin{figure*}[htp]
\caption{Radio images overlapped with the Chandra contours (blue lines).  Central crosses indicate the position of the BCG 
obtained from the HST optical image. The FOV is  $1' \times 1'$. The small panel in the top right corner shows the enlarged 
central region with a FOV $10"$ across. The beam size is shown as a gray ellipse in the bottom left corner. The color scale 
ranges from 3$\sigma$ to the maximum flux in each field with a logarithmic step. 
These images are generated with APLpy \citep{2012Robitaille}.}
\label{fig:img}
\includegraphics[width=0.49\textwidth]{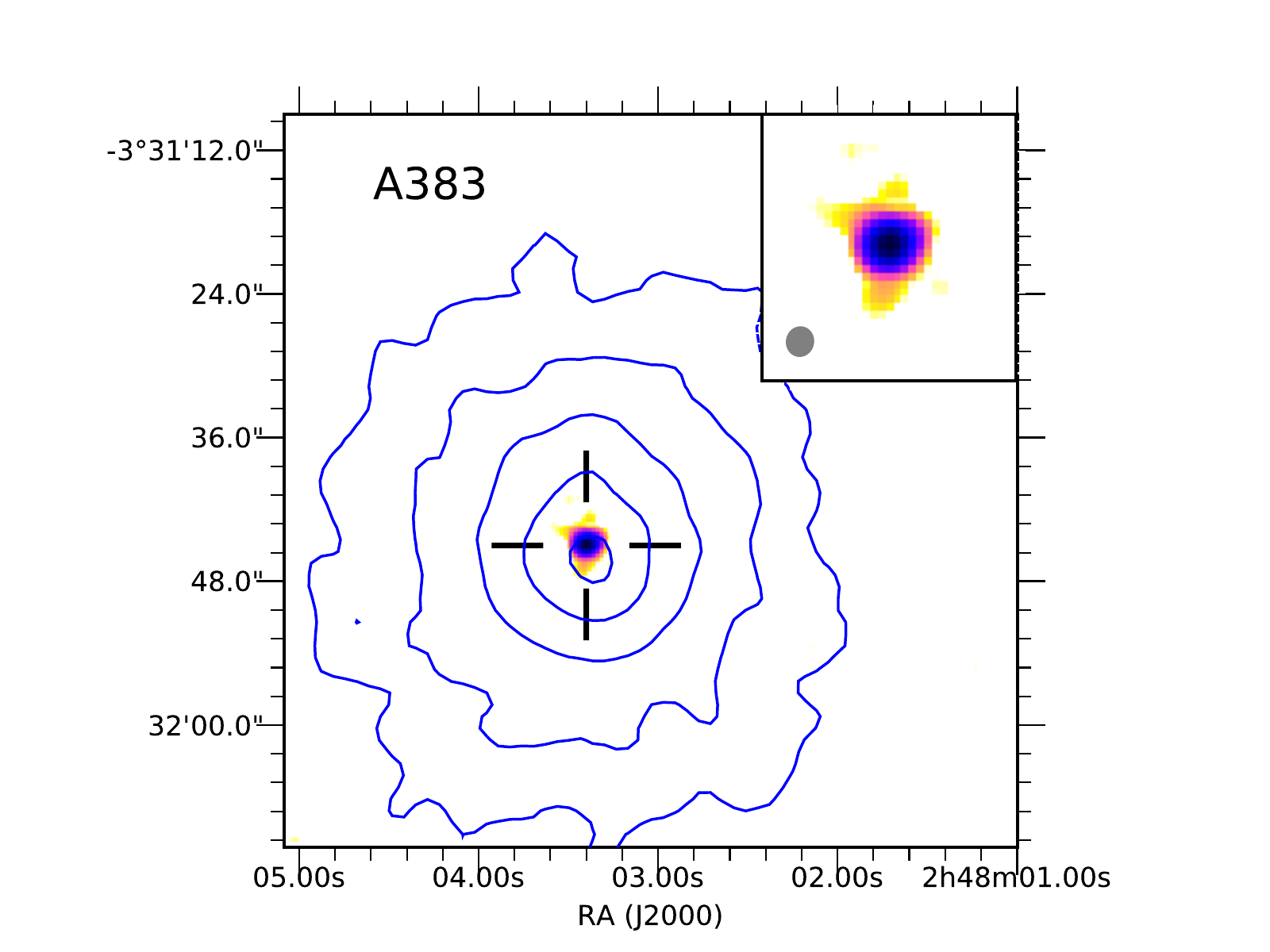}
\includegraphics[width=0.49\textwidth]{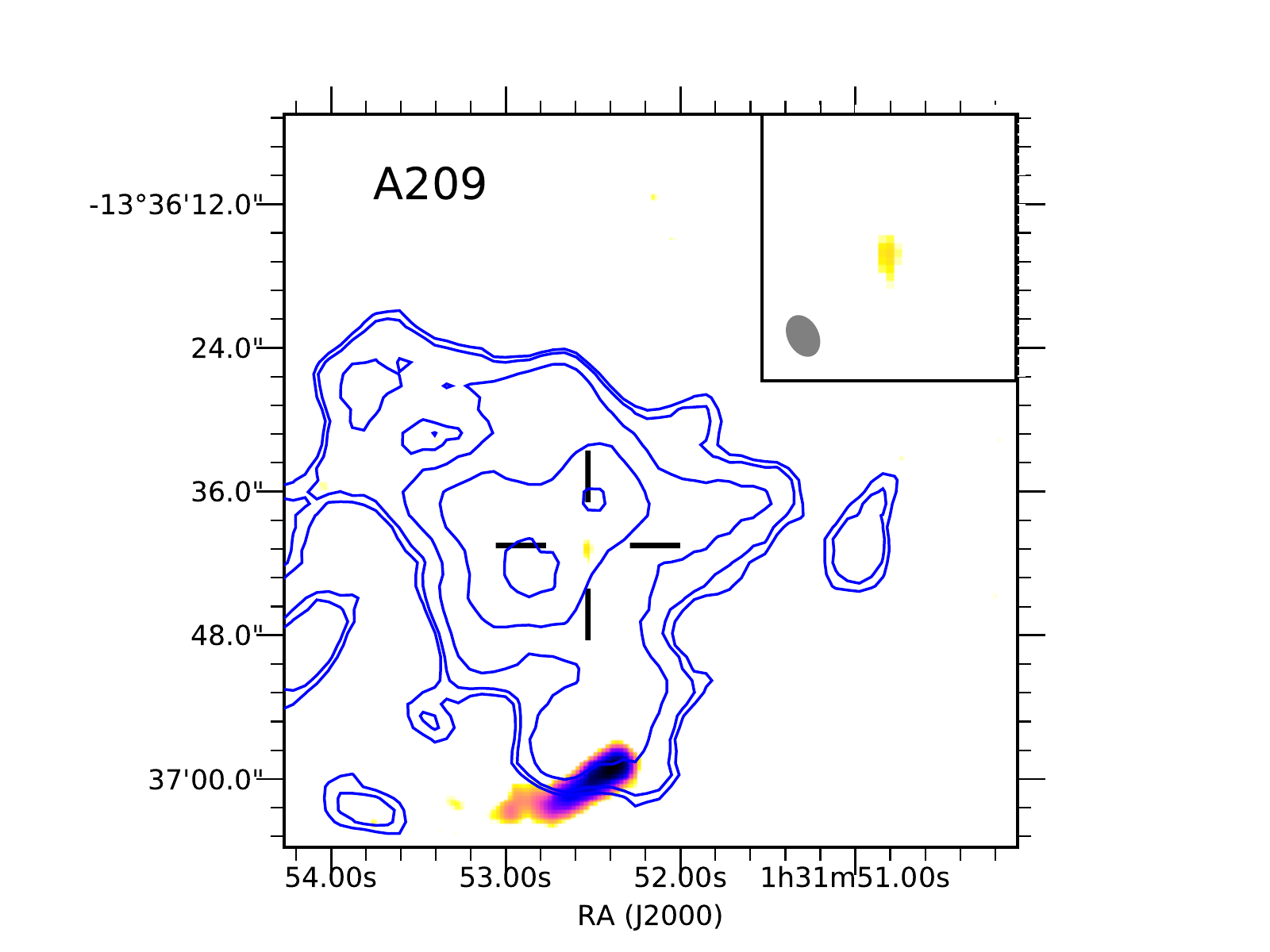}\\
\includegraphics[width=0.49\textwidth]{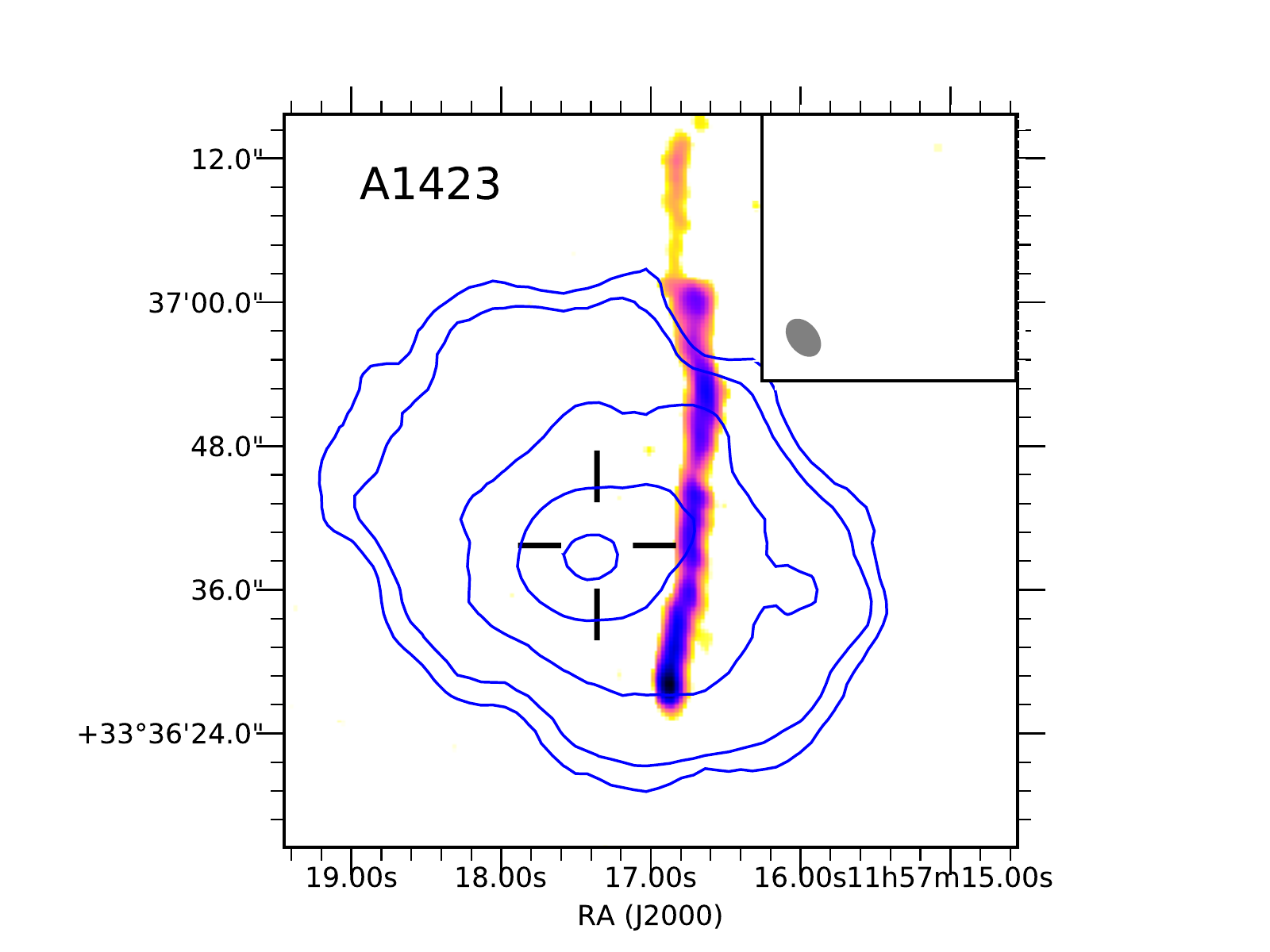}
\includegraphics[width=0.49\textwidth]{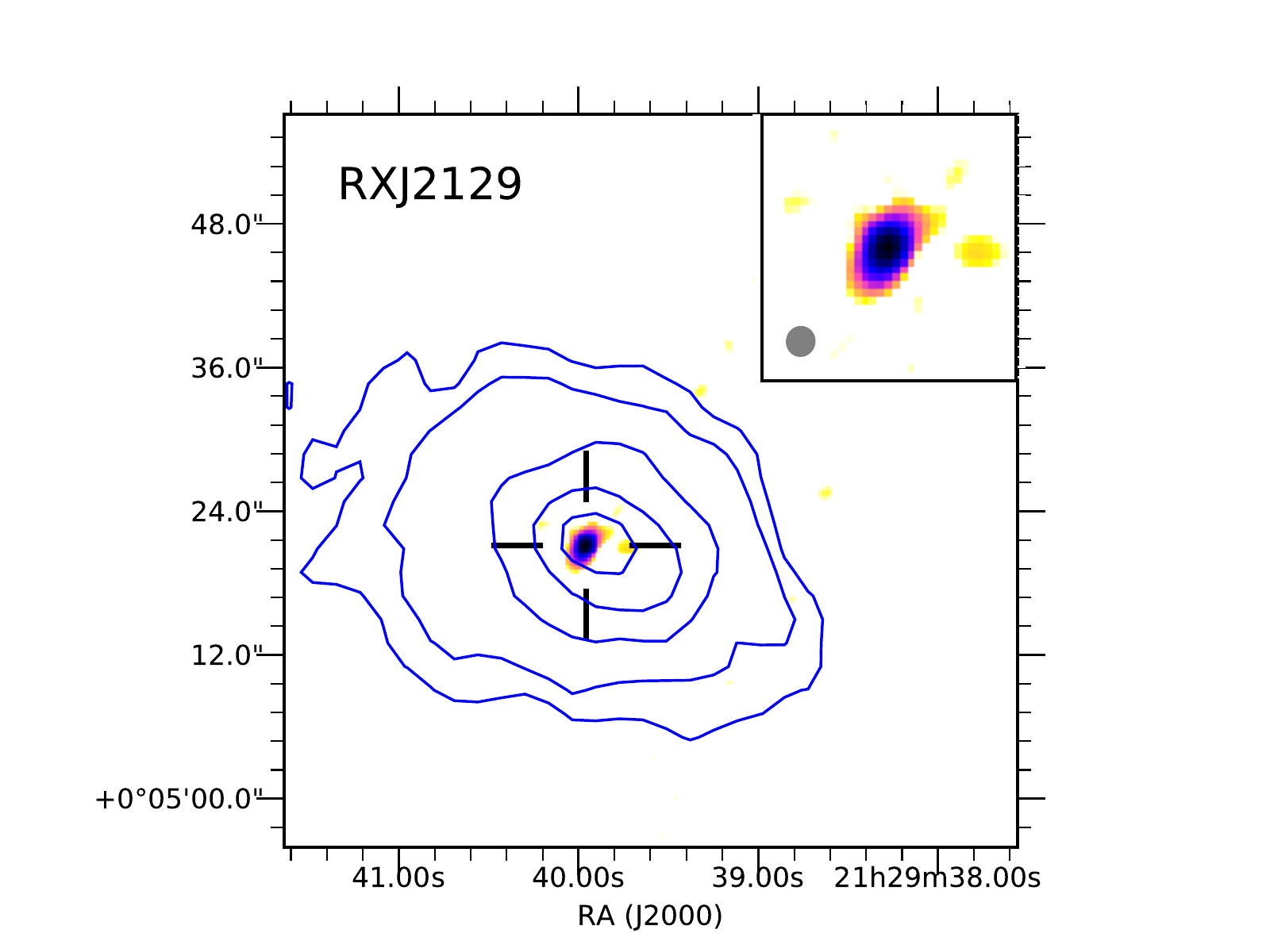}\\
\includegraphics[width=0.49\textwidth]{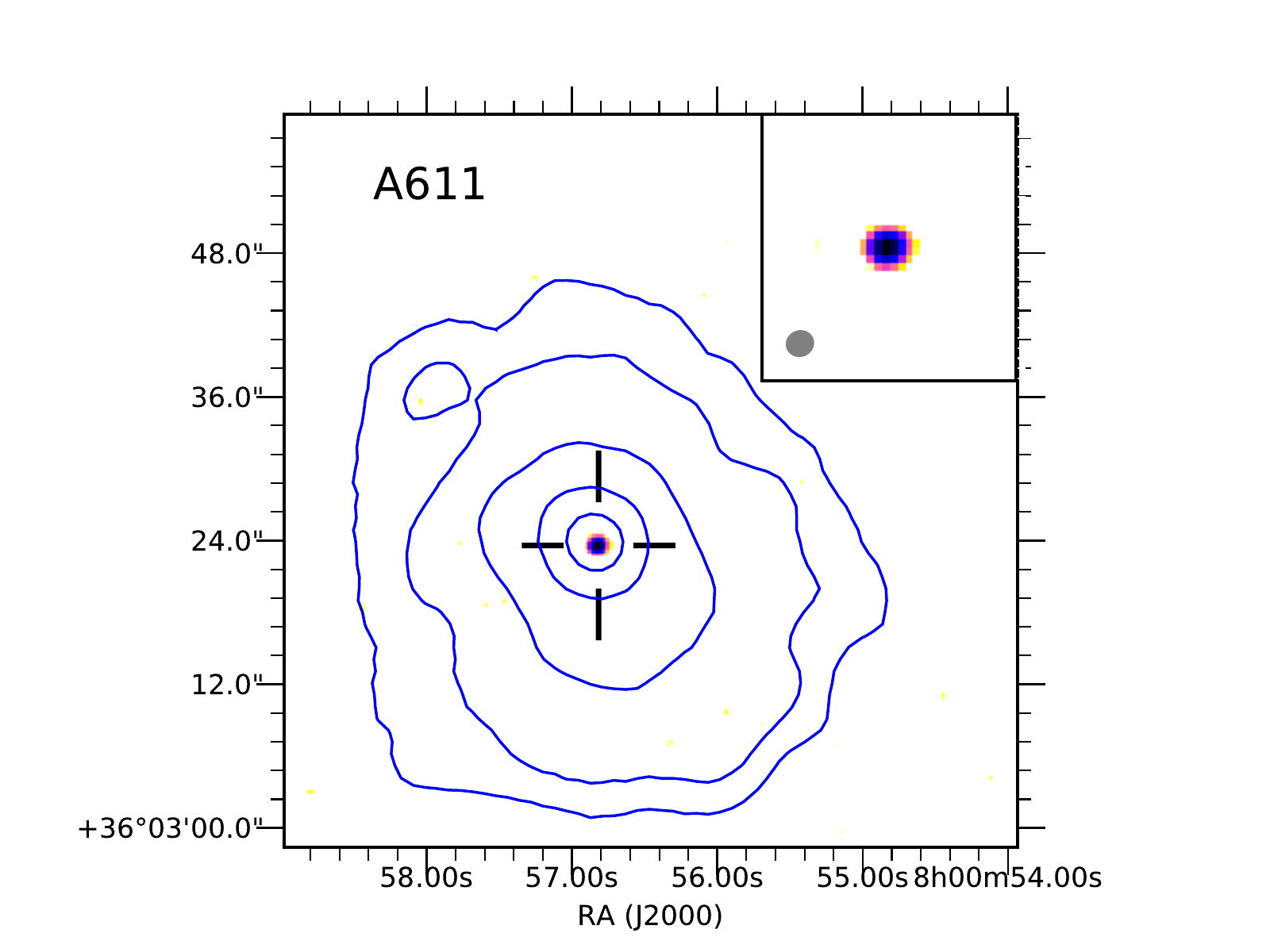}
\includegraphics[width=0.49\textwidth]{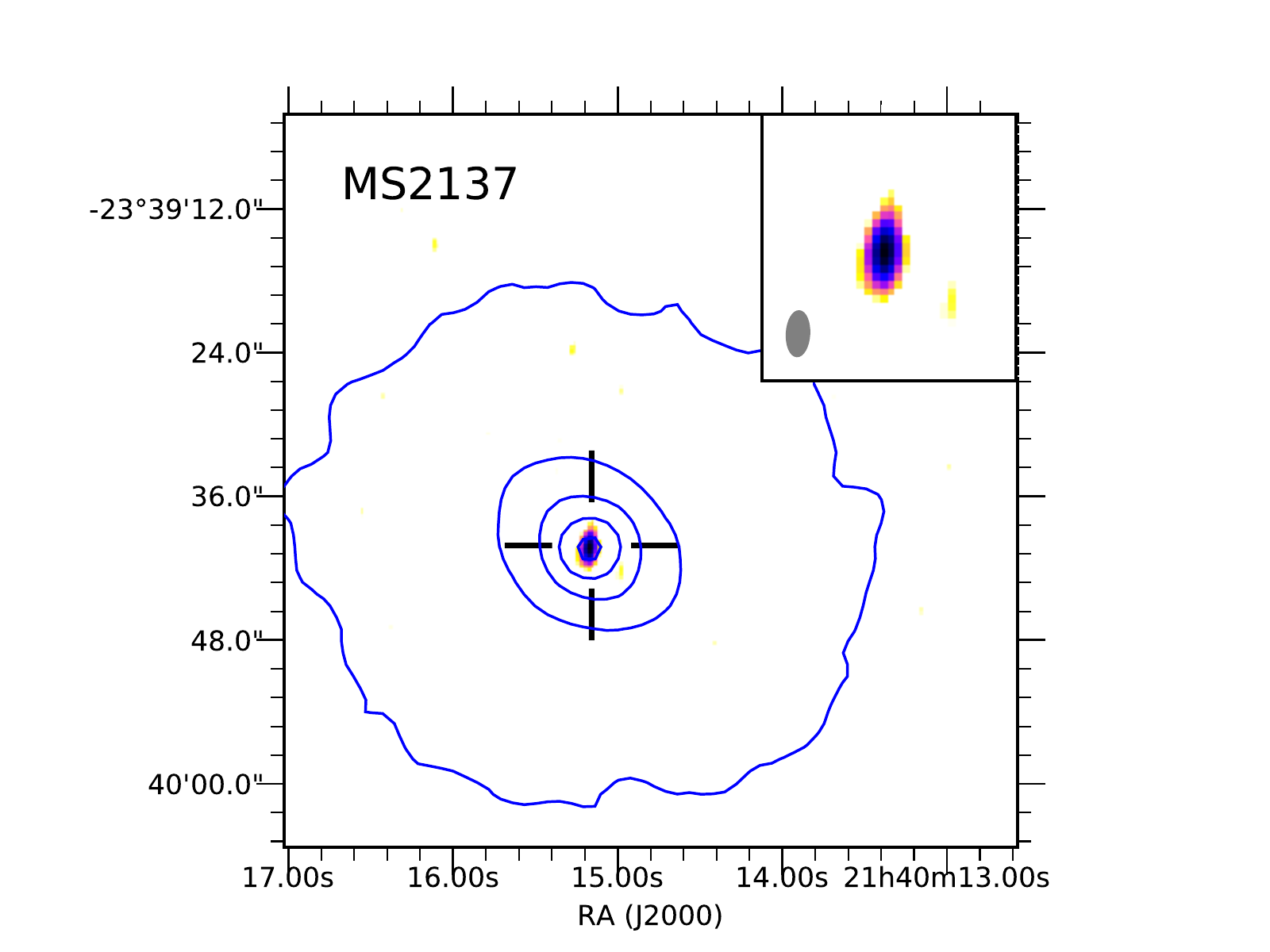}
\end{figure*}

After at least 3 self-calibrations, the final images are generated with the wide-field multi-frequency synthesis algorithm 
and are cleaned by interactive deconvolution. In Figures \ref{fig:img} we show the central $1'\times 1'$ fields, 
centered on the optical position of the BCGs, shown as a cross. The color scale varies logarithmically from 3$\sigma$ 
to the maximum flux density of each field. X-ray surface brightness contours from {\sl Chandra} are also shown with solid blue 
lines.  A direct visual inspection shows that in 11 out of 13 cases the peak of the X-ray emission overlaps with the position 
of the radio emission within the positional errors, while in 2 cases (Abell 209 and Abell 1423) no radio emission is detected 
at the optical position of the BCG.  In both clusters a strong radio source is found nearby, but clearly displaced from the X-ray 
peak \citep[as also noticed by][]{2015aHogan}. The full-field images will be presented and discussed in a future paper focused on the member  galaxy population (H. Yu et 
al. 2018, in preparation).

\renewcommand{\thefigure}{\arabic{figure} (Cont.)}
\addtocounter{figure}{-1}

\begin{figure*}[htp]
\caption{ }
\includegraphics[width=0.49\textwidth]{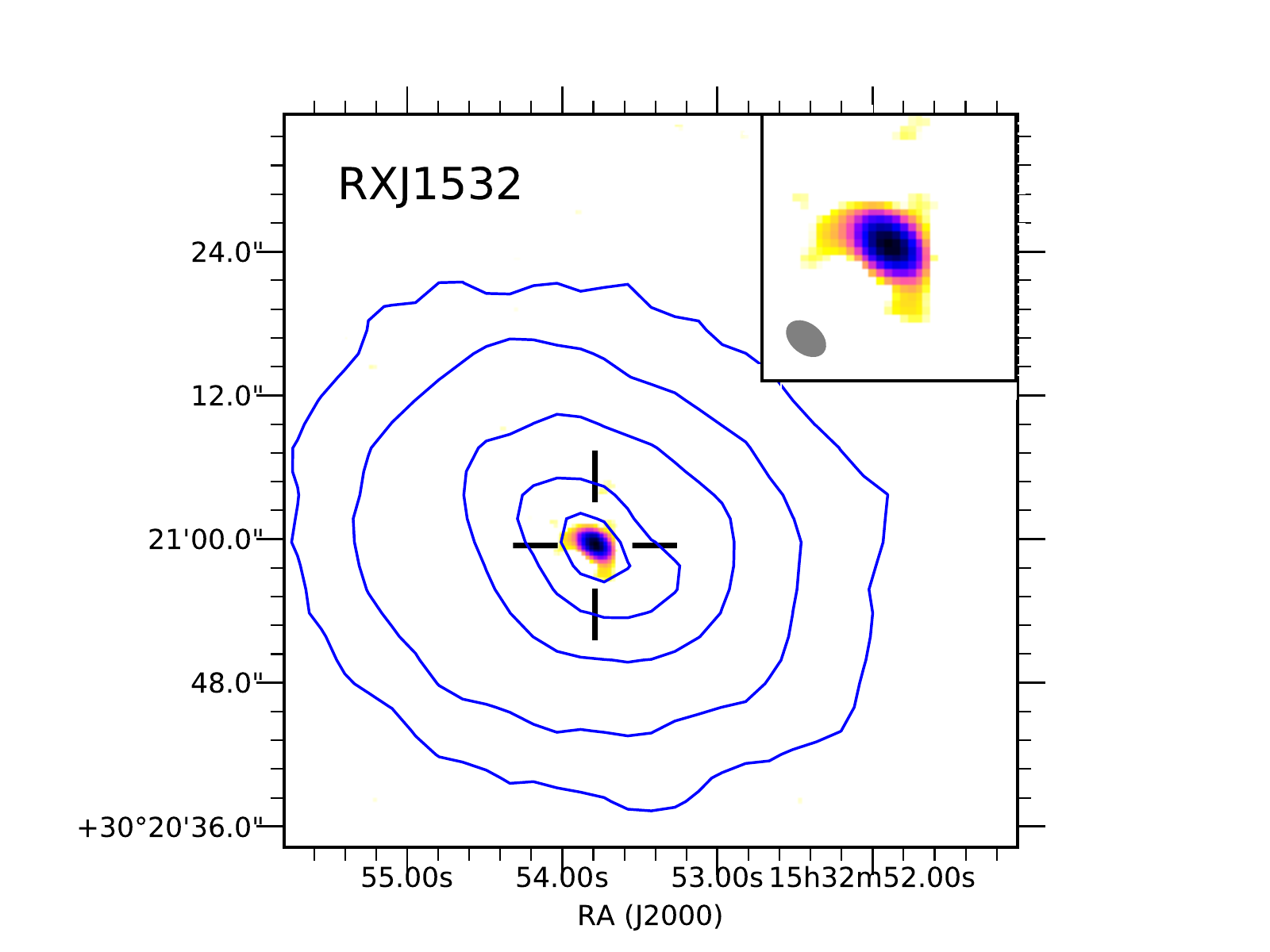}
\includegraphics[width=0.49\textwidth]{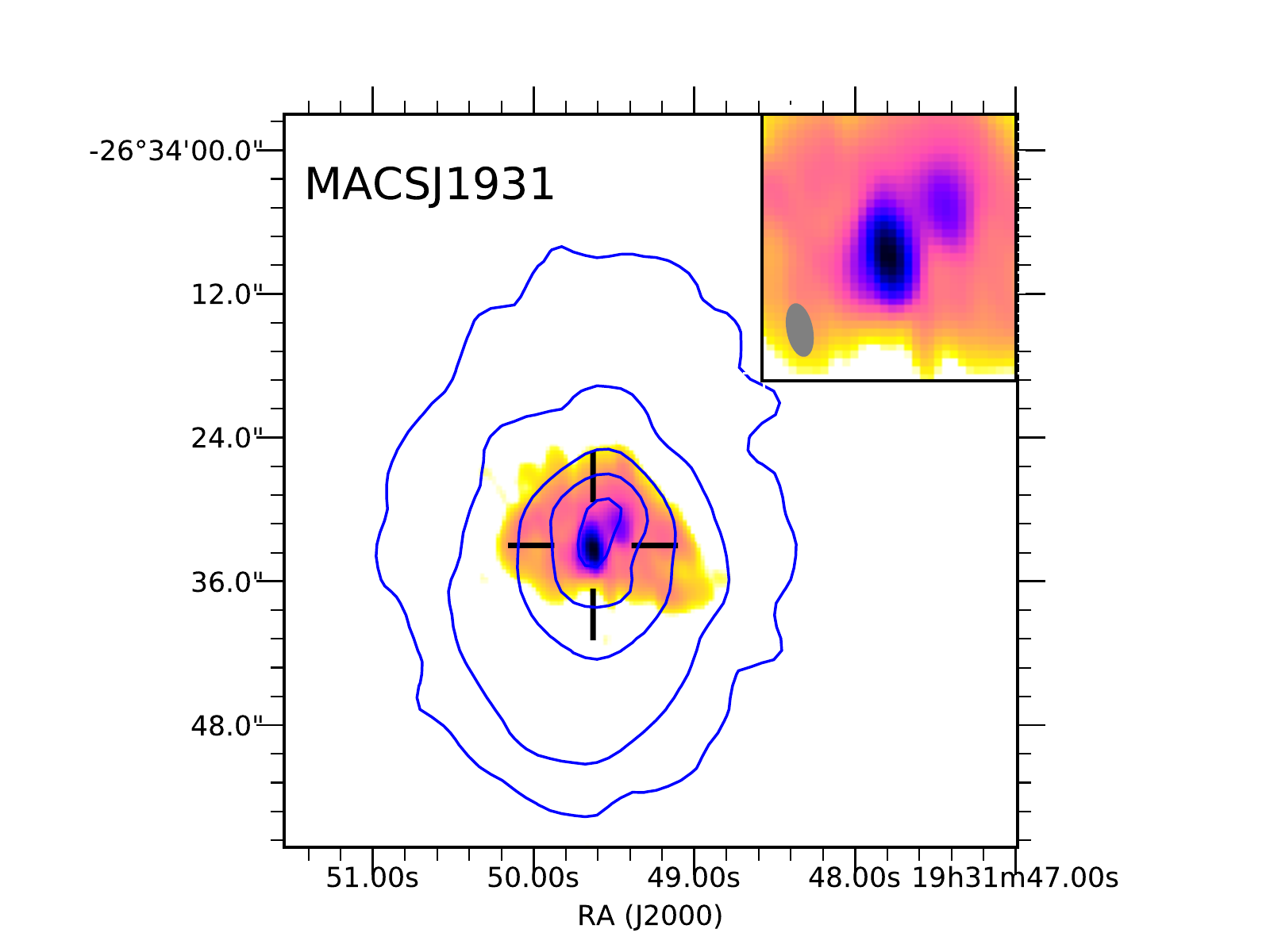}\\
\includegraphics[width=0.49\textwidth]{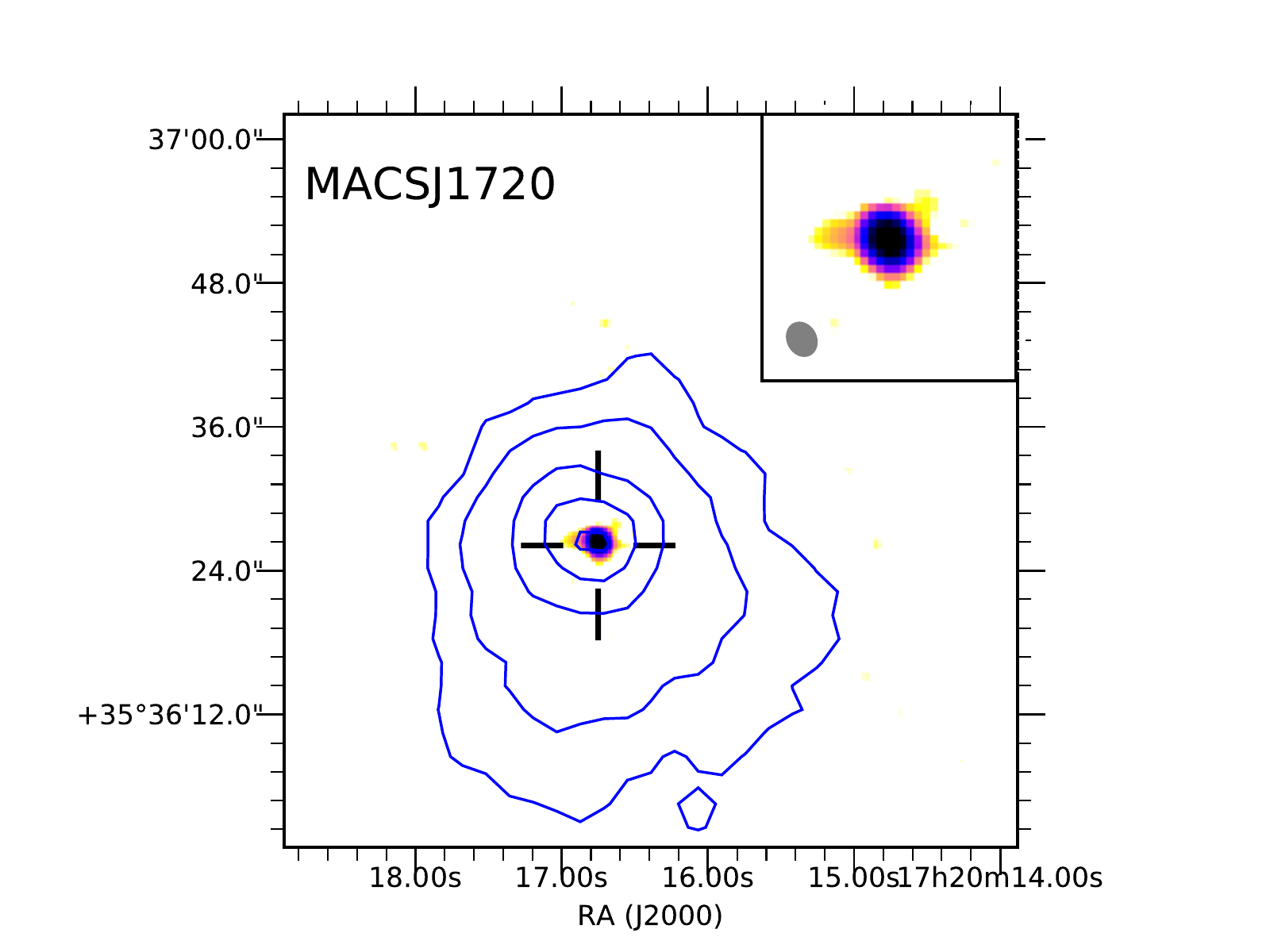}
\includegraphics[width=0.49\textwidth]{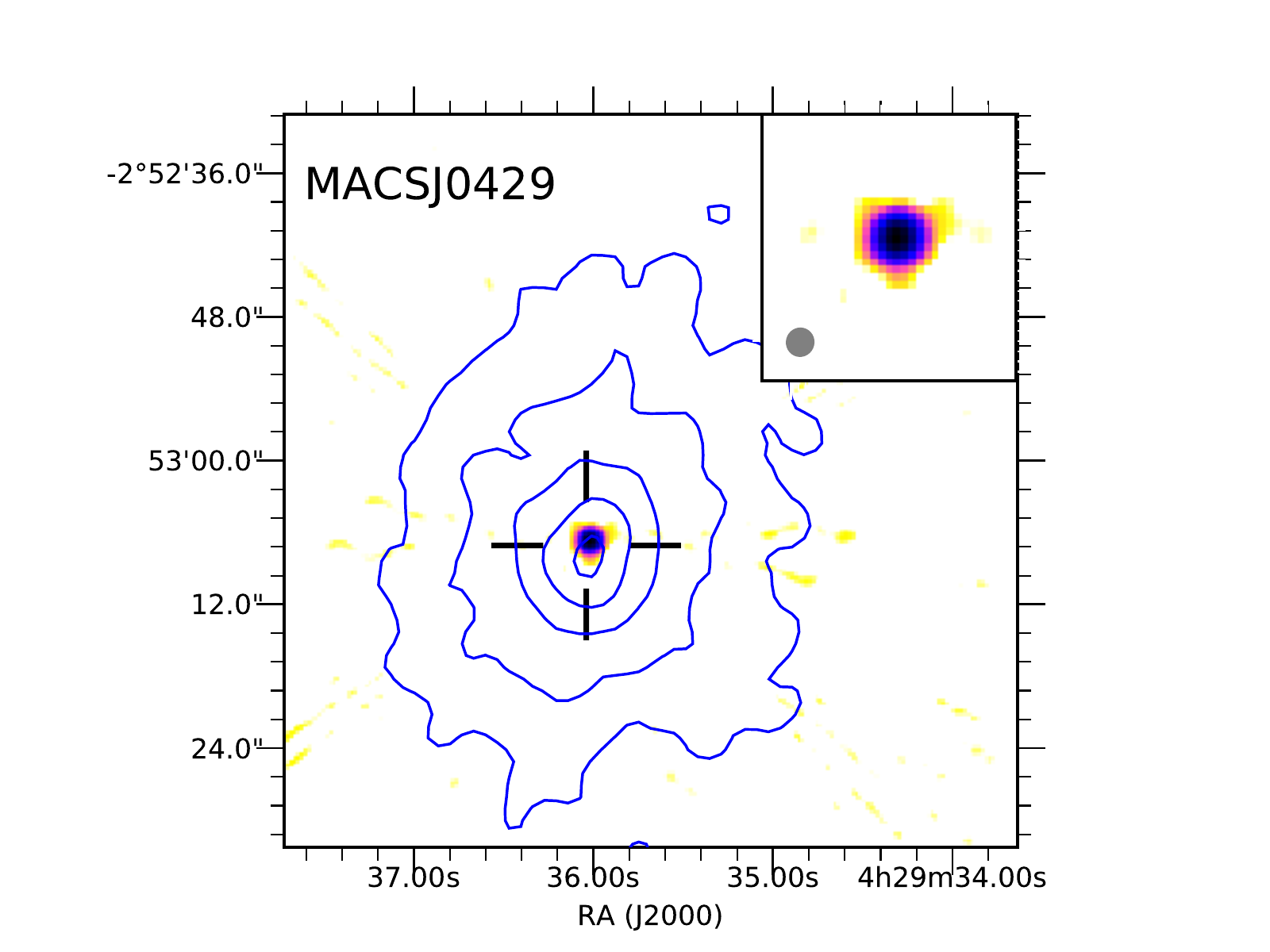}\\
\includegraphics[width=0.49\textwidth]{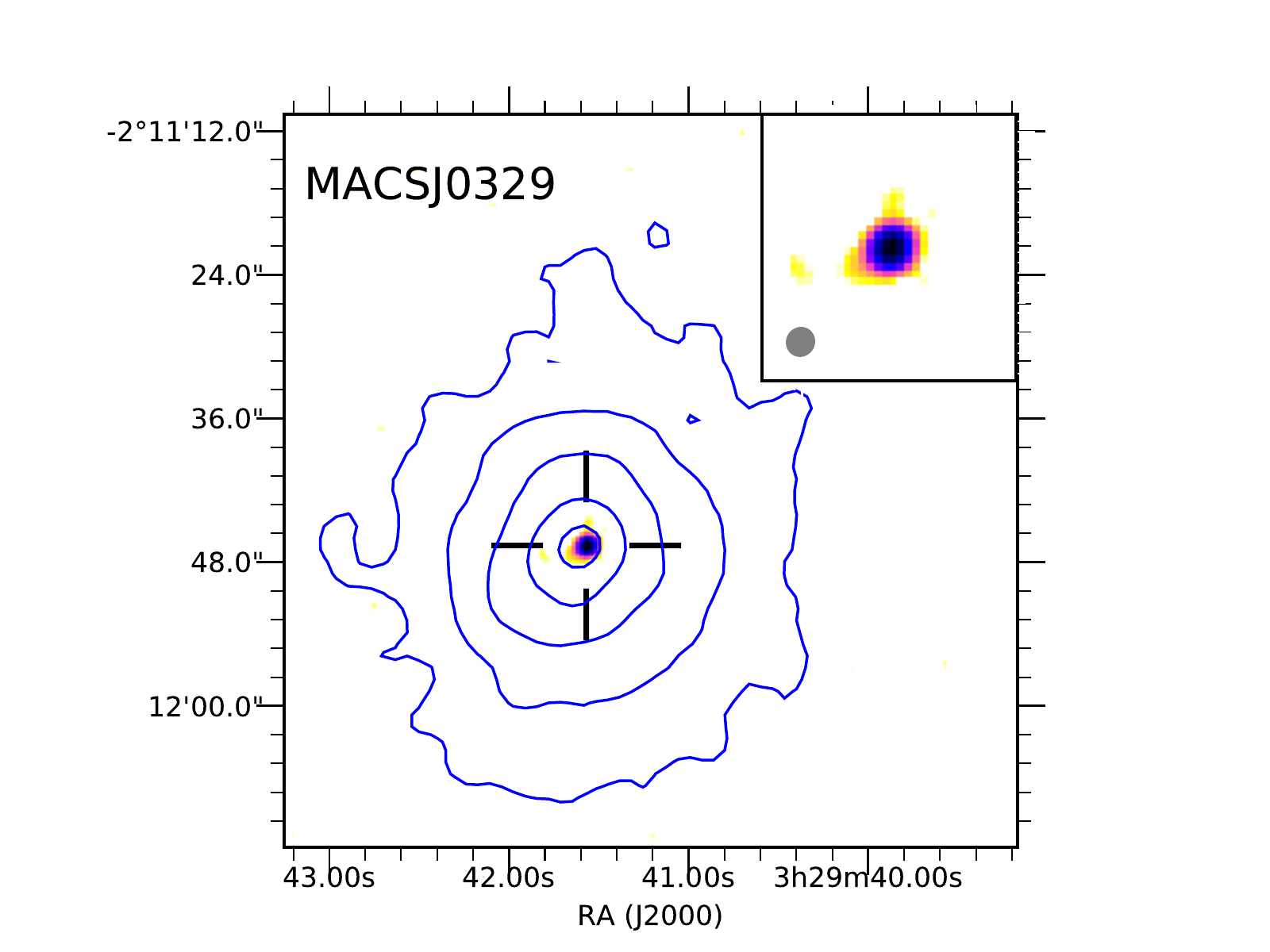}
\includegraphics[width=0.49\textwidth]{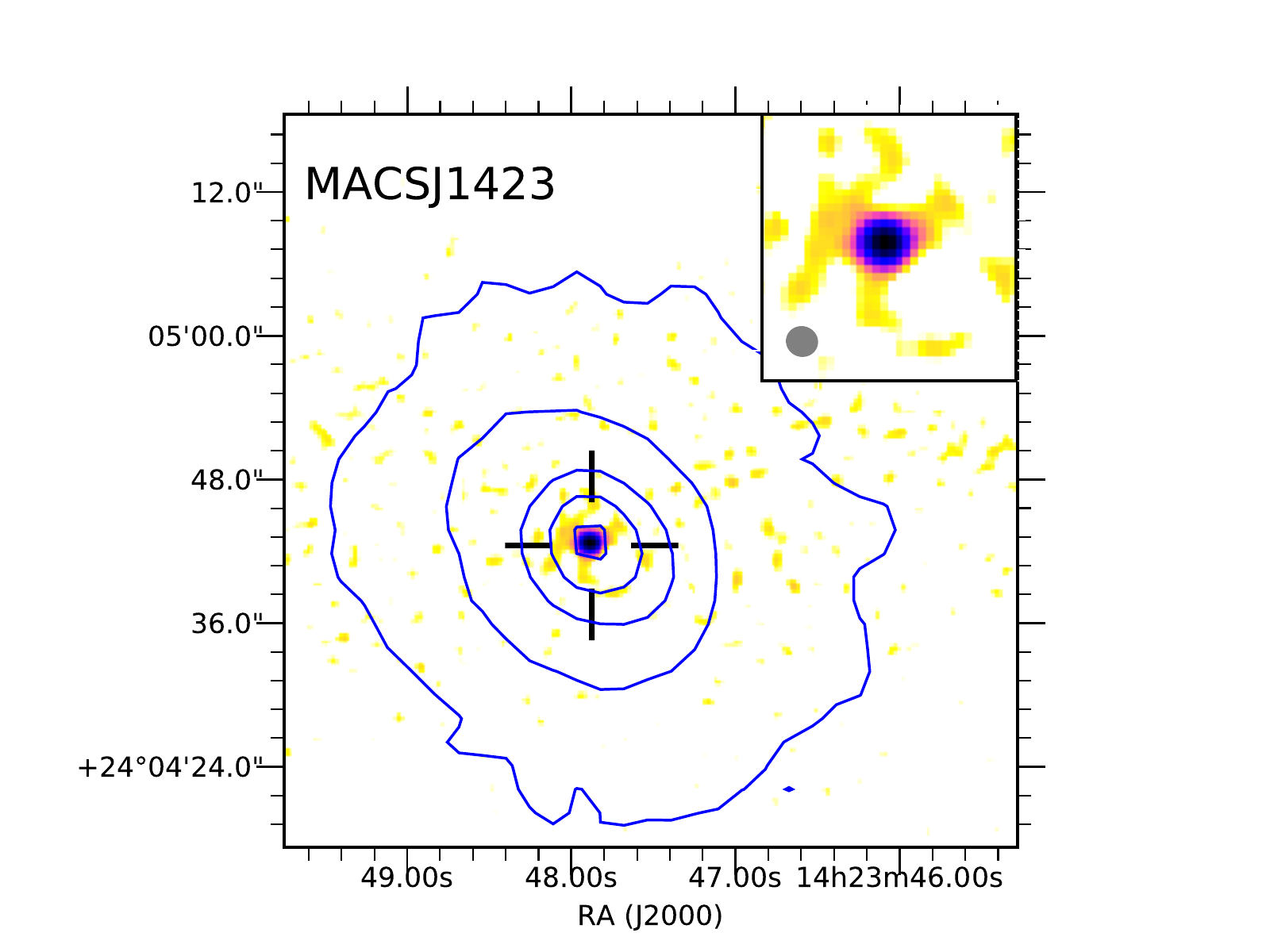}
\end{figure*}

\renewcommand{\thefigure}{\arabic{figure} (Cont.)}
\addtocounter{figure}{-1}

\begin{figure}[htp]
\caption{ }
\includegraphics[width=0.49\textwidth]{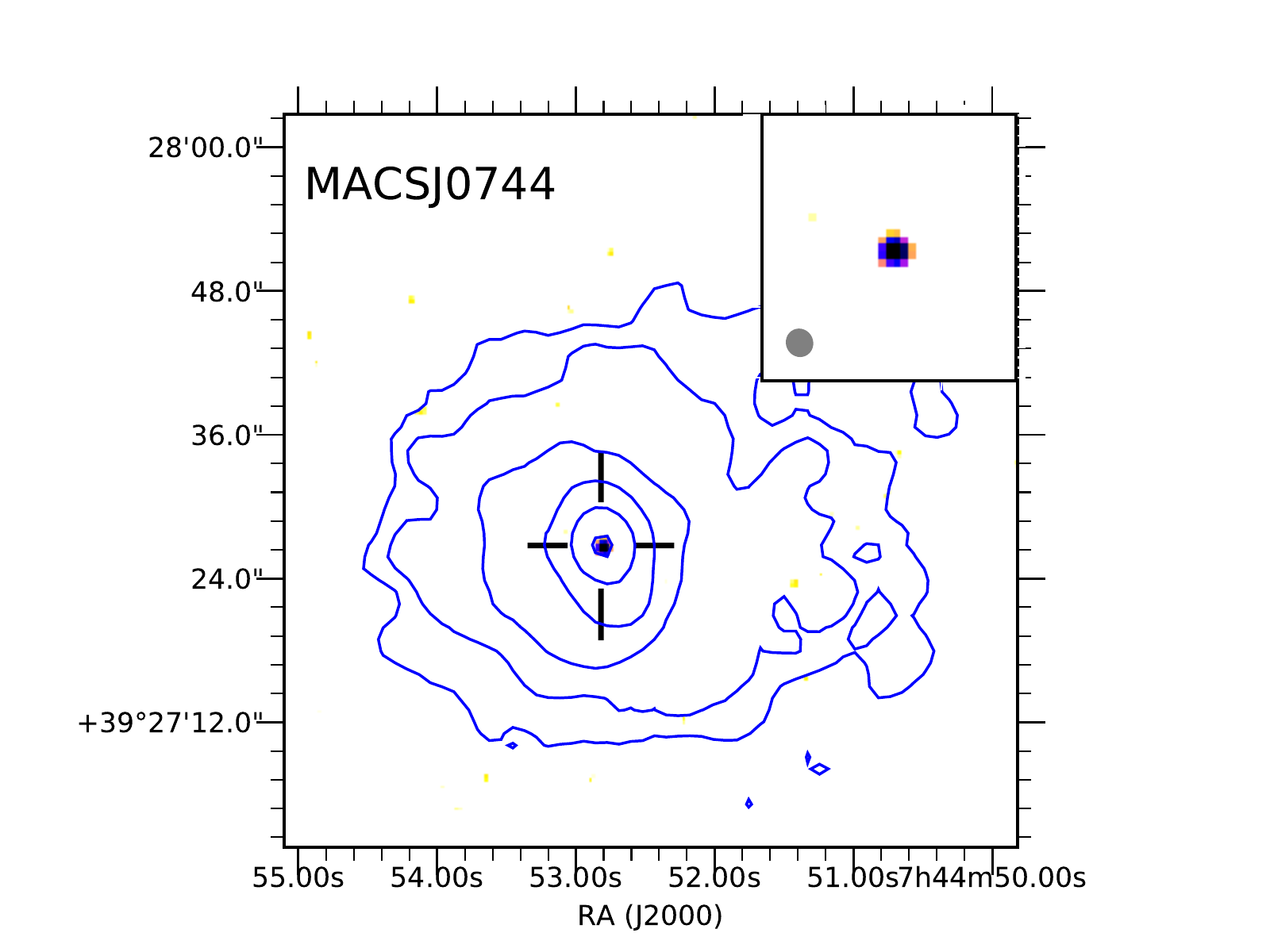}
\end{figure}

\renewcommand{\thefigure}{\arabic{figure}}

\section{Results}\label{sec:results}

In this section we present the results of our data analysis.  We start from the identification of the counterparts of the 
central radio sources, then we measure flux, source extent, spectral slope, and luminosity for each BCG.  
Finally, the average radio properties of our BCG sample are compared with other quantities derived from the literature, 
such as star formation rate (SFR) in the BCG, and free enthalpy measured from the cavity size in the X-ray images.

\subsection{Identification of the BCG}
\label{BCG}

Our observations are all centered at the position of the peak of the X-ray emission.  As previously mentioned, in 
each of the relaxed CLASH clusters the X-ray emission is centered on the position of the optical BCG. 
\citet{2016Donahue} also show that offsets are usually within a couple of arcseconds or less. 
In Section 2 we already identified radio sources  in NVSS and FIRST catalogs as potential candidate counterparts of 
the BCG and listed them in Table \ref{tab:sam}.

We now reconsider the potential candidate counterparts of Table \ref{tab:sam} on the basis 
of our radio images.
First, we cross-correlate the position of the radio sources at the pointing centers with the position of the optical
BCG using HST data.  Given the subarcsecond positional error of our radio data and of the 
HST images, we are able to unambiguously associate the central radio source of our images to the 
nucleus of the BCG.
There are 11 clusters containing radio galaxies in their center.  The typical offset between 
the radio and optical positions of the center of the BCG is 0.2 arcsec, consistent
with the positional error.  For Abell 383, RXJ2129, MS2137, RXJ1532, MACSJ1729, MACSJ0429, 
MACSJ0329, and MACSJ1423, we confirm the unique counterpart found in the NVSS and/or FIRST catalogs
and listed in Table \ref{tab:sam}.  

Only two clusters (Abell 209 and Abell 1423) do not show any radio counterpart for the BCG, and
we were able to put only upper limits on the flux and luminosity of these BCGs.  
In both cases we find two bright radio sources with head-tail morphologies at a distance of 
few arcsec from the BCG. Each source can be easily identified with satellite galaxies in the HST images.  
Both of them are cluster members, confirmed by spectroscopic data. 
In particular, in Abell 1423, the head-tail galaxy is at a projected distance of $12$'' (corresponding to 41.2 
kpc), while in Abell 209, it is found at a projected distance of  $17.8$'' (corresponding to 59.5 kpc).  These 
radio sources would have been mistakenly assumed to be the radio counterpart of the BCG in NVSS data without a 
careful screening of each single case and a refined analysis, as 
shown by our preliminary search for radio counterparts (see Table \ref{tab:sam}).
Bright head-tail, or wide-angle tail radio galaxies have been found in relaxed clusters, 
for example in the case of Abell 194 \citep{2008Sakelliou}, 
although they are thought to be more frequent in merging
clusters \citep[see Abell 562 and Abell 2634,][]{2011Douglass,2005Hardcastle}.  The presence of head-tail radio galaxies may be 
a tracer of an unrelaxed dynamical state, as already suggested by \citet{1998Bliton}.  We plan to investigate the nature 
of these galaxies on a forthcoming paper on the member galaxy population.

Finally, in the case of MACSJ1931 we find that an NVSS source at a distance of $41.4"$ (corresponding to 205.4 
kpc and therefore not included as a preliminary candidate counterpart) is actually the sum of the BCG 
radio emission and another nearby, bright head-tail galaxy.  In this case the use of NVSS data would have assigned 
an incorrect flux for the BCG.  Overall, we find $3/11\sim 30$\% wrong or partially misidentified associations 
would be made if based on a direct cross-correlation with the NVSS.  Studies on radio properties of BCGs at 
medium and high redshift may be significantly improved by the use of high resolution data, because of the possible
contamination by bright radio sources in the inner regions of the clusters.

\subsection{Centroid offset}
\label{xoffset}

\begin{figure}[htbp]
\includegraphics[width=0.49\textwidth]{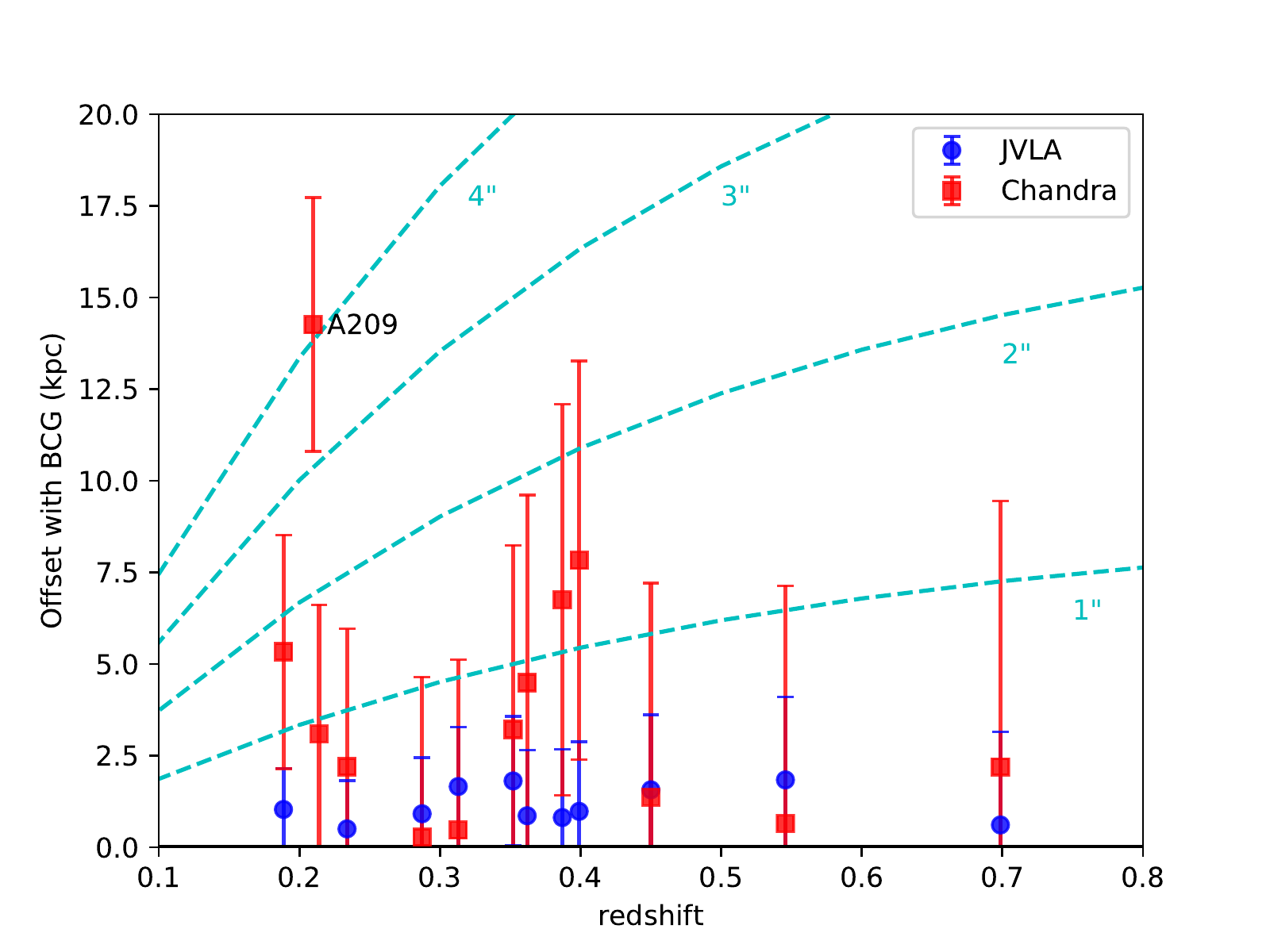}
\caption{The redshift distribution of the offsets in kpc between the optical position of the BCG and radio source 
position (blue circles ) and the X-ray peak (red squares).  
Typical uncertainties in the X-ray peak and radio centroids are $\sim 1$ arcsec and $\sim 0.35$
arcsec, respectively.  The dashed lines show constant offset in the plane of the sky. The small average displacement 
between radio and optical position confirms that the radio sources are always consistent with the nucleus of the BCGs.}
\label{fig:bcg_centroid}
\end{figure}

We compute the projected distance of the optical center of the BCG from the radio source 
associated with the BCG nucleus, and the distance of the optical center from the X-ray peak in the soft band (0.5-2 keV).
Uncertainties in the X-ray peak and radio centroids are almost constant and equal to $\sim 1$ arcsec and $\sim 0.35$
arcsec, respectively.  Errors on the optical positions are always below $0.1$ arcsec and therefore negligible.
Due to the relaxed status of our cluster sample, the X-ray peak is coincident within less than 1 arcsec with the
10 kpc X-ray centroid, defined as the center of the highest S/N circle with a fixed radius of 10 kpc.  
In Table \ref{tab:position} we list the equatorial coordinates (epoch J2000.0) 
of BCGs (optical and radio bands) and of the  cluster X-ray centroids.  
In Figure \ref{fig:bcg_centroid} we show the distribution of the displacement 
between the optical position and the radio position (blue circles), and
between the optical position and the X-ray centroid (red squares).

The narrow distribution of the optical-radio displacements (less than 0.5 arcsec) 
is consistent with the radio positional error and confirms the unambiguous
identification of the radio source with the BCG nucleus.  On the other hand, the distribution of 
the optical-X-ray displacements is slightly wider than expected from the positional errors.
\citet[][]{2016Donahue} have shown a similar result with the whole CLASH sample.
Offsets are known to be the signature of an unrelaxed dynamics, and are often found in clusters
with no or weak cool cores and a radio-silent BCG \citep{2009Sanderson}.  On the other hand, the X-ray centroid and
the $H_\alpha$ line emission region are tightly linked, sometimes despite an offset between the X-ray centroid and the 
BCG \citep{2012Hamer}, showing that the cooling process is not immediately switched off when the dynamics in 
the core is disturbed.

We note that larger cluster samples (several hundreds) show an average projected spatial offset between the optical 
position of the BCG and the X-ray center of about 10 kpc, with only 15\% of the BCGs lying more than 100 kpc  
from the X-ray center of their host cluster \citep[see][]{2014Lauer}.  In addition, the BCG position relative to the cluster 
center is correlated with the degree of concentration of X-ray morphology \citep{2014Hashimoto}.
However, the offset of the optical position of the BCGs with respect to the X-ray peaks in our sample is consistent with the 
measurement uncertainties in most cases, so that we do not draw any conclusion on the dynamical state of the 
cluster from this measurement. The largest offset is observed in Abell 209, where the distance of the optical/radio position 
from the X-ray centroid is $4.1"$, corresponding to 14.3 projected kpc.  
A deeper X-ray observation of Abell 209, which has so far been observed only for 20 ks with {\sl Chandra}, 
is needed to further investigate this aspect and possibly identify the origin of the offset\footnote{The XMM observation of 
Abell 209, carried out by the EPIC pn and MOS detector, can also be used to investigate the ICM dynamics, but with a poor 
angular resolution corresponding to an Half Energy Width $\geq 15$''.}.  However, the optical study in \citet{2016Annunziatella} already shows that Abell 209 is not fully 
relaxed.  This interpretation is also supported by the observation of a radio halo in Abell 209, which may be regarded 
as the signature of a strong ongoing merger \citep{2009Giovannini,2015Kale}.  Overall, 
the CLASH clusters discussed in this work are expected to be dynamically relaxed, 
while we expect to find much larger BCG-X-ray peak displacements in the five CLASH clusters
in the high-magnification subsample, not included in this study.

\begin{table*}
\centering
\caption{BCG Coordinates and Cluster Centroids}
% \footnotesize
\label{tab:position}
\begin{tabular}{|l|r|r|r|r|r|r|}
 \hline
 cluster & RA$_{\rm HST}$ & DEC$_{\rm HST}$  & RA$_{\rm JVLA}$ & DEC$_{\rm JVLA}$ 
 & RA$_{\rm Chandra}$ & DEC$_{\rm Chandra}$ \\
\hline
Abell 383  &  2:48:03.38 &  -3:31:45.27 &  2:48:03.4 &  -3:31:45.1 &  2:48:03.4 & -3:31:46.7 \\
Abell 209  &  1:31:52.55 & -13:36:40.49 &  -           &  -           &  1:31:52.9 & -13:36:41.7 \\
Abell 1423 & 11:57:17.36 & +33:36:39.57 &  -           &  -           & 11:57:17.3 & +33:36:38.8 \\
RXJ2129    & 21:29:39.96 &  +0:05:21.19 & 21:29:40.0 &  +0:05:21.1 & 21:29:40.0 & +0:05:21.8 \\
Abell 611  & 8:00:56.82 & +36:03:23.63 &  8:00:56.8 & +36:03:23.5 &  8:00:56.8 & +36:03:23.6 \\
MS2137     & 21:40:15.16 & -23:39:40.12 & 21:40:15.2 & -23:39:40.4 & 21:40:15.2 & -23:39:40.2 \\
RXJ1532    & 15:32:53.78 & +30:20:59.45 & 15:32:53.8 & +30:20:59.6 & 15:32:53.7 & +30:20:58.8  \\
MACSJ1931  & 19:31:49.63 & -26:34:33.16 & 19:31:49.6 & -26:34:33.5 & 19:31:49.6 & -26:34:33.8   \\
MACSJ1720  & 17:20:16.75 & +35:36:26.22 & 17:20:16.8 & +35:36:26.4 & 17:20:16.8 & +35:36:26.9 \\
MACSJ0429  &  4:29:36.01 &  -2:53:06.72 & 4:29:36.0 & -2:53:06.8 &  4:29:36.0 &  -2:53:08.2 \\
MACSJ0329  &  3:29:41.57 &  -2:11:46.45 &  3:29:41.6 &  -2:11:46.7 & 03:29:41.6 &  -2:11:46.7 \\
MACSJ1423  & 14:23:47.88 & +24:04:42.44 & 14:23:47.9 & +24:04:42.6 & 14:23:47.9 & +24:04:42.4 \\
MACSJ0744  &  7:44:52.80 & +39:27:26.65 &  7:44:52.8 & +39:27:26.6 &  7:44:52.8 & +39:27:26.4   \\
\hline
\end{tabular}
\tablecomments{BCG coordinates in the optical band from HST, position of the radio source from JVLA and 
 centroid of the ICM X-ray emission from {\sl Chandra}  for the 13 targets observed with our JVLA program.}
\end{table*}

\begin{table*}[htbp]
\caption{Radio flux of the BCGs}
\begin{center}
\begin{tabular}{lrrrc}
 \hline
\multirow{2}{*}{cluster}  & \multicolumn{2}{c}{\underline{JVLA (mJy, 1.5 GHz)}} &  
        \multicolumn{2}{c}{\underline{ $F_{int}$ (mJy, 1.4 GHz)}} \\
 &  $F_{peak}$ & $F_{int}$ & NVSS & FIRST \\
\hline
Abell 383 & 27.52 $\pm$ 0.02 & 36.75 $\pm$ 0.07 & 40.9 $\pm$ 1.3 & 41.37 $\pm$ 0.12 \\ 
Abell 209   & $<$0.10 & $<$ 0.08 & no detection & no coverage \\
Abell 1423   & $<$ 0.04 & $<$ 0.05 & no detection  & no detection \\
RXJ2129  & 14.94$\pm$ 0.04 & 22.52 $\pm$ 0.12 & 25.4 $\pm$ 1.2  & 24.27 $\pm$ 0.10  \\
Abell 611  & 0.80 $\pm$ 0.01 & 0.85 $\pm$ 0.02 &  no detection & no detection \\
MS2137  & 1.24 $\pm$ 0.02 & 1.39 $\pm$ 0.03 & 3.8 $\pm$ 0.5  & no coverage \\
RXJ1532$^*$  & 15.33 $\pm$ 0.05 & 16.19 $\pm$ 0.11 & 22.8 $\pm$ 0.8 & 17.11 $\pm$ 0.14 \\
MACSJ1931   & 11.57 $\pm$ 0.03 & 19.38 $\pm$ 0.05 & no detection & no coverage \\ 
MACSJ1720$^*$  & 21.14 $\pm$ 0.07 & 24.08 $\pm$ 0.17 & 18.0 $\pm$ 1.0 & 16.75 $\pm$ 0.24 \\
MACSJ0429  & 124.27$\pm$ 0.03 & 126.16 $\pm$ 0.07  & 138.8 $\pm$ 4.2 & no coverage \\ 
MACSJ0329  & 2.92 $\pm$ 0.02 & 3.33 $\pm$ 0.03 & 6.9 $\pm$ 0.6 & no coverage\\ 
MACSJ1423  & 3.55 $\pm$ 0.03 & 4.28 $\pm$ 0.05 & 8.0 $\pm$ 1.1 & 5.22 $\pm$ 0.15 \\ 
MACSJ0744   & 0.27 $\pm$ 0.01 & 0.27 $\pm$ 0.03 & no detection &  no detection \\
\hline
Abell 2261 &  &  & 5.3 $\pm$ 0.5 & 3.40 $\pm$ 0.15 \\
RXJ2248   &  &   & no coverage & no coverage\\
MACSJ1115  &  &  & 16.2 $\pm$  1.0 & 8.27 $\pm$ 0.15 \\
MACSJ1206  &  &  &  160.9  $\pm$ 6.3 & no coverage \\
RXJ1347   &  &   & 45.9 $\pm$ 1.5 & no coverage \\
MACSJ1311  &  &  &  no detection & no detection \\
ClJ1226   &  &  & 4.3  $\pm$ 0.5  & 3.61 $\pm$ 0.13 \\
% MACSJ0647 &  &  &  &    no detection & no coverage \\
% MACSJ2129 &  &  &   &   no detection  & no detection \\
% MACSJ1149 &  &  &   &   no detection & no detection \\
% MACSJ0416 &  &  &   &   no coverage & no coverage \\
% MACSJ0717 &  &  &   &   90.9 $\pm$ 3.7& no detection \\
\hline
\end{tabular}
\end{center}
\tablecomments{Columns 2 and 3  show peak and integrated flux densities measured with our JVLA data at 1.5 GHz,
in units of mJy.  In columns 4 and 5 we report the integrated flux densities of the confirmed radio counterparts in the NVSS 
and FIRST catalogs, respectively.  The sources listed in the second part of the table are not observed in the current dataset, 
and therefore have only NVSS or FIRST candidate counterparts. } 
\vspace{1ex}
\raggedright {\footnotesize *: MACSJ1720 and RXJ1532 may have errors in the flux larger than quoted, 
due to the use of the phase calibrator also as a flux calibrator.  We plan to refine the estimate of the errors
when investigating the full source sample in the two fields.}
\label{tab:resu}
\end{table*}

\begin{figure}[htbp]
\includegraphics[width=0.49\textwidth]{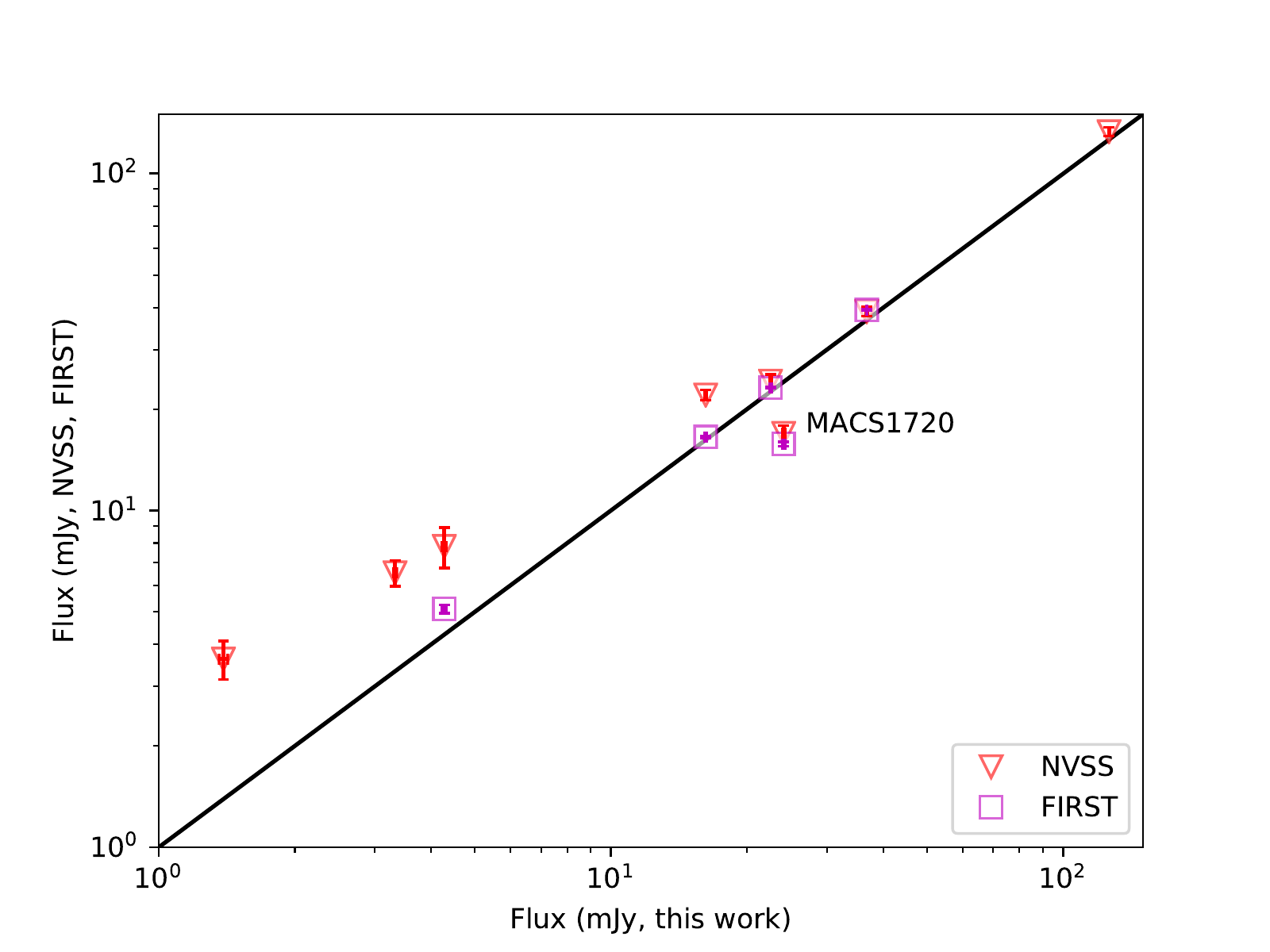}
\caption{JVLA integrated flux densities of the BCG galaxies compared to the integrated flux densities of the confirmed
counterpart in NVSS (red triangles) and FIRST (magenta squares).  Error bars correspond to 1 $\sigma$ confidence level.  }
\label{fig:flux_bcg}
\end{figure}

\subsection{Radio fluxes}

The redshifts, coordinates, peak fluxes, and integrated fluxes for the radio counterparts of the 
BCGs observed in our program are listed in Table \ref{tab:resu}.
All peak fluxes and integrated fluxes are measured with the software PyBDSF
\citep[the Python Blob Detector and Source Finder\footnote{Also named PyBDSM, see http://www.astron.nl/citt/pybdsm/.},][]{PyBDSF}.

As a first step, we compare the radio fluxes of the BCGs in our data with data from NVSS and FIRST, whenever
a clear counterpart is identified in one of these two surveys and confirmed by our data\footnote{For the sake of 
comparison, fluxes computed at  $1.4$ GHz are corrected by the factor $(1.5/1.4)^{\alpha}$, where the spectral index is 
discussed in Section \ref{alpha}.  This correction amounts to a maximum of 5\%.}.  
By comparing Table \ref{tab:resu} with Table \ref{tab:sam}, we find that three NVSS counterparts are dropped
completely or partially (Abell 209, Abell 1423 and MACSJ1931), and in all the cases this is due to 
contamination by radio galaxies close to the BCG, which are unambiguously identified as cluster members 
in our data.  In Figure \ref{fig:flux_bcg}, we plot the 
JVLA integrated fluxes of the BCG versus the NVSS and FIRST integrated fluxes.  We find
overall a good agreement for the  5 sources in FIRST, with some discrepancy that can be
ascribed to variability \citep[see][for a discussion on the variability at high frequencies]{2015bHogan}. 
On the other hand, fluxes from NVSS are systematically higher, particularly at low fluxes.
This excess may be explained with the presence of extended radio emission that is not detected in our data.  
At bright fluxes the emission is likely to be dominated by the nucleus, so that 
measured fluxes do not depend significantly on the angular resolution.  In addition, despite the limited statistics, this 
result is consistent with the comparison of NVSS and FIRST fluxes with previous VLA data \citep[see][]{2012Wold}.  
Therefore, we conclude that radio fluxes measured in our data show no obvious discrepancy with previous
measurements. This also allows us to consider FIRST and NVSS fluxes for the CLASH targets not included in this work.  In 
particular, Abell 2261, CLJ1226 and MACSJ1115 are in FIRST, while MACSJ1206 and RXJ1347 have only NVSS data, but are bright 
enough ($>$20 mJy) to be considered dominated by the nuclear emission. Although the angular resolution of NVSS
data does not allow a secure identification by itself, we refer to \citet{2009Ebeling} and \citet{2015aHogan} for a detailed discussion on the likely association of the radio emission with the BCG in both cases. Recently, high resolution JVLA 5GHz 
observations of MACSJ1206 confirmed the presence of a compact double source associated with the BCG (A. Edge 2017, private
communication). Therefore we consider NVSS and FIRST counterparts as reliable for the sources not observed in our JVLA program.

\begin{table}
\caption{Observed and Deconvolved Size of BCGs}
\begin{center}
\begin{tabular}{|c|c|c|c|}
 \hline
\multirow{2}{*}{Cluster} & FWHM size & Deconvolved size  &
\multirow{2}{*}{$F_{int}/F_{peak}$}\\
\cline{2-3}
  &  $ '' \times '' $ , $^\circ$ &  $ '' \times '' $ , $^\circ$ &       \\
\hline
Abell 383 & 1.28 $\times$ 1.21, 115  &  0.75 $\times$ 0.51, 94 & 1.34
$\pm$ 0.01\\
Abell 209 & - & -          & -\\
Abell 1423 & -     & -         & - \\
RXJ2129 & 1.64 $\times$ 1.01, 150 &  1.20 $\times$ 0.22 , 148 & 1.51
$\pm$ 0.01  \\
Abell 611  & 1.08 $\times$ 0.97, 102 & unresolved & 1.06
$\pm$ 0.05\\  % 0.34 $\times$ 0.13 , 85
MS2137-2353 & 1.88 $\times$ 0.91, 175 & 0.66$\times$0.24 , 165 & 1.12
$\pm$ 0.04\\
RXJ1532  & 1.72 $\times$ 1.13, 55 & unresolved & 1.06 $\pm$ 0.01\\
MACSJ1931$^{\dagger}$ & 7.38 $\times$ 5.58, 76 & 7.28 $\times$ 5.24, 79
& 1.68 $\pm$ 0.04 \\
MACSJ1720 & 1.39 $\times$ 1.20, 60 & unresolved  & 1.14 $\pm$ 0.01\\
MACSJ0429 & 1.05 $\times$ 1.03, 160 & unresolved  & 1.02 $\pm$ 0.01\\
MACSJ0329 & 1.20 $\times$ 1.09, 145 & 0.48 $\times$ 0.32, 139 & 1.14 $\pm$ 0.02\\
MACSJ1423 & 1.31 $\times$ 1.20, 90 & 0.61 $\times$ 0.42, 103 & 1.21 $\pm$ 0.02\\
MACSJ0744 & 1.03 $\times$ 0.96, 73 & unresolved & 1.00 $\pm$ 0.15 \\ \hline
\end{tabular}
\end{center}
\tablecomments{the BCG size (major and minor axis, and orientation of the elliptical fit) 
as measured directly in radio images (second column) compared to the deconvolved size (third column). 
The ratio $F_{int}/ F_{peak}$ is also listed in the last column.}
\vspace{1ex}
\raggedright {\footnotesize $\dagger$: In the case of MACSJ1931 the deconvolution algorithm includes also the
minihalo, while the flux ratio refers only to the nuclear fluxes listed in  Table
\ref{tab:resu}.  The integral flux including the minihalo would be F$_{int} =55.16$
mJy.}
\label{tab:size}
\end{table}

\subsection{Extent of BCG radio emission}

In our radio images, we are not able to identify clear extended emission, despite the fact that jets and radio lobes are
expected in any cool core, independently of the detection of X-ray cavities.  Our peak and
integrated fluxes are representative of the nuclear power, with the inclusion, if any, of some extended emission
corresponding to the base of the jets, or to compact extended radio emission not directly associated with the 
nuclear BCG emission such as minihalos.  In a systematic study based on the VLA archive \citep{2014Giacintucci},
minihalos have been detected in two clusters of our sample: RXJ1532 \citep[see also][]{2013Hlavacek} and MACSJ1931,
with an additional candidate found in MACSJ0329. Despite the A configuration of JVLA being less sensitive to 
extended sources, we present here a very preliminary investigation of the source sizes based on our high resolution data.

The existence of extended structures can be estimated by comparing the beam size with the deconvolved size of our 
sources, obtained by PyBDSF.  The deconvolved (DC) sizes and the ratio $F_{int}/F_{peak}$ are listed in Table 
\ref{tab:size}. Roughly we find that the deconvolved size correlates with the ratio  $F_{int}/F_{peak}$, as expected.
Formally, the measurement errors on the deconvolved size are negligible (of the order of $\sim 1$\%) 
but they do not include possible smearing of the image due to small errors in the phase calibration.  Therefore, 
we should use a conservative criterion to asses the extent of a source.  

We notice that the highest $F_{int}/F_{peak}$ values (above 1.3) are associated with deconvolved sizes typically larger than half 
the beam size.  Based on this criterion, we classify three sources (Abell 383, RXJ2129 and MACSJ1931) to be clearly 
resolved\footnote{For Abell 383 and RXJ2129 the presence of a non core component has already been shown 
in \citet{2015aHogan}}.  MACSJ1931 has the largest size and flux ratio, mostly because of its minihalo, 
which lies 2.8 arcsec offset from the BCG and with a peak flux of 2.1 mJy, as shown in \citet{2014Giacintucci}. 
The deconvolved size of MACSJ1931 also includes the minihalo.  

There are three other sources with  $1.1< F_{int}/F_{peak}<1.2$, whose deconvolved sizes are about half of the beam. 
We classify these sources (MS2137, MACSJ0329, MACSJ1423) as tentatively resolved.  Finally, the remaining
5 sources (Abell 611, RXJ1532, MACSJ1720, MACSJ0429 and MACSJ0744) are unresolved with present data.  A discussion on 
the presence of non core emission for some of the sources not observed in our program (namely  MACSJ1115, Abell 2261, MACSJ1347 and MACSJ1206) can be found in the Appendix of \citet{2015aHogan}.

\subsection{BCG Spectral properties}
\label{alpha}

\begin{table*}[htbp]
\centering
\caption{Flux densities in the frequency range 150 MHz-30 GHz and Spectral indexes for the CLASH BCGs}
\footnotesize
\begin{tabular}{lr@{$\pm$}lrrrrrcc}
 \hline
 \multirow{2}{*}{Cluster}  & \multicolumn{2}{c}{$F_{\rm 150MHz}^{1}$}  & $F_{\rm 330MHz}^{2}$ & $F_{\rm 5GHz}^3$ &  $F_{\rm 10GHz}^3$ & $F_{\rm 28.5GHz}^4$ & $F_{\rm 30 GHz}^5$ & $\alpha_{fit}$  & $\alpha_{1.5}^{30}$ \\
    &  \multicolumn{2}{c}{(mJy)}  &  (mJy)    &  (mJy)    &   (mJy)    &     (mJy)   &     (mJy)   &     &         \\
\hline
Abell 383 & 224.8 & 23.0 & - & - & - & 4.40 $\pm$ 0.50 & 4.3 $\pm$ 0.2 & -0.72 $\pm$ 0.01 & -0.72 $\pm$ 0.02  \\ 
Abell 209 & \multicolumn{2}{c}{-} & - &  - & - & - & - & - & -  \\
Abell 1423 & \multicolumn{2}{c}{-} & &  - & - & - & - & - & -   \\
RXJ2129 & 114.8  & 12.4 & - &  9.05 $\pm$ 0.07 & 4.2 $\pm$ 0.1 & 2.33 $\pm$ 0.14 & 2.6 $\pm$ 0.2 & -0.79 $\pm$ 0.03  & -0.72 $\pm$ 0.03 \\
Abell 611 & \multicolumn{2}{c}{-} & - &  $0.45\pm 0.05$& -  & - & - &  - & -  \\
MS2137 &  \multicolumn{2}{c}{-} & - &  $1.0^7$ & -  & - & - &  - & -  \\
RXJ1532 & 52.2  & 7.1 & 71 $\pm$ 3.6 &  8.82 $\pm$ 0.08 & 6.30 $\pm$ 0.1  & 3.25 $\pm$ 0.22$^8$ & 3.2 $\pm$ 0.3 &  -0.52 $\pm$ 0.03 &-0.54 $\pm$ 0.02 \\
MACSJ1931 & 6315.0  &  631.7& - &  - & - & -  & - & -0.72$^{5}$ & -\\ 
MACSJ1720 & 119.7  & 13.9& 103 $\pm$3.0 &  - & - & - & 1.8 $\pm$ 0.4 & -0.89 $\pm$ 0.07 & -0.87 $\pm$ 0.08  \\
MACSJ0429 & 106.2  & 12.1 	& - &  - & -  & - & 18.2 $\pm$ 0.2 & -0.61 $\pm$ 0.11 & -0.65 $\pm$ 0.01 \\ 
MACSJ0329 & \multicolumn{2}{c}{-} & - &  - & -  & - & 0.3 $\pm$ 0.4$^4$ &  - & -0.80 $\pm$ 0.45  \\ 
MACSJ1423 & 78.5  & 12.7& 27 $\pm$ 2 $^{6}$&  - & -  & 1.49 $\pm$ 0.13 & 2.0 $\pm$ 0.2 & -0.37 $\pm$ 0.10  & -0.25 $\pm$ 0.04  \\ 
MACSJ0744 & \multicolumn{2}{c}{-} & - &  - & -  & - & - &  - & -  \\
\hline
Abell 2261 & 33.0  & 5.9 & 36 $\pm$3.4&  0.59 $\pm$ 0.05  & - &  0.20 $\pm$ 0.30 & - &-1.24 $\pm$ 0.13 &-0.95 $\pm$ 0.52\\
RXJ2248  & \multicolumn{2}{c}{-} & - & - & - & - & - & - & - \\
MACSJ1115  & 138.1  & 14.4& - & - & $<3.8$ & - &   1.4 $\pm$0.4 &  -1.21 $\pm$ 0.14 & -0.59 $\pm$ 0.03 \\ 
MACSJ1206 & 2154.3  & 215.7& - & - & - & - & - &  - & -  \\
RXJ1347  & 215.2 & 22.3& - & - & $17.8 \pm 3.0$ & 10.38 $\pm$ 0.47 & 8.7 $\pm$ 0.2 &  -0.56 $\pm$ 0.02 & -0.56 $\pm$ 0.02  \\
MACSJ1311 & \multicolumn{2}{c}{-} & - & - & - & - &  - & - & - \\
ClJ1226  & \multicolumn{2}{c}{-} & - & - & - & - & $0.3 \pm 0.2^4$  & - & -0.83 $\pm$ 0.23 \\
% MACSJ0647 & \multicolumn{2}{c}{-} & - & - & - & - & - & - & - \\
% MACSJ2129 & \multicolumn{2}{c}{-} & - & - & - & - & - & - & - \\
% MACSJ1149 & \multicolumn{2}{c}{-} & - & - & - & - & - & - & - \\
% MACSJ0416 & 47.7  & 6.2 & - & - & - & - & - & - & - \\
% MACSJ0717 & \multicolumn{2}{c}{-} & - & - & - & - & - & - & - \\
\hline
\end{tabular}
\label{tab:alpha}
\tablerefs{~[1]~ TGSS \citep{2017Intema} ~[2]~ WENSS\citep{1997WENSS}  ~[3]~ \citet{2015aHogan}, 
~[4]~\citet{2013Sayers}, ~[5]~\citet{2012Bonamente}, ~[6]~\citet{2008Birzan}, ~7\citet{1994Gioia} (error for this source is not 
listed),~8 \citet{2007Coble}.}
\end{table*}

The spectral energy distribution (SED) of a BCG in the radio band is usually decomposed into a 
nuclear component and an extended one. The nuclear component is directly linked to the AGN and 
shows a rather flat spectral energy distribution with an energy index $\alpha <0.5$ \citep[see][]{2015aHogan}.  
The core component may show synchrotron self-absorption, or, in some cases, free-free absorption, at around few 
GHz, but usually it remains flat to frequencies up to several GHz.  The extended component, on the other hand, is mostly 
associated with lobe emission, and therefore is generated by an older, relativistic electron population accelerated during past 
nuclear activity.  Other forms of emission surrounding the BCG may be due to processes not related to the nuclear activity, as
in the case of minihalos, appearing as spherically symmetric, small scale (a few 10$^2$ kpc), with a steep radio spectrum, 
probably originating from electrons accelerated {\sl in-situ} by the turbulent motion of the ICM in the core 
\citep[hence, indirectly due to the nuclear activity, see][]{2014Giacintucci}. In general, this steeper component is less 
prominent at 1.5 GHz.  

Usually, the SED of BCGs can be modeled with two components corresponding to the different central activities.  However,  
modeling two components goes beyond our capability given the present data, and therefore that effort is postponed to a future 
work, which will include also our 2-4 GHz data.  To achieve a preliminary characterization, we model the spectra of our BCGs 
with a single power law defined as $S_{\nu} \propto \nu^{\alpha}$, where $S_\nu$ is the flux energy density as a function of the 
frequency $\nu$.  Our goal is to derive an effective spectral index
that can be used to apply the k-correction when computing the radio power at different redshifts.  Therefore, we 
collect all the radio measurements in the frequency range 150 MHz to 30 GHz from the literature (the data coverage
above 30 GHz is too sparse to be useful). The radio SED of our BCGs are shown in Table \ref{tab:alpha}, where the flux densities 
are sparsely sampled at six different frequencies to complement the 1.5 GHz flux densities measured in this work.  

% collected from the 150MHz IFR–GMRT Sky Survey (TGSS), the 330MHz Westerbork Northern Sky Survey (WENSS), ...
% In principle, we can measure the spectral slope around 1.5 GHz in a 1 GHz bandwidth. However, the spectral slope 
% measured in this way strongly depends on the calibration of spectral windows, and, given the relatively faint  fluxes, the 
% local slope is affected by large measurement errors. 

We fit the SED with a single power law when at least three points are present, deriving an average 
spectral slope $\alpha_{fit}$ when the $\chi^2$ is acceptable.  Then we compute the index 
$\alpha_{1.5}^{30}\equiv {\tt log}(F_{30 GHz}/F_{1.5 GHz})/{\tt log}(30 GHz/1.5 GHz)$ as a proxy of the average
spectral slope.  We note that the values of $\alpha_{1.5}^{30}$ and $\alpha_{fit}$ are always consistent when $\alpha_{fit}$ is 
available (see Table \ref{tab:alpha}).   In the few cases where we have no means to compute a proxy for the spectral index, we 
simply assume $\langle \alpha \rangle = -0.7$ to compute the k-correction. In the Appendix we show the SED in the 
150 MHz-30 GHz range for 7 BCGs observed in our JVLA program and for 4 with FIRST counterparts for which we are able to
measure  $\alpha_{1.5}^{30}$.  We also show the lines corresponding to the index $\alpha_{1.5}^{30}$, the reference slope 
$\langle \alpha \rangle = -0.7$, and when possible, the best-fit power law with slope $\alpha_{fit}$.

\begin{figure}[htbp]
\includegraphics[width=0.46\textwidth]{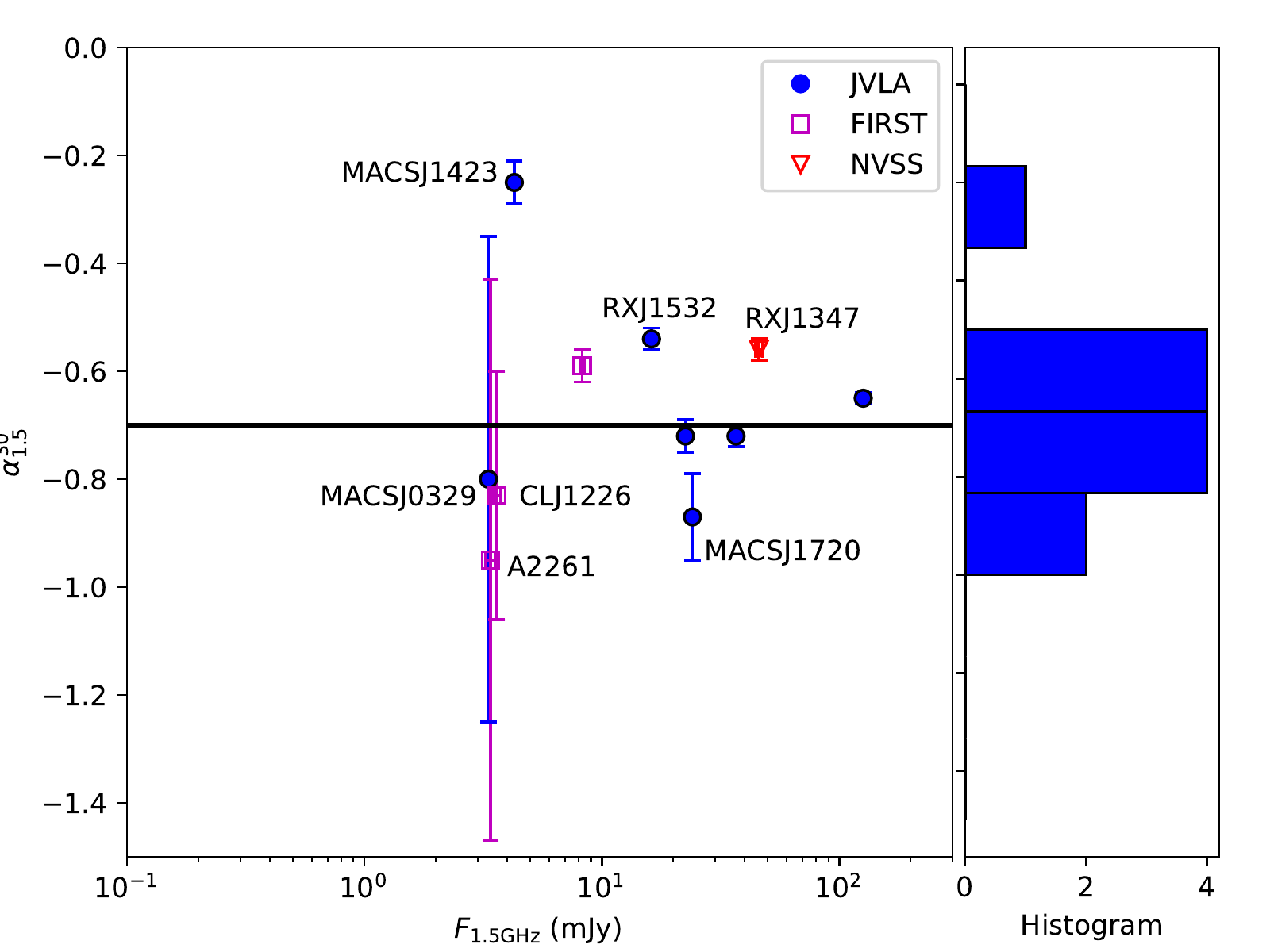}
\caption{Spectral index proxy $\alpha_{1.5}^{30}$ versus flux density and  histogram for seven BCGs 
observed with JVLA and presented in this work (solid circles).  Empty squares and triangles correspond to FIRST and
NVSS, respectively.  The black horizontal line marks the reference value $\alpha = -0.7$.}
\label{fig:ahist}
\end{figure}

Despite the broad agreement among the three spectral indices, we can still identify some sources whose spectra are clearly 
not well fitted by a single power law.  In particular, MACSJ1423 shows a hint of a steep component at low frequencies;
%  which is usually associated to lobe emissions 
% or diffuse emission caused by turbulence acceleration in the core regions.  
MACSJ0429 shows a GHz-peaked SED, possibly due to a self-absorbed core; finally, we are not 
able to distinguish the core and the minihalo emission in MACSJ1931 in the flux measurement at low frequencies
(the TGSS counterpart J193149.6-263432 has a size of $40" \times 33"$).  For these sources we are not able to derive a 
meaningful $\alpha_{fit}$.  For MACSJ1931, in Table \ref{tab:alpha}  we report the value of 
the spectral slope measured by \citet{2013Sayers}.  
The histogram of the spectral index $\alpha_{1.5}^{30}$ for the sources observed with JVLA or with FIRST counterpart 
is shown in Figure \ref{fig:ahist}.  Values of $\alpha_{1.5}^{30}$ range from -0.25 to $\sim -1$, with an average 
$\langle \alpha \rangle = -0.68$.   
% Also the index
% $\alpha_{1.5}^{30}$ computed for the sources with only FIRST detection is in line with our measurements. 
We find that the distribution of $\alpha_{1.5}^{30}$ is consistent with results obtained for the spectral shape of BCGs in 
NVSS \citep{2007Lin} and in the more recent work by \citet{2015aHogan}.

As discussed in the previous section, our high-resolution data are not 
sensitive to extended, low surface brightness emission and therefore mainly sample the
nuclear emission, with no possibility of separately identifying and analyzing an extended component.  
Therefore our average estimate of the spectral slope may be somehow affected by diffuse emission.
Despite this, the distribution of our measured average spectral slope is consistent with 
radio emission dominated by nuclear emission.  Therefore, for the sake of computing radio power,
we assume $\alpha = -0.7$ as the default choice when we are not able to 
derive a value for the spectral index, or rely on measurements presented in \citet{2013Sayers} in the case of MACSJ1931. 
We are aware that these results on the spectral shape are merely an approximation of the real spectral shape in the 
relevant frequency range, given the significant variety in the 
spectral shape of BCGs.  However, we conclude that $\alpha_{1.5}^{30}$ is still a useful quantity for estimating the 
k-correction, also considering the low redshift leverage of our sample.  We will improve our measurements of spectral slope when 
the 2-4 GHz data are fully analyzed.

\begin{table*}[htbp]
\centering
\caption{Radio power, central ICM entropy, UR-IR color, estimated SFR and total cavity power associated to the CLASH BCGs.}
\footnotesize
\begin{tabular}{lccccc}
 \hline
 \multirow{2}{*}{Cluster}  & $P_{\rm 1.5 GHz}$ & K$_0$ &   UV-IR  & SFR   & P$_{cav}$\\
       &  ($ 10^{24} W~Hz^{-1}$) & (keV ~cm$^{2}$) & (mag) & ($M_\odot~yr^{-1}$) & (10$^{44}$ erg/s) \\
\hline
Abell383 & 3.62 $\pm$ 0.02 & 13.0 $\pm$ 1.6 & 4.36 $\pm$ 0.04 & 3.29 $\pm$ 0.40  & 19 $\pm$ 7 \\
Abell209 & $<$ 0.010 & 105.5 $\pm$ 26.9 & 5.5 $\pm$ 0.1 & 1.2 $\pm$ 1.1  &  -\\
Abell1423 & $<$ 0.005 & 68.3 $\pm$ 12.9 & 4.96 $\pm$ 0.13 & 2.2 $\pm$ 0.4  & - \\
RXJ2129 & 3.55 $\pm$ 0.04 & 21.1 $\pm$3.7 & 4.98 $\pm$ 0.09 & 2.9 $\pm$ 0.4  &  -\\
Abell611 &  0.211 $\pm$ 0.005 & 124.9 $\pm$ 18.6 & 5.69 $\pm$ 0.14 & 0.90 $\pm$ 1.7  &  -\\
MS2137 & 0.418  $\pm$ 0.009  & 14.7 $\pm$ 1.9 & 4.07 $\pm$ 0.03 & 5.6 $\pm$ 0.7  &  -\\
RXJ1532 & 6.44 $\pm$ 0.08 & 16.9 $\pm$ 1.8 & 2.83 $\pm$ 0.04 & 48.6 $\pm$ 2.6  &  54 $\pm$ 22 \\
MACSJ1931 & 7.65 $\pm$ 0.02 & 14.6 $\pm$ 3.6 & 2.04 $\pm$ 0.04 & 83.1 $\pm$ 2.3  & 5 $\pm$2 \\
MACSJ1720 & 12.4 $\pm$ 0.4 & 24.0 $\pm$ 3.4 & 4.54 $\pm$ 0.05 & 6.1 $\pm$ 0.7  & 16 $\pm$7 \\
MACSJ0429 & 64.7 $\pm$ 0.3 & 17.2 $\pm$ 4.3 & 3.75 $\pm$ 0.05 & 20.1 $\pm$ 2.1  &  -\\
MACSJ0329 & 2.37 $\pm$ 0.42 & 11.1 $\pm$2.5 & 3.3 $\pm$ 0.03 & 31.0 $\pm$ 2.4  &  52 $\pm$ 20\\
MACSJ1423 & 3.77 $\pm$ 0.11 & 10.2 $\pm$ 5.1 & 3.14 $\pm$ 0.02 & 16.7 $\pm$ 1.2  &  15 $\pm$ 6 \\
MACSJ0744 & 0.51 $\pm$ 0.06 & 42.4 $\pm$ 10.9 & 4.6 $\pm$ 0.13 & 8.5 $\pm$ 3.1  & 85 $\pm$ 39 \\
\hline
Abell 2261   & 0.48 $\pm$ 0.07 & 61.1 $\pm$ 8.1 & 5.47 $\pm$ 0.07 & 3.3 $\pm$ 2.8 & - \\
RXJ2248  & -  & $42.0 \pm 10.0$ & $4.91\pm 0.04$ & $2.29\pm 0.05$ & - \\
MACSJ1115 & 3.01 $\pm$ 0.08 & 14.8 $\pm$ 3.1 & 3.38 $\pm$ 0.02 & 6.4 $\pm$ 0.5 &  - \\
MACSJ1206 & 99.9 $\pm$ 3.9  & 69.0 $\pm$ 10.1 &  4.5 $\pm$ 0.05 & 6.8 $\pm$ 3.0 & -  \\
RXJ1347  & 28.72 $\pm$ 1.8 & 12.5 $\pm$ 20.7 & 3.81 $\pm$ 0.03 & 16.5 $\pm$ 1.8  & - \\
MACSJ1311 & - &  - & - & -  & - \\
ClJ1226 &  12.43 $\pm$ 2.27 & 166.0 $\pm$ 45.0 & 5.37 $\pm$ 0.17 & 2.7 $\pm$ 1.5 &  - \\
\hline
\end{tabular}
\label{tab:bcg_p}
\end{table*}

\subsection{Radio luminosity and correlation with SFR and ICM entropy}

\begin{figure}
\includegraphics[width=0.46\textwidth]{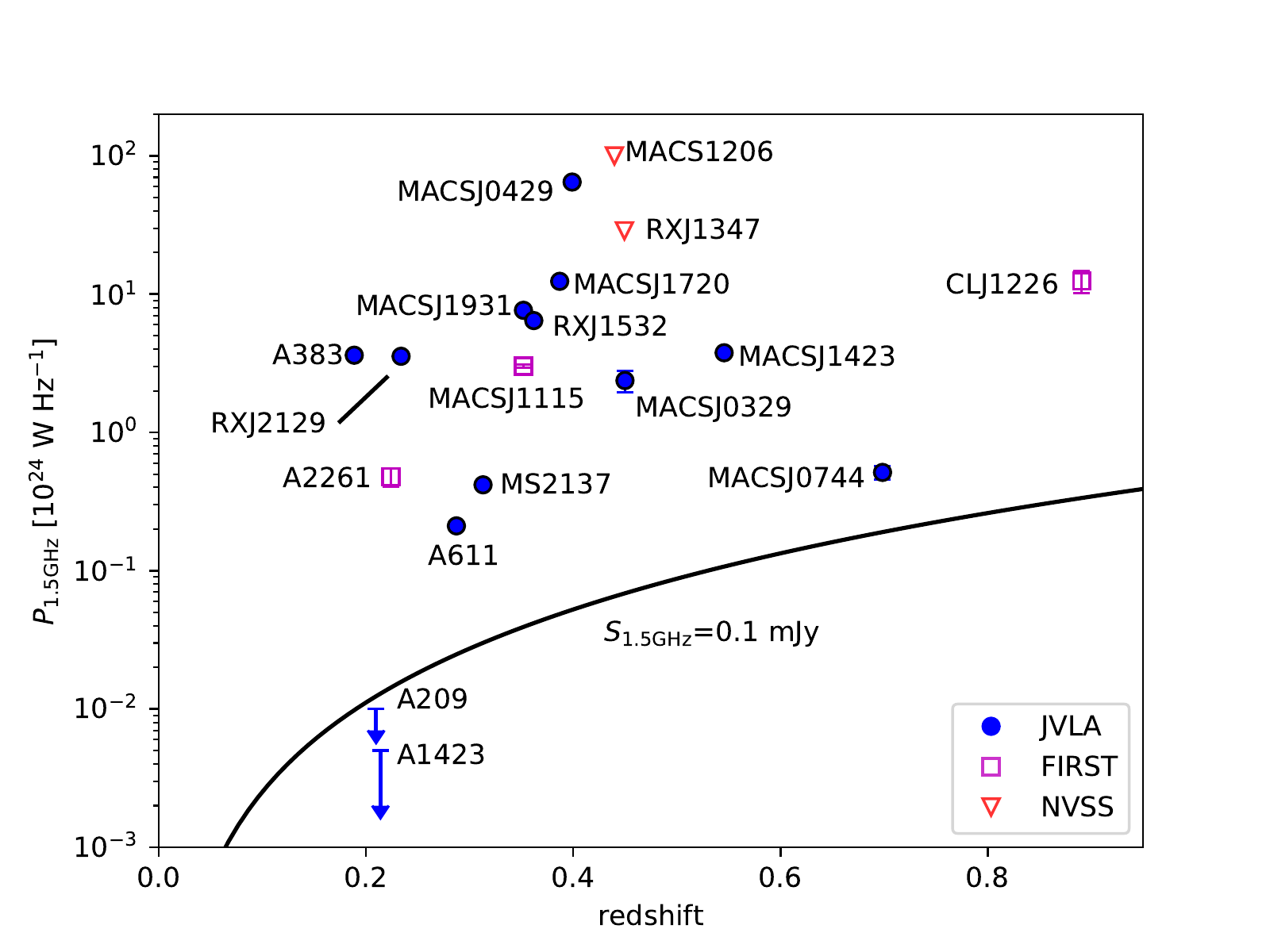}
\caption{Density distribution of the 1.5 GHz rest-frame absolute luminosity density distribution of the BCGs.  
Error bars are too small to be visible here. 
% Three red squares and two red triangles are CLASH clusters which have not been observed by JVLA yet.
The continuous line represents the luminosity corresponding to the observed flux density of 0.1 mJy with an average
spectral slope of $\alpha = -0.7$, and it is the average limit of our detection (corresponding to $S/N=5$ and 
assuming a noise of 0.02 mJy per beam).}
\label{fig:rlf_bcg}
\end{figure}

The emitted power density at 1.5 GHz in the rest frame of a source is derived from its flux density and spectral slope.
We compute the radio power at 1.5 GHz as

\begin{equation}
 L = 4 \pi D_l(z)^2 F_{int} \times (1+z)^{-(1+\alpha)} \, \, \,  {\rm W~Hz^{-1}}   \, ,
\end{equation}

\noindent
where the k-correction is computed as $(1+z)^{-(1+\alpha)}$ and $D_l(z)$ is the luminosity distance assuming the 
cosmological parameters quoted in Section \ref{sec:intro}.  The distribution of radio power of the 11 BCGs whose radio 
emission is  detected in our data is shown in Figure \ref{fig:rlf_bcg}, where we also include the five BCGs with FIRST and NVSS 
fluxes. The range of radio luminosities of our BCGs spans more than two and a half orders of magnitude.  We have 
$23.29 < \log(L_R) < 24.85$ for 11 BCGs, and $\log(L_R) > 25.3$ (therefore above the knee of the BCG radio luminosity 
function) for 3 BCGs, with MACSJ1226 reaching the highest luminosity of $\log(L_R) = 26.0$.  
We note that the detection of a few very bright sources in a small sample of cool core clusters is consistent with 
the radio luminosity function of BCGs in a comparable X-ray sample.  In particular, cool-core clusters have a frequency of 
BCGs with radio power $>10^{25}$ W Hz$^{-1}$ at least $3-5$ times larger than non-cool-core clusters 
\citep[see][]{2015aHogan}. 
% Similar as recently obtained from a large sample
% of more than 7000 BCG from NVSS \citep[see][]{2016Yuan}.

We present a preliminary comparison of radio luminosity with properties of the surrounding ICM and star formation 
rate (SFR) measured in the BCGs \citep[see][]{2015Donahue}.  We also obtain the central X-ray gas entropy of our clusters
from the cluster sample in the Archive of Chandra Cluster Entropy Profile Tables (ACCEPT)  \citep{2009Accept}, updated with the 
revised values in \citet{2015Donahue} when needed.
All these quantities are listed in Table \ref{tab:bcg_p}. We expect to find a clear difference in the radio properties of BCGs 
depending on the cluster core properties, as already shown in the literature.
As \citet{2008Cavagnolo} already pointed out on the basis of a lower redshift sample, and 
also confirmed by \citet{2008Rafferty}, high-power BCG radio sources only inhabit clusters with low central gas entropy, 
with a threshold at $K_0= 30$ keV cm$^2$, roughly corresponding to a cooling time of $5\times 10^8$ yr.  
Also, star formation activity appears to be ubiquitous in BCG hosted by a cool core with  $K_0 < 30$ keV cm$^2$
\citep{2015Fogarty}. More comprehensive studies also showed that all BCGs with a low central entropy (with emission lines
linked to ongoing star-formation events) are detected as radio sources \citep[][]{2015aHogan} and as star forming 
galaxies \citep{2017Fogarty}, pointing toward a common fueling source from the hot ICM 
for both nuclear activity and star formation.

\begin{figure}[htbp]
\includegraphics[width=0.46\textwidth]{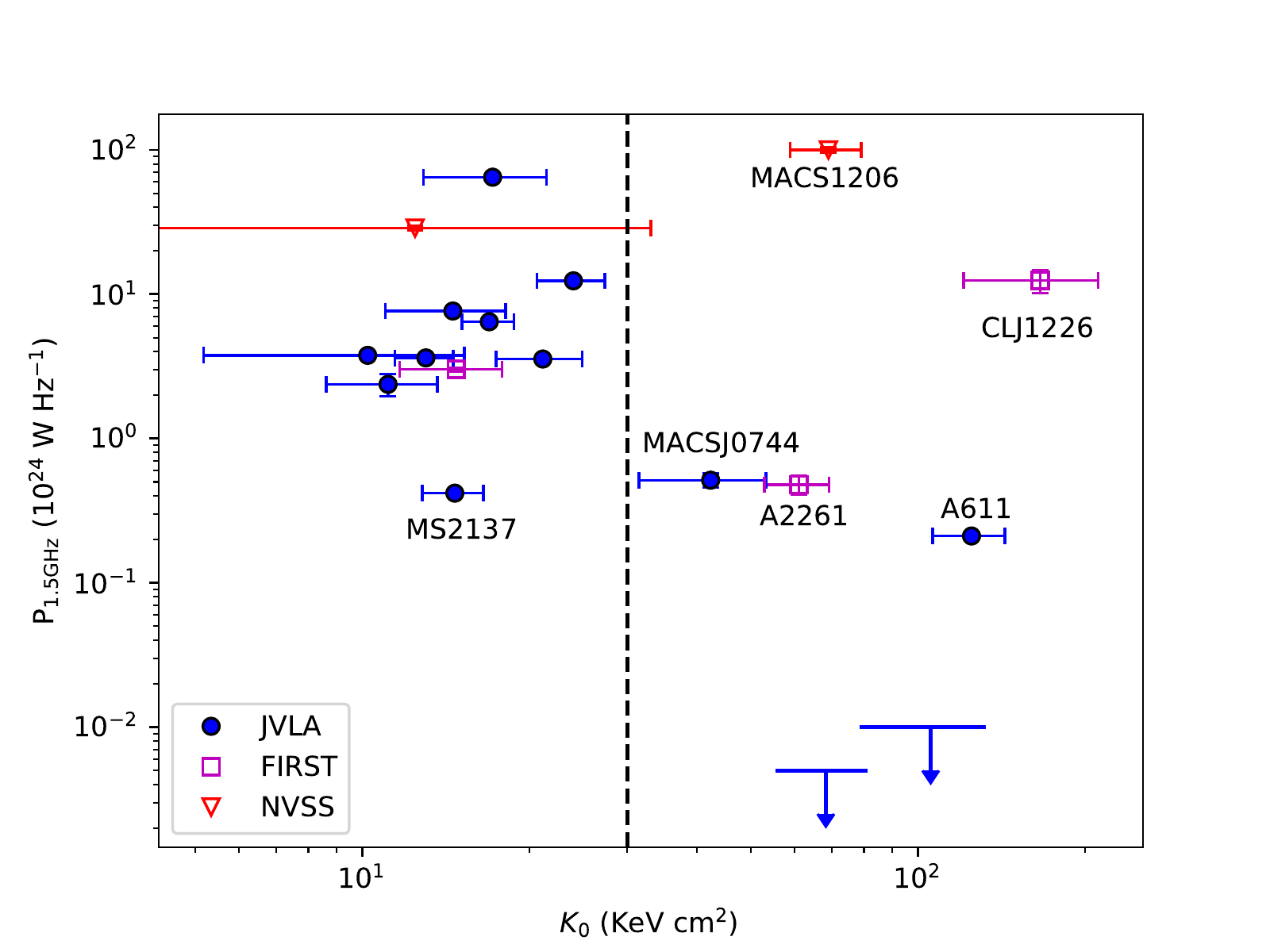}
\caption{Nuclear radio power of BCG measured in this work versus the central X-ray gas entropy as estimated in ACCEPT 
\citep{2009Accept} and \citet{2015Donahue}. The dashed line corresponds to the threshold $K_0 = 30$ keV cm$^2$ as indicated by 
\citet{2008Cavagnolo}  as the transition between clusters hosting BCGs with multi-phase gas, 
radio sources, and star formation, and clusters hosting quiescent BCGs. Solid circles correspond to the sources observed with JVLA in this work, while empty squares and triangles are obtained from FIRST and NVSS, respectively.}
\label{fig:rk}
\end{figure}

The relation between the central ICM entropy and the radio luminosity of BCGs in our sample  is shown in Figure 
\ref{fig:rk}.  In particular, the threshold $K_0= 30$ keV cm$^2$ efficiently identifies the radio-luminous BCGs. 
For values $K_0 < 30$ keV cm$^2$  we find luminosities mostly in the range $10^{24}-10^{25}$ W Hz$^{-1}$, with three
sources equal to or above $10^{25}$ W Hz$^{-1}$.  Five of the seven BCGs above $30$ keV cm$^2$ have radio power density of a 
few $\times 10^{23}$ W Hz$^{-1}$ or lower.  However, two of them (MACSJ1206 at $z=0.44$  and CLJ1226 at $z=0.89$,
 with fluxes from NVSS and 
FIRST, respectively) are in strong  contrast with this picture.  To better quantify the presence of high radio power sources 
in high entropy cores, we consider the cumulative luminosity function presented in \citet{2015aHogan}, where line-emitting BCGs
can be associated with low entropy ($K_0<30 $ keV cm$^{2}$) cores, and non-line-emitters with high entropy cores.  The fraction of 
sources with radio power larger than $10^{25}$ W Hz$^{-1}$ at $K_0<30 $ keV cm$^{2}$ is 20-30\%, in line with our value
of 3/11.  On the other hand, the fraction of luminous sources at $K_0>30 $ keV cm$^{2}$ is 5-10\%, lower than our value 2/7.  
Clearly our results, based only on two sources, and on a limited sample (we do not consider the five dynamically disturbed 
CLASH clusters in this work) do not allow to draw any conclusions.
If this is due to some evolution with redshift in the ICM properties in 
the core or in the radio properties of BCG, is a topic that must be investigated with a refined analysis of the {\sl Chandra} 
X-ray data and high resolution JVLA data.  In particular, MACSJ1206 is the target of an approved {\sl Chandra} proposal in AO19
for a deep exposure of 180 ks (PI S. Ettori).

Finally, the radio emission in Abell 2261 has been discussed 
extensively in \citet{2017Burke}, where it has been found to be associated with a compact radio relic, with a steep spectrum, 
and with a significant offset from the BCG nucleus.  Although this relic is most probably associated with nuclear activity
recently switched off, this source is definitely different from that expected from a radio active nucleus, and therefore it may 
not share the same properties of our sample.

\begin{figure}[htbp]
\includegraphics[width=0.46\textwidth]{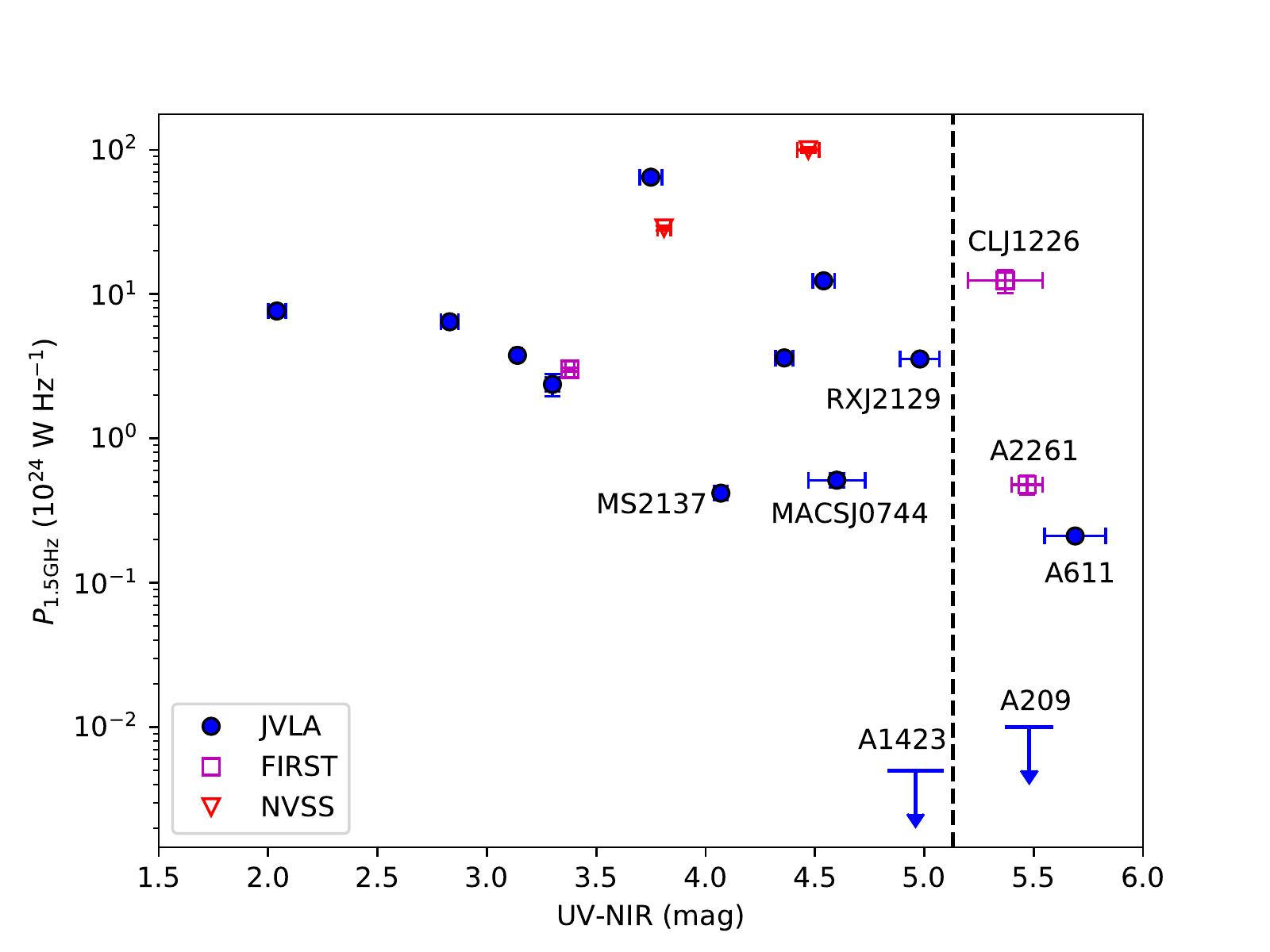}
\caption{Radio power of BCGs measured in this work plotted vs the rest frame UV (280 nm) - NIR (1 $\mu$) color 
of BCGs \citep[from][]{2015Donahue}. The dashed mark the threshold $UV-NIR= 5.13$ which  
is the average color of quiescent BCGs in CLASH sample.   Solid circles correspond to the sources observed with JVLA in this work, while empty squares and triangles are obtained from FIRST and NVSS, respectively.}
\label{fig:uvir}
\end{figure}

Below the $30$ keV cm$^{2}$ threshold, BCGs are observed to have ongoing star formation and multiphase gas,
as already pointed out by \citet{2015Donahue}. The UV-NIR color is a reliable proxy of the instantaneous star 
formation activity of a galaxy, by comparing the rest-frame 280 nm UV emission contributed by young hot stars
to the 1 $\mu$m peak of the stellar-light spectrum from evolved stars.  Note that the excess UV luminosity does not take 
into account obscured star formation.  So the quoted SFRs should be considered lower limits to the total star formation rate for 
these galaxies. The relation between the UV-NIR colors and the
SFRs in CLASH cluster BCGs is discussed in \citet{2015Donahue}.  In Figure \ref{fig:uvir} we plot the radio
luminosities of our BCGs vs the UV-NIR colors. The average color of quiescent BCGs in 
CLASH sample is 5.13 $\pm$ 0.35 \citep{2015Donahue} and it is shown as a vertical dashed line. 
We note a trend of higher radio luminosities associated with bluer UV-NIR colors, showing a significant presence
of star formation activity in radio luminous BCGs. As for the  two outliers in the radio power-entropy plot, 
CLJ1226 stands out in the upper right corner of the plot, with UV-NIR color larger than 5.2, while MACSJ1206 
has a UV-NIR color$\sim 4.5$.

\begin{figure}[htbp]
\includegraphics[width=0.46\textwidth]{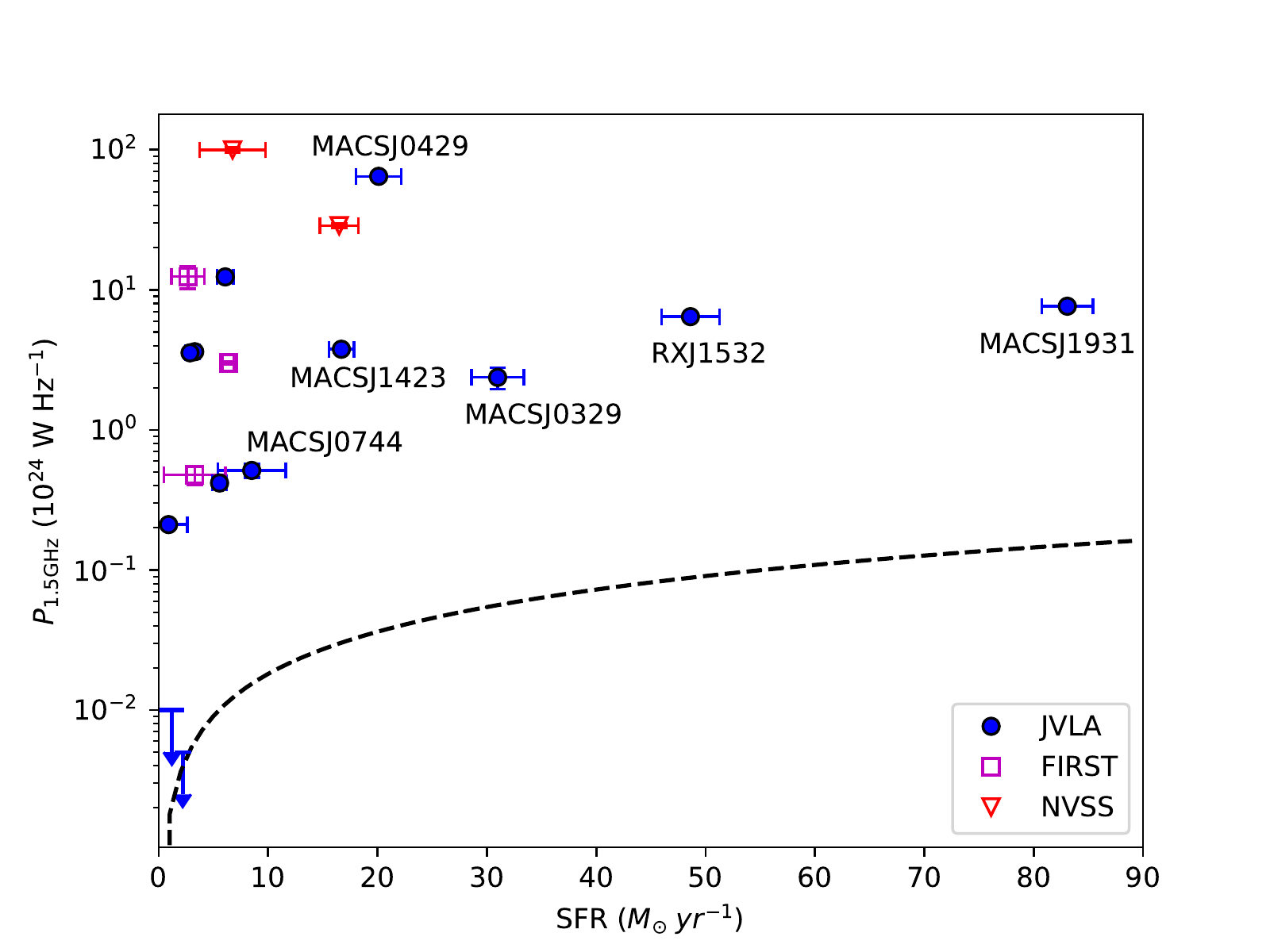}
\caption{Radio luminosity vs star formation rate as measured from the excess UV luminosity, after \citet{2015Donahue}. 
The dashed line shows the radio luminosity associated with star formation derived} by \citet{2003Bell}.
\label{fig:sf}
\end{figure}

An estimate of the SFR based on the UV luminosity is provided by \citet{2015Donahue}, where they used the conversion from the 
excess UV luminosity to an unobscured SFR following \citet{1998Kennicutt}.  This estimate has several sources of uncertainty:
the initial mass function of stars in BCGs may be different from that of the star forming galaxies used by 
\citet{1998Kennicutt}; in addition, the star formation events in the BCGs may be shorter and thus the BCGs younger than expected; 
finally, they applied no correction for dust-obscured star formation, for which IR-based measurements 
are required.  With these uncertainties in mind, we use these values to compare the radio power with 
the estimated SFR, finding that  the measured radio power is always more than one order of magnitude larger than that 
expected from star formation alone. This confirms the general assumption that the radio emission in BCGs is dominated
by nuclear emission, as also shown by \citet{2016Cooke}.  Only in two cases (Abell 611 and MACSJ0744) the contribution of 
the SFR at the 1.5 GHz flux density can be as high as 10\%.  This is clearly shown in Figure 
\ref{fig:sf} where we compare the radio power versus star formation rate of our BCGs with the average radio luminosity-SFR 
relation found by \citet{2003Bell}: 
\begin{equation}
 SFR = 5.52 \times 10^{-22} L_{1.5 {\rm GHz}}~ {\rm M_{\odot}~yr^{-1}} \, .
\end{equation}

The same conclusion is reached if we use a SFR measurement based on IR luminosity, and therefore not significantly affected
by obscuration.  For example, in the case of our strongest star forming BCG (in MACSJ1931) 
the SFR derived from Herschel data is $\sim 150\,  {\rm M_{\odot}~yr^{-1}}$ \citep{2016Santos}, as opposed to the 
value of $83\,  {\rm M_{\odot}~yr^{-1}}$ from \citet{2015Donahue}.  Even in this case, the expected contribution of the SFR 
to the radio emission is not larger than 5\% of the total flux.  We remark that the association 
of  higher star forming rates with the largest radio power, while the weakest radio sources
appear in BCGs with no detectable star formation in the UV \citep{2015Donahue}, does not imply that quenching is not 
happening.  In fact, if these radio sources were not dumping energy into the surrounding gas, the star formation rates would be 
much higher, as seen in simulations that do not include AGN feedback. 
% Finally, we remark that the correlation between nuclear radio emission (hence 
% mechanical-mode feedback) and high star formation rate may seem counterintuitive, as star formation is eventually
% quenched when the cooling process is halted, despite with a significant time-delay \citep[see][]{2016Molendi}.  
In addition, mechanical feedback is better traced by the extended emission from 
jets, while the nuclear radio emission is linked to the feeding of the SMBH, which, together with star formation events, 
is due to the cooling and condensation of the surrounding gas, as expected in top-down multiphase condensation models 
\citep[see][]{2017Gaspari}. 

The two sources with the faintest radio power density, Abell 611 and MACSJ0744, are both above the
entropy threshold $K_0=30$ kev cm$^{2}$, but are too faint to qualify as counter-examples to the pattern we see at 
low z. Being hosted by a weak cool core, they may not be accreting efficiently enough to be bright radio sources.  Still, 
it would be important to understand whether they are fading AGN or burgeoning AGN.  In any case, we can guess that they may 
be accreting at the Bondi rate from the hot gas, while the more luminous radio sources are fueled by cold gas, ultimately 
supplied via thermal instabilities in the hot gas  \citep[on this issue see][]{2013Russell,2006Allen}.  Abell 611 show a clear unresolved X-ray emission in the hard band, and the
BCG of MACSJ0744 is also a candidate X-ray AGN.  These are the only two detections of unresolved X-ray emission
in our sample together with MACSJ1931, which hosts a bright obscured AGN \citep[see][]{2016Santos}.  This may suggest
different modes of accretion marked by the presence of nuclear X-ray emission, as discussed in a forthcoming paper by our
team (Li-Lan Yang et al. 2018, in preparation).  

\subsection{Radio power and energetics of X-ray cavities}

A significant fraction of the feedback energy budget is stored in mechanical energy 
associated with large cavities carved into the ICM.  These cavities can be detected as circular or
ellipsoidal-shaped depressions in the projected X-ray surface brightness.
The energetics required to inflate the X-ray cavities may be approximated with 
a standard technique \citep[see][]{2004Birzan,2015Hlavacek-Larrondo} which consists
in computing the enthalpy of each bubble as $E_{bubble} = 4pV$,
where $p=n_e kT$ is the thermal electron (only) pressure of the ICM at the radius of the
bubble, and the electron density $n_e$ and the ICM temperature $kT$ are derived
from spatially-resolved spectral analysis.  Here, $V$ is the volume of the cavity, computed as
$V = 4\pi R_w^2R_l/3$, where $R_l$ and $R_w$ are the semi-major axes projected 
along directions parallel and perpendicular, respectively, to the
the jet (i.e., the direction connecting the BCG nucleus with the center of the cavity).  

Several CLASH clusters have already been searched for cavities.  
We consider the measurements of the cavity sizes presented in \citet{2016Shin} for a sample of
133 clusters with sufficient X-ray photons for their analyses.  Ten of the clusters in our sample are included in
the list of \citet{2016Shin}. The missing three are Abell 209, Abell 1423, and Abell 611.  Interestingly, 
the first two show no radio emission from the BCG nucleus, and Abell 611 is the second least luminous among
our BCGs. Abell 611 has only an upper limit to the cavity power from \citet{2013Hlavacek}.
Among the ten clusters in  \citet{2016Shin}, three have no cavities in their analysis (RXJ2129, MS2137, MACSJ0429), 
while there is at least one cavity for the remaining 7 clusters.  We measure $E_{bubble}$ using the projected values 
of $n_e$ and $kT$ from the ACCEPT cluster sample \citep{2009Accept} and \citet{2015Donahue}.  Clearly this is an approximation to the 
actual enthalpy of the bubble; however, the largest source of uncertainty is associated with the 
size of the bubbles (typically 20\% of the linear size).  In the case of multiple bubbles, the total value is obtained simply 
by summing the values of $E_{bubble}$ for each cavity.  A more
meaningful quantity is the average mechanical power, which is obtained by dividing the mechanical energy in each cavity
by the age of the cavity itself, approximated by the buoyancy time $t_{buoy} \sim R \sqrt{3C/8g r}$
\citep[see][]{2004Birzan}.  Here, $R$ is the distance between  the cluster core and the center of the bubble, $C$ is a 
drag coefficient, usually assumed to be $C\sim 0.75$, $g$ is the acceleration $\sim GM/R^2$, where $M$ is the 
total mass within $R$ \citep[taken from][]{2014Donahue}, and $r$ is the bubble size with uncertainties of 20\%.  
However, the uncertainty in these diagnostics may be severely underestimated, since the total mechanical power
depends on the number of detected cavities, and therefore depends also on the depth of the X-ray data or specific
properties of the surface brightness distribution of the clusters. 

\begin{figure}[htbp]
\begin{center}
\includegraphics[width=0.49\textwidth]{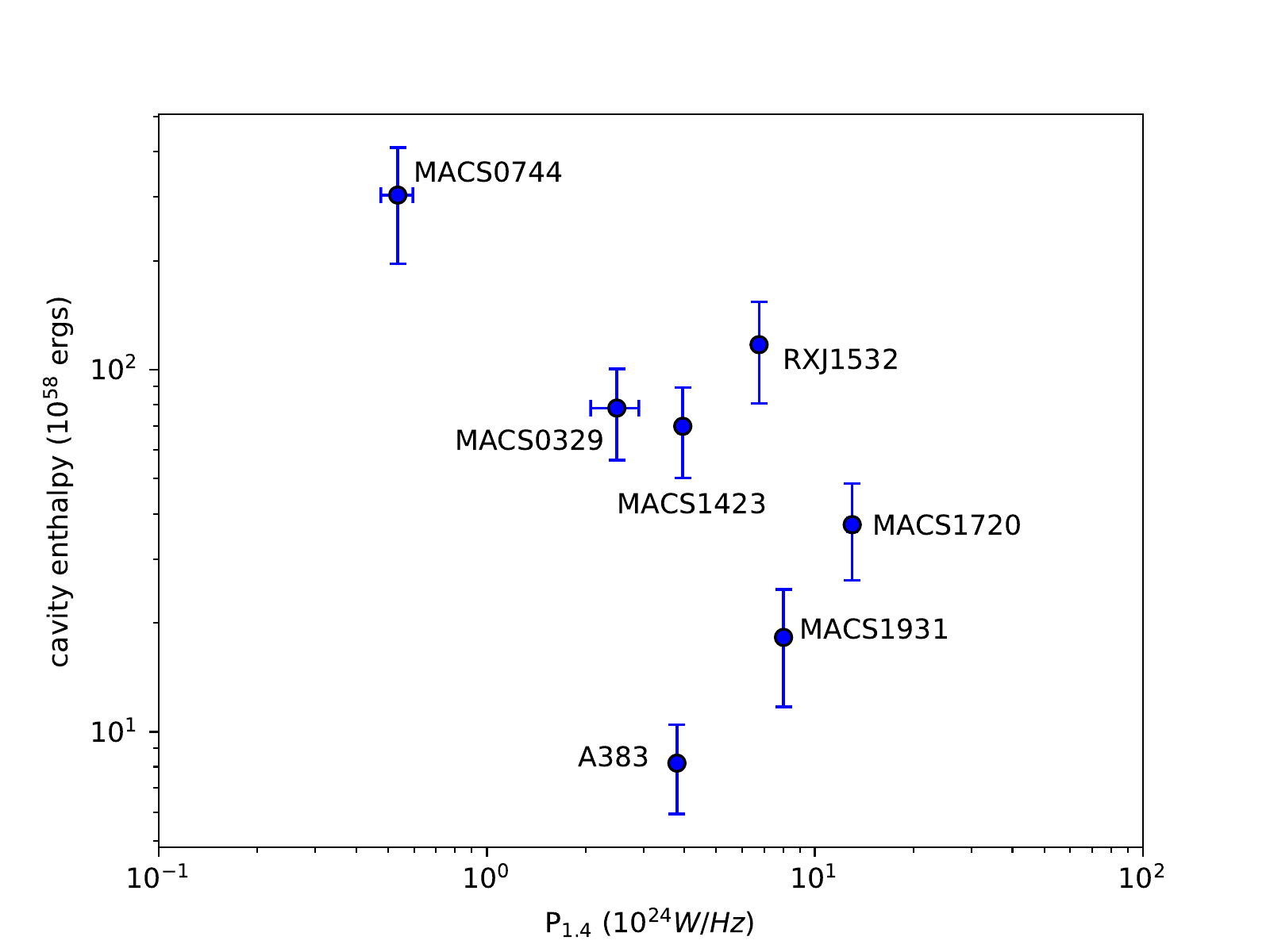}
\includegraphics[width=0.49\textwidth]{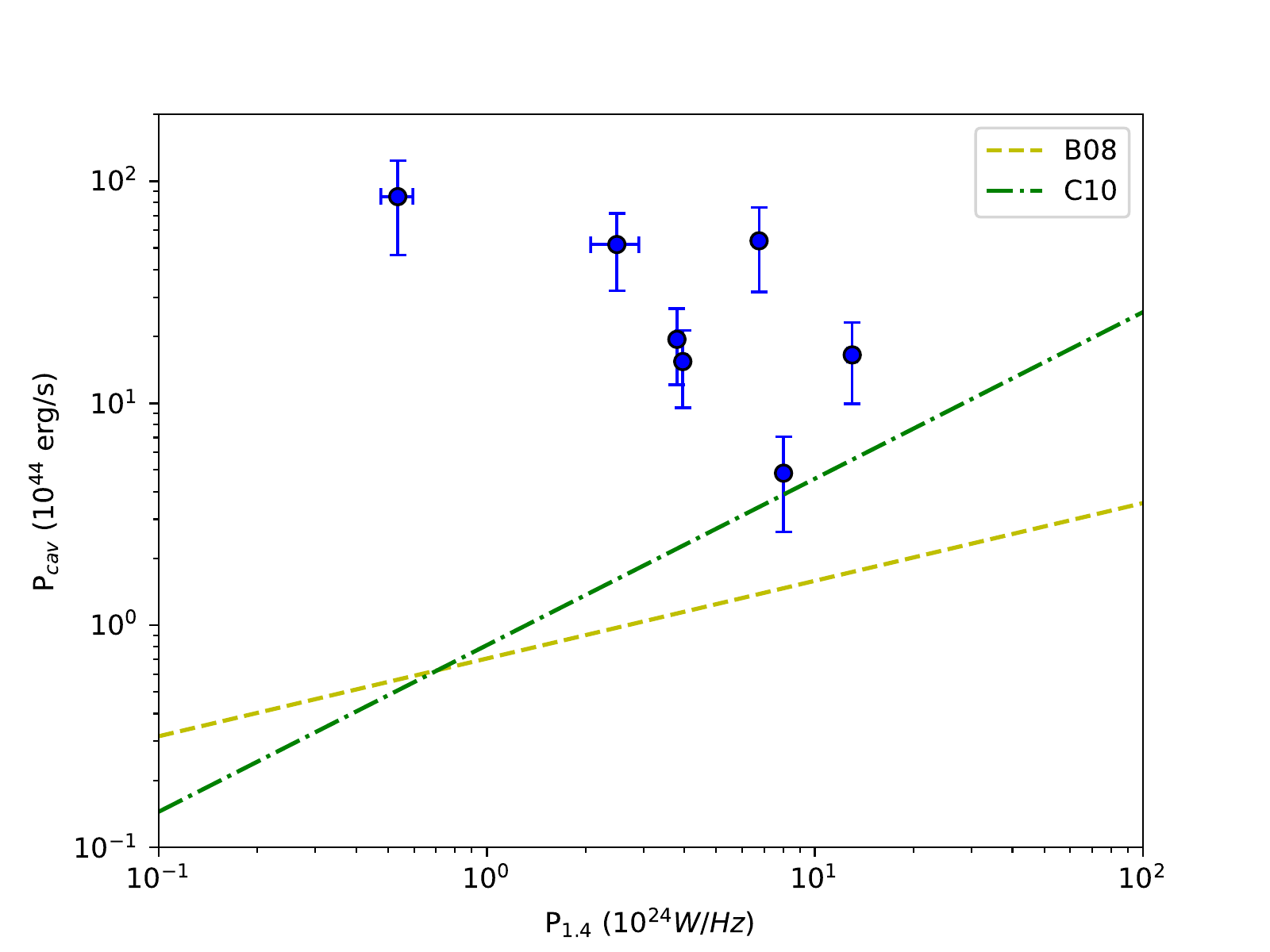}
\caption{Upper panel: the total enthalpy as measured from the size of the cavities, taken from \citet{2016Shin}, 
vs the radio power of the BCGs.  Lower panel: the average mechanical power, computed by dividing 
the enthalpy of each cavity by the buoyancy time, according to \citet{2004Birzan}.  The yellow dashed line represents
the best-fit power-law relation presented in \citet{2008Birzan}, 
while the green dotted-dashed line is the relation from \citet{2010Cavagnolo}.}
\label{fig:cavities}
\end{center}
\end{figure}

Despite these uncertainties, we compare the radio nuclear emission with the energy and the mechanical power stored in the 
ICM as observed in current X-ray data. In the upper panel of Figure \ref{fig:cavities} we plot the 
mechanical energy of the seven clusters in which cavities have been detected versus the 
nuclear radio luminosity of their BCG.  In the lower panel of Figure \ref{fig:cavities} we also plot nuclear radio power versus
the mechanical power obtained from the cavity size and position, for the same seven clusters. 
At first glance, our sources are not described by the average relations found in the literature
\citep[see, e.g.,][]{2008Birzan,2010Cavagnolo}, shown in the second panel. We observe a large intrinsic scatter between the 
average mechanical energy injected into the ICM and the instantaneous nuclear power of the BCG, and an average
mechanical power 
higher than in local clusters hosting BCGs with comparable radio power.  However, we are not able to draw any conclusions 
% on the correlation of instantaneous nuclear power and the average mechanical energy or power estimated from the cavities, 
mainly because of the small size and the limited luminosity range of our sample.  In addition, the sensitivity of X-ray 
observations of the CLASH sample does not guarantee a uniform sampling of cavities, particularly at low power (therefore
smaller size) and medium-high redshift. In fact, a large component of the observed 
scatter may be due to the difficulty in identifying and measuring ICM cavities in current data.  For example, the most discrepant 
cluster in Figure \ref{fig:cavities} is MACSJ0744, which is not listed by \citet{2013Hlavacek} among the MACS clusters
with cavities, but turns out to be the one with the largest mechanical power in our sample according to
\citet{2016Shin}, despite the large errors.  The cluster MACSJ0744 does not host an extremely strong cool core on the basis of its 
central entropy value $K_0\sim 42$ keV ~cm$^{2}$, so it can be interpreted as a case in which the cooling in the core has been recently 
quenched, while the outer halo still retain the imprint of the past mechanical-feedback activity. On the other hand, a 
positive correlation between the radio power and the average mechanical power is found in a much larger sample across four 
orders of magnitude in luminosity, despite the large scatter \citep[see][]{2008Birzan,2015aHogan}.  In general, we conclude 
that the nuclear power should be considered only an approximation of the past history of the central radio source within at least 
an order of magnitude, which possibly indicates that feedback may occur also as outflows and winds not associated with 
energetic radio jets.

\section{Conclusions}
\label{sec:conclusions}

In this work we present new high-resolution, medium-deep 1.5 GHz continuum  JVLA  observations of the BCGs 
of 13 CLASH clusters of galaxies at $0.18<z<0.69$.  Our results can be summarized as follows:

\begin{itemize}

\item We are able to characterize the radio properties of the nucleus in 11 BCGs, while 2 BCGs do not 
show radio emission in our data.  

\item We find a head-tail galaxy close to the BCG in the two non-detections (Abell 209 and Abell 1423).  
The fact that at least one of the clusters (Abell209) appears to be unrelaxed, as discussed in Section 4.2, suggests that the 
presence of head-tail radio galaxies may be a tracer of an unrelaxed dynamical state.

\item We find nuclear luminosities for the CLASH BCGs in the range from $10^{23}$ to $10^{26}$ $W~Hz^{-1}$; 
all our sources are consistent with being powered by an AGN, since their radio power is significantly larger
than the value associated with the measured star formation rate in the BCG.

\item Average radio spectral slopes are estimated with the index $\alpha^{30}_{1.5}$, defined as the flux density ratio 
between 1.5 and 30 GHz, and are found in the range from $\alpha^{30}_{1.5} \sim -1$ to $ -0.25$, with an average 
$\langle \alpha^{30}_{1.5} \rangle = - 0.68$, therefore consistent with synchrotron radiation from relativistic electrons in the 
nucleus.  

\item Most of our sources are consistent with being unresolved in our high-resolution data. Only for three cases (Abell 383, 
RXJ2129, and  MACSJ1931), the radio emission from the BCG is resolved with a high confidence level, suggesting a 
contribution from the base of jets.  The remaining sources are unresolved (5 sources) or marginally resolved (3 sources).

\item BCGs with high radio power in JVLA data are associated with low-entropy hot gas and higher SFR, indicating that stronger 
AGN activity may be correlated with more intense star formation. This correlation is consistent with the standard scenario in 
which the nuclear activity of the BCG is fueled by cooling of gas from the hot ICM, which also provides the reservoir for star 
formation.

\item We also investigate five sources in the CLASH sample not yet observed with JVLA, but with reliable counterparts in FIRST 
and NVSS.  Two of these sources (MACSJ1026 at $z=0.44$ and CLJ1226 at $z=0.89$) are unexpectedly found 
to have high nuclear radio power associated with a high-entropy core. This calls for a more in-depth 
multiwavelength analysis to investigate the nature of these sources.

\item We confirm a significant scatter between nuclear radio luminosity and average mechanical power derived
from the cavity size and ICM pressure.  However, we do not have the dynamic range nor the statistics to 
further investigate this correlation.
% property, and eventually, the duty cycle for the mechanical-mode feedback. 
\end{itemize}

Further progress in understanding the complex scenario of the baryon cycle in and around BCGs requires a massive 
and multiwavelength analysis, from the radio to the X-ray band. In our effort to provide a radio coverage of 
one of the best studied cluster samples such as CLASH, we are planning to extend our observations in the A configuration, L band, to 
the CLASH clusters  not included in this work and to use the 2-4 GHz data already acquired in a previous program by our group 
(VLA/13B-038, PI M. Aravena).  We also plan to propose for JVLA in the B and C configurations to search for extended radio 
emission like jets and lobes or cavity-filling, relativistic plasma.  In the meantime, we are currently mapping the 
entire field of view for our observations (30 arcmin on a side) to investigate the radio properties of CLASH member galaxies, 
exploiting the extensive spectroscopic follow-up of CLASH fields.  

\acknowledgments
We sincerely thank the referee Alastair Edge for his constructive suggestions and valuable comments, which greatly improved the manuscript.
We thank Heidi Medlin for her help in preparing the JVLA observing runs, and Maite Beltran, Julie Hlavacek-Larrondo, Hui Shi, 
and Marcella Massardi for help with the radio data reduction.  We thank Massimo Gaspari and Luisa Ostorero for useful 
discussions.  We also thank the referee, Alastair Edge, for his comments and suggestions, which significantly improved the 
quality of the paper.  This work was supported by the National Natural Science Foundation of 
China under Grants No. 11403002, the Bureau of International Cooperation of the Chinese Academy of Sciences under Grants No. GJHZ1864, the Fundamental Research Funds for the Central Universities 
and Scientific Research Foundation of Beijing Normal University. P.T. is supported by the Recruitment Program of High-end 
Foreign Experts and he gratefully acknowledges hospitality of Beijing Normal University.

\bibliography{clashradio.bib}

\newpage
\clearpage

\appendix

\section{Spectral energy distribution}

\begin{figure*}[htbp]
\caption{Radio SED of BCGs observed with our JVLA program obtained complementing our 1.5 GHz measurement with 
measurements at other frequencies available in the literature. Only sources with a measured $\alpha_{1.5}^{30}$ are shown.  
The dashed black line shows the reference slope normalized to the 1.5 GHz flux density, 
while the red solid line shows the slope corresponding to $\alpha_{1.5}^{30}$.  The magenta dotted line, when present, shows 
the best-fit power law obtained using all the available flux measurements.}
\includegraphics[width=0.48\textwidth]{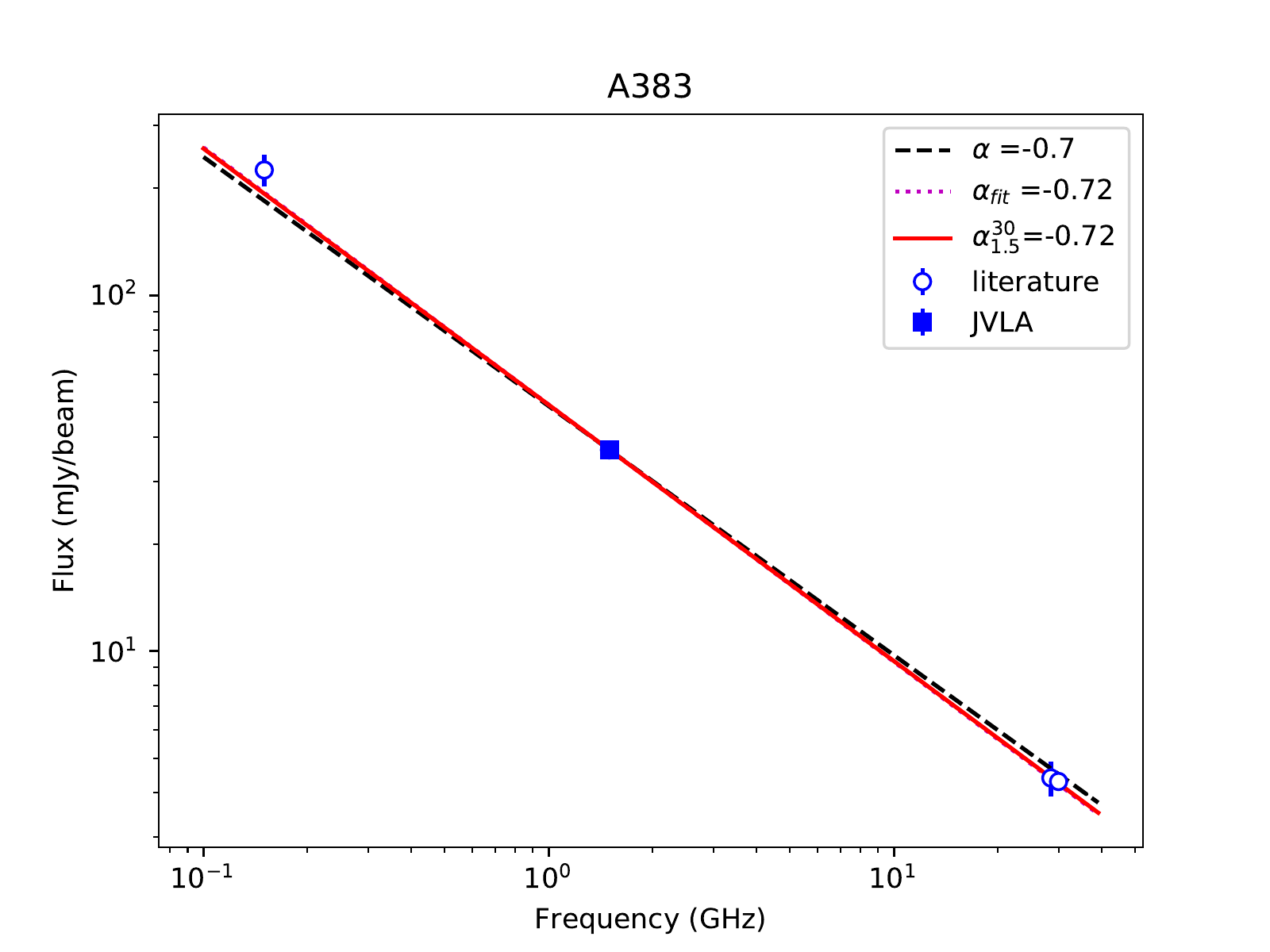}
\includegraphics[width=0.48\textwidth]{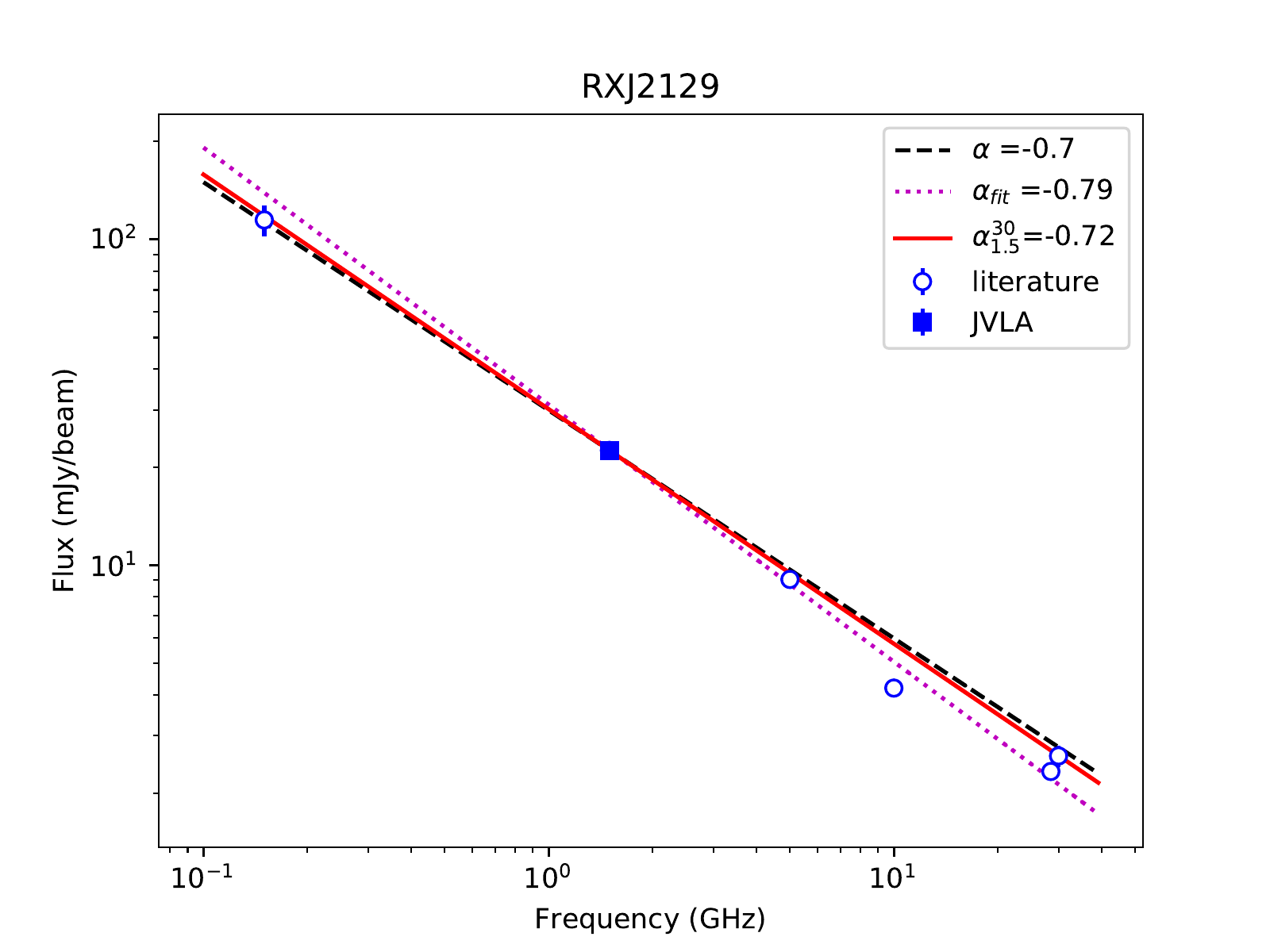}
\includegraphics[width=0.48\textwidth]{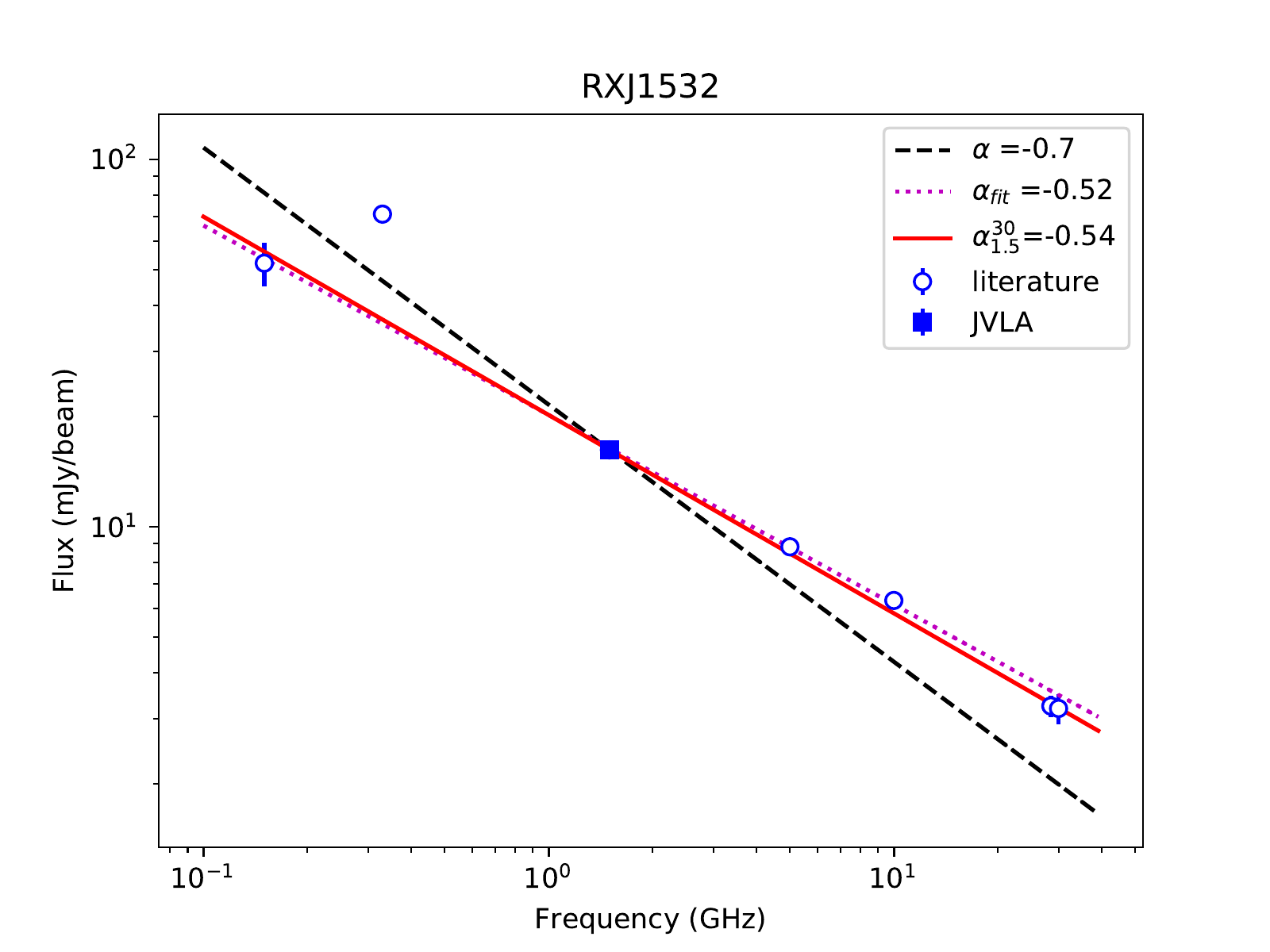}
\includegraphics[width=0.48\textwidth]{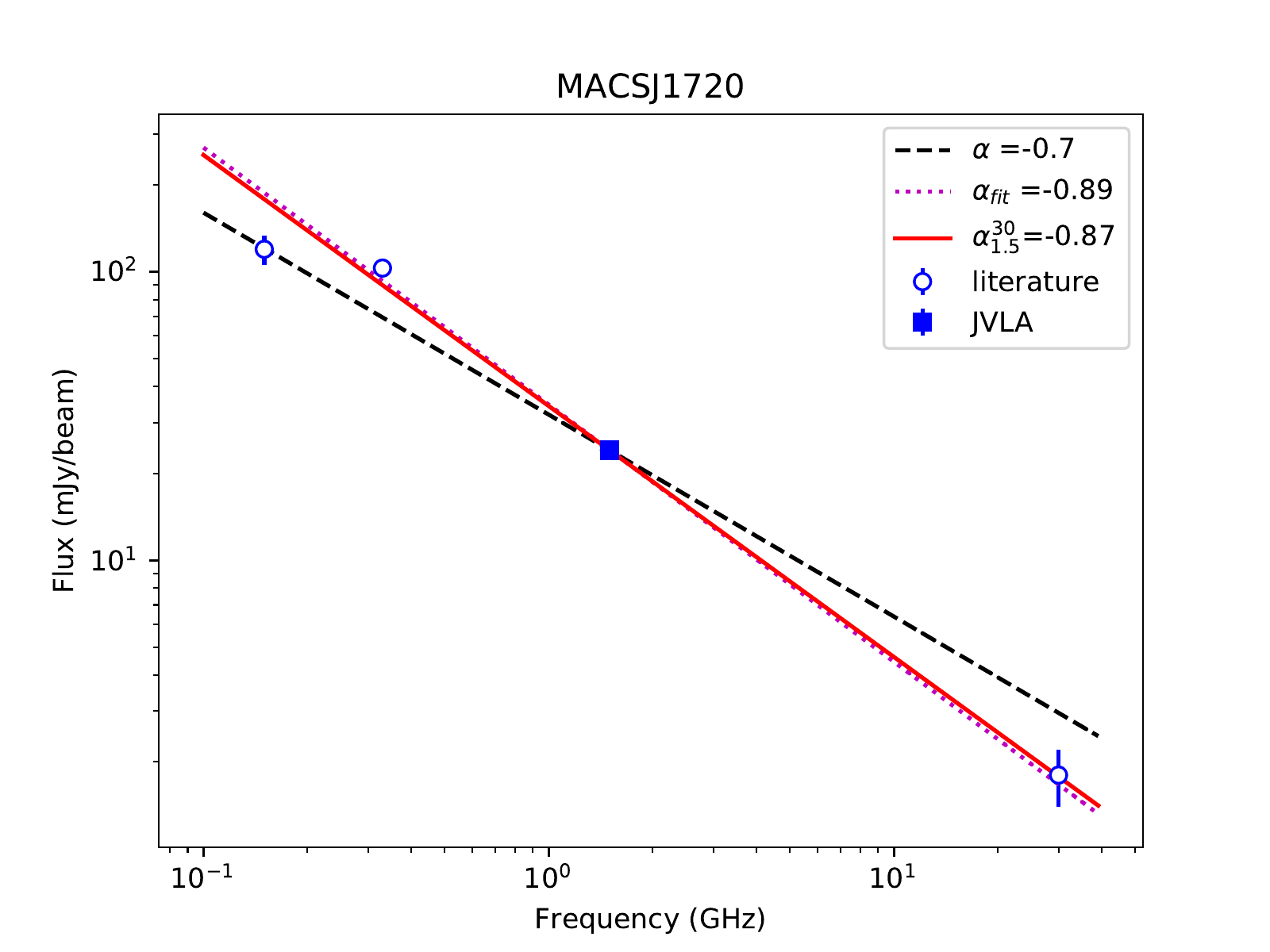}
\includegraphics[width=0.48\textwidth]{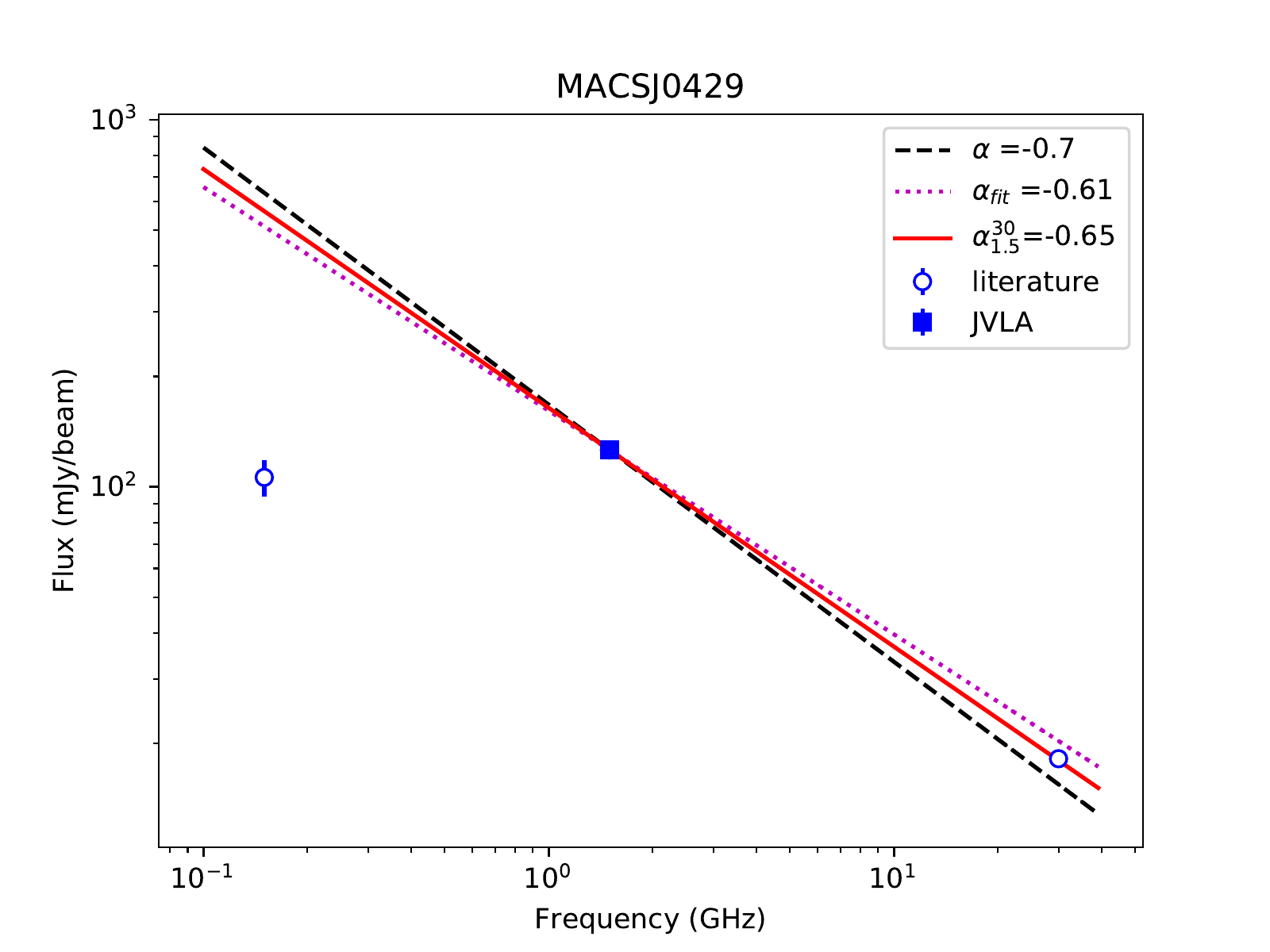}
\includegraphics[width=0.48\textwidth]{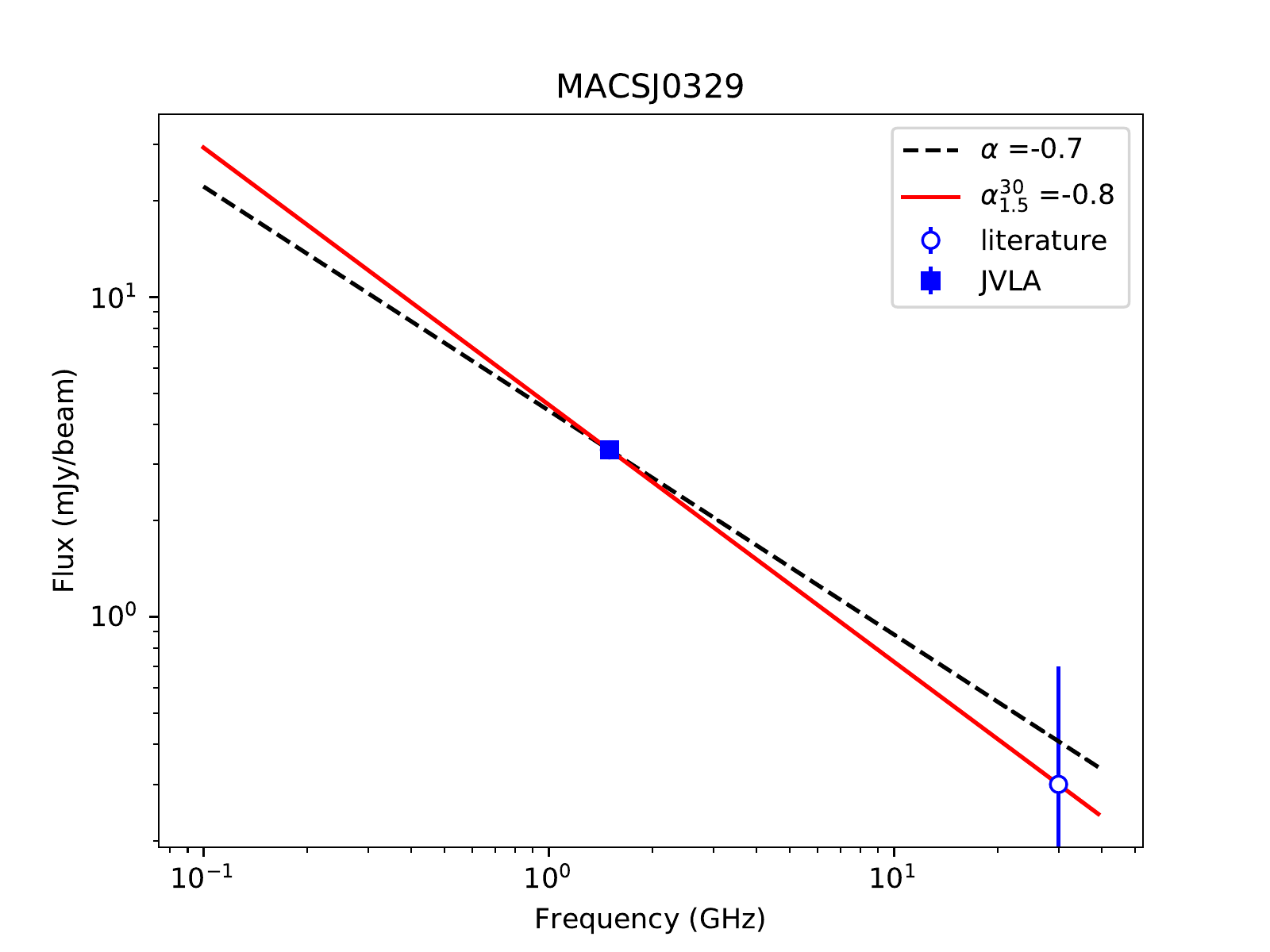}
\label{sed_1}
\end{figure*}

In this Appendix we show the radio SED of our BCGs including all the flux density values published in the literature in the 
150 MHz-30 GHz range, complementing the 1.5 GHz JVLA measurements presented in this work.  We show the comparison 
of our 1.5 and 30 GHz ratio to the slope of the best-fit power law including all the flux measurements.  We do not aim at a 
comprehensive description of the radio SEDs, given the uneven frequency sampling of the different sources and the lack of a 
uniform angular resolution at different frequencies.  Our goal here is simply to show the level of accuracy of our spectral index 
$\alpha^{30}_{1.5}$ as a proxy of the average spectral slope.   In Figure \ref{sed_1} we show the radio SEDs of BCGs observed 
with our JVLA program, while in Figure \ref{sed_2} we show the radio SEDs of BCGs with FIRST or NVSS detection only. Only BCGs 
with a measured $\alpha_{1.5}^{30}$ are shown.

\renewcommand{\thefigure}{\arabic{figure} (Cont.)}
\addtocounter{figure}{-1}

\begin{figure*}[htp]
\caption{ }
\includegraphics[width=0.48\textwidth]{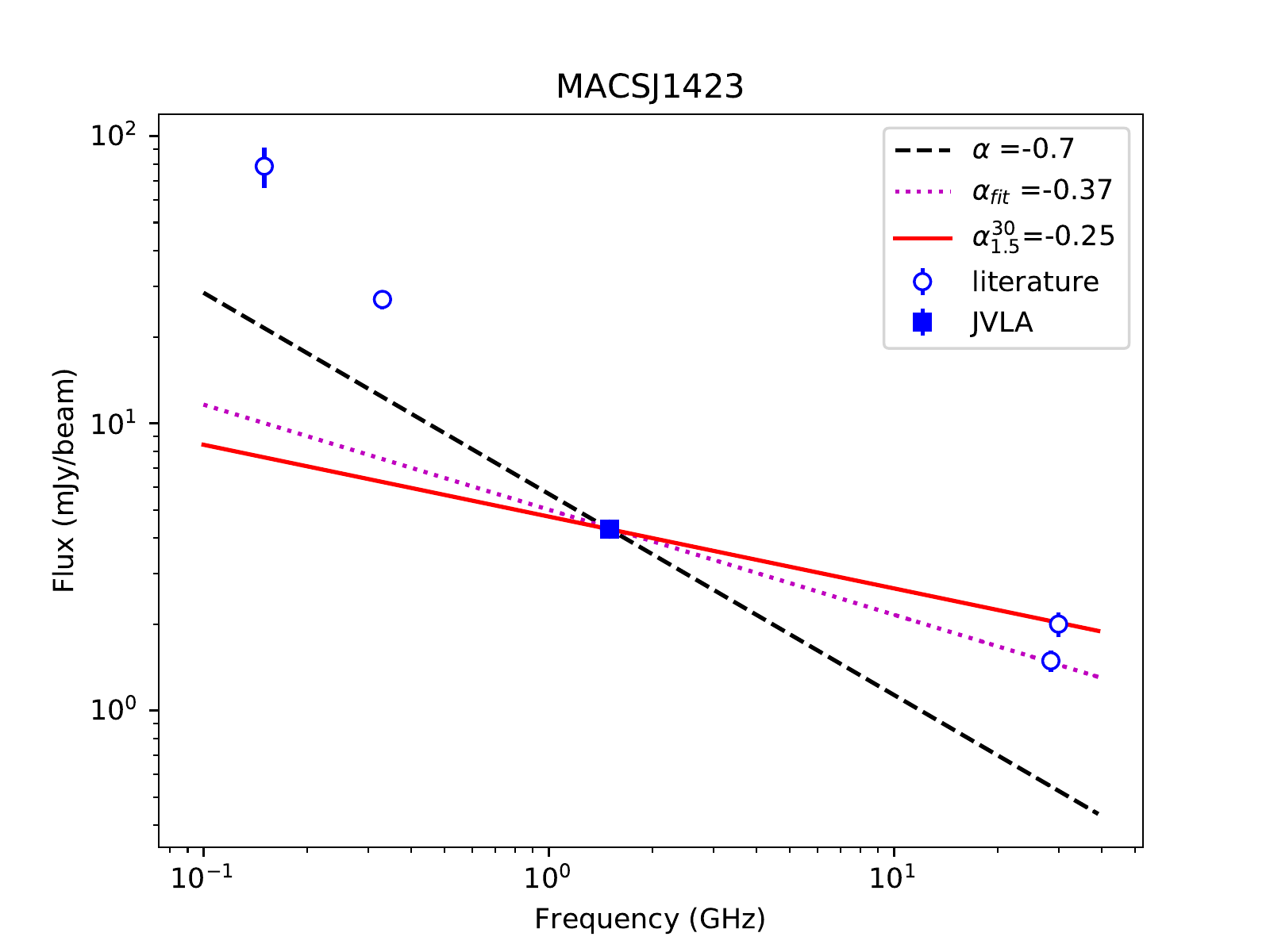}
\end{figure*}

\renewcommand{\thefigure}{\arabic{figure}}

\begin{figure*}[htp]
\caption{Radio SED of BCGs in the CLASH relaxed sample not observed with our JVLA program.  The dashed black line shows the 
reference slope normalized to the 1.5 GHz flux density, while the red solid line shows the slope corresponding to 
$\alpha_{1.5}^{30}$.  The magenta dotted line, when present, shows the
best-fit power law obtained using all the available flux measurements.}
\includegraphics[width=0.48\textwidth]{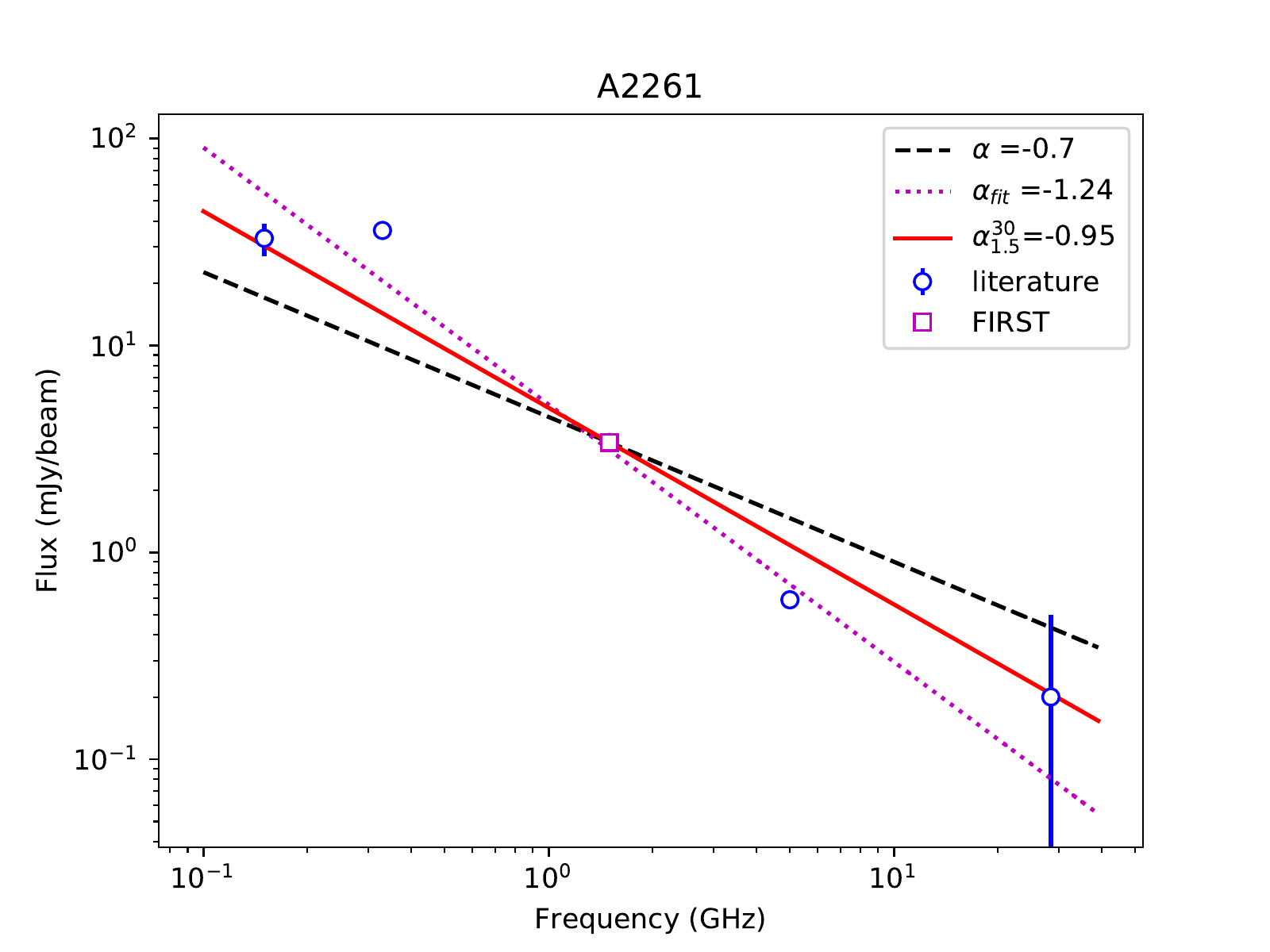}
\includegraphics[width=0.48\textwidth]{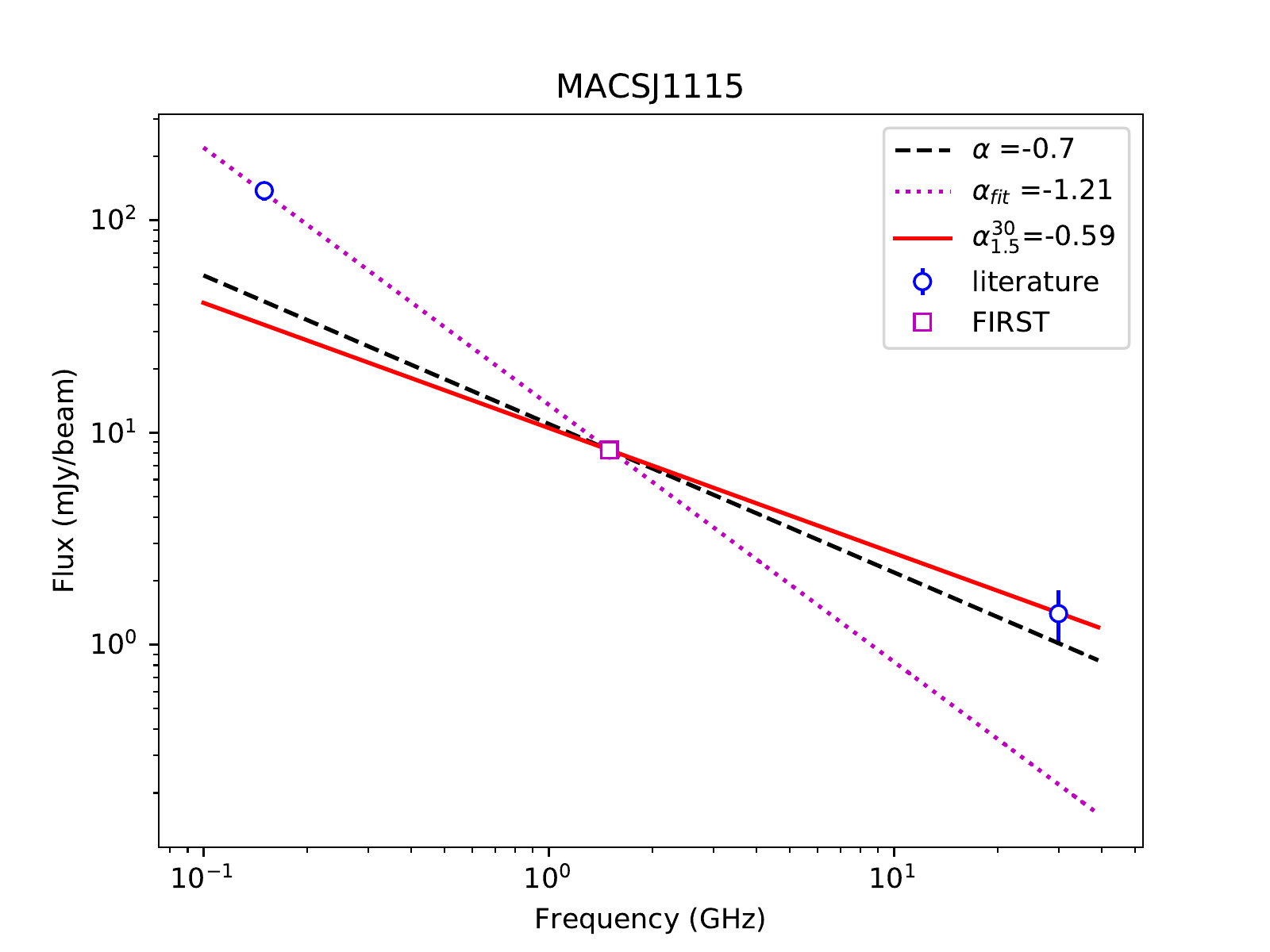}
\includegraphics[width=0.48\textwidth]{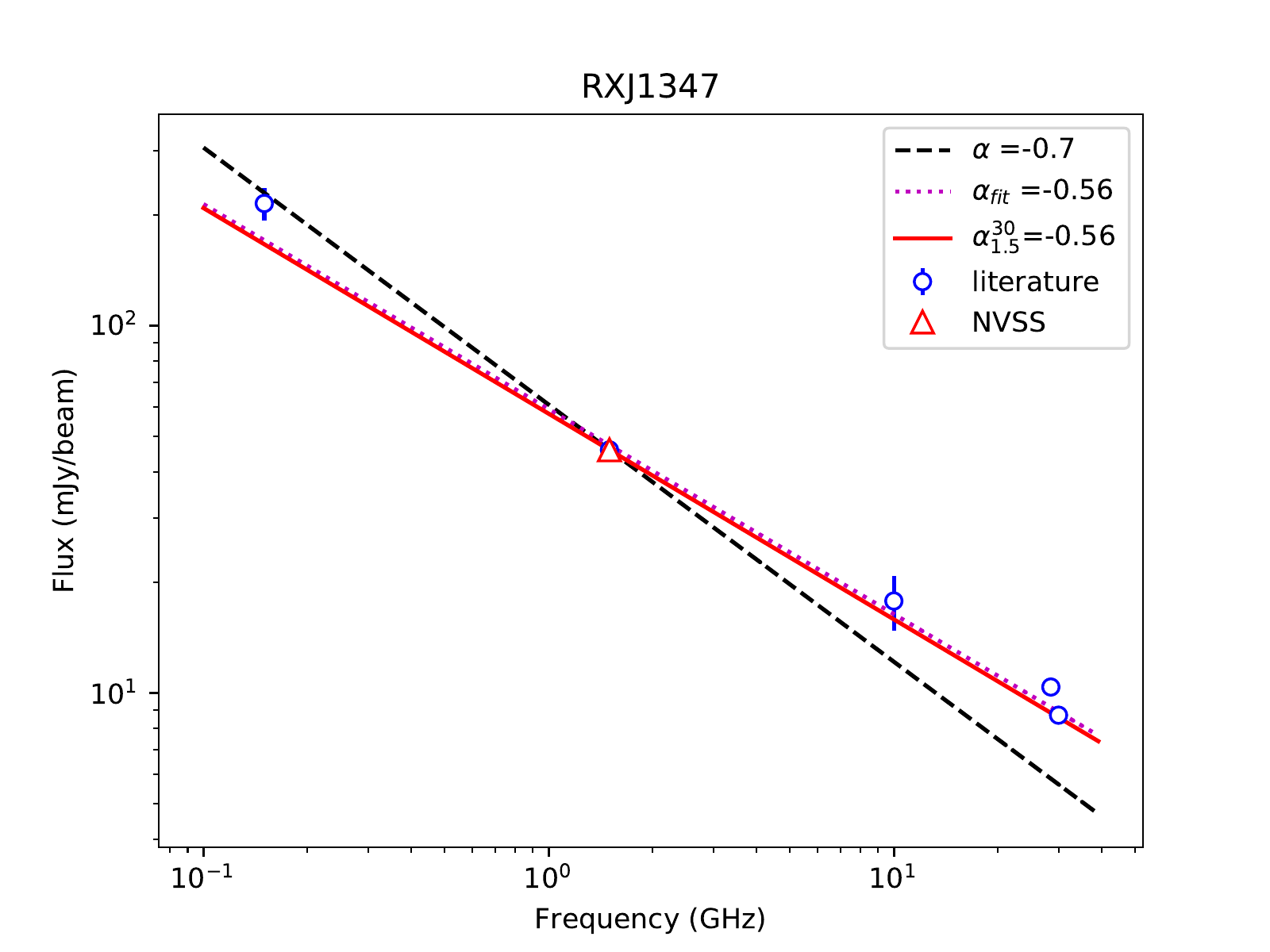}
\includegraphics[width=0.48\textwidth]{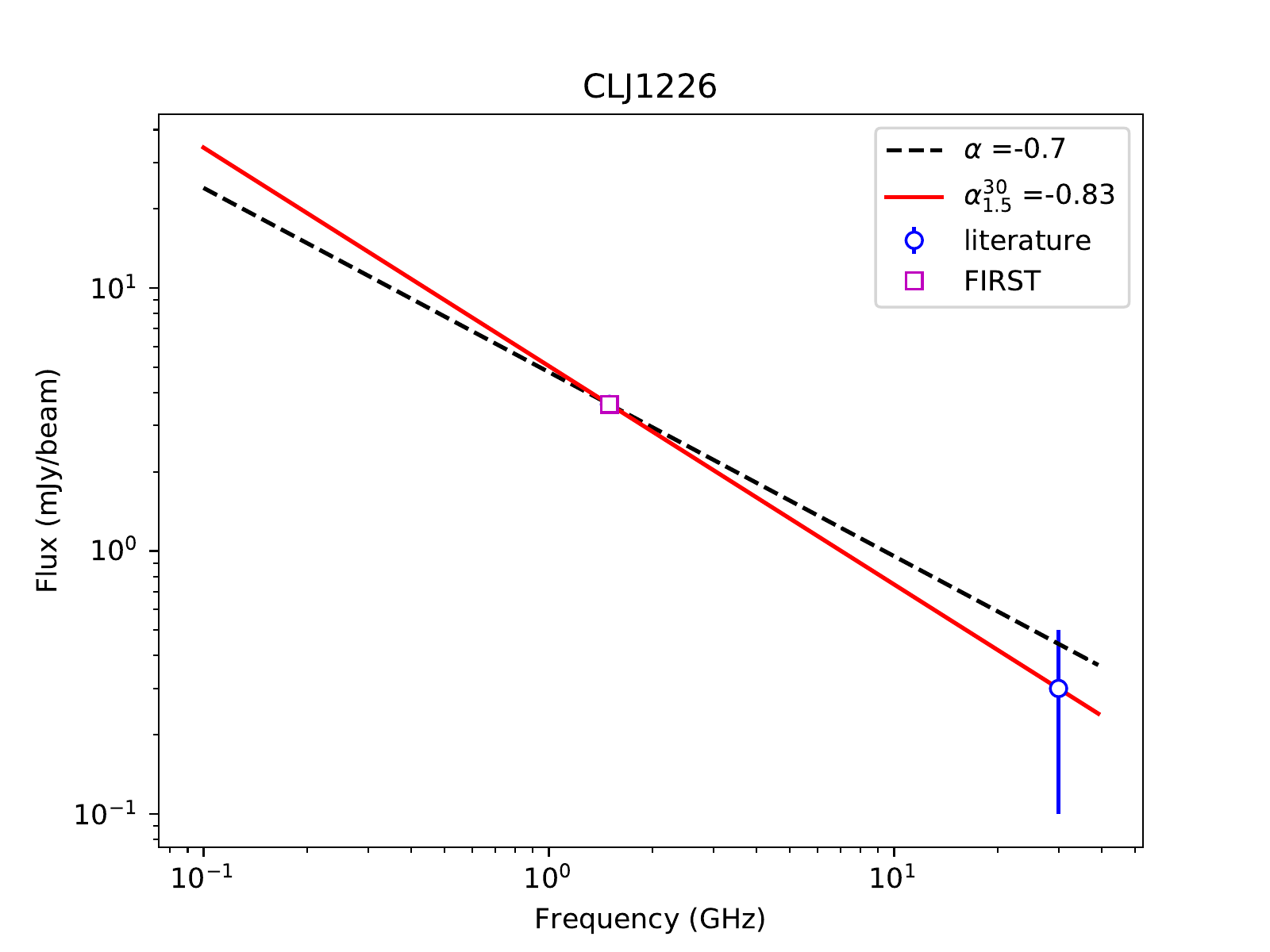}
\label{sed_2}
\end{figure*}

\end{document}